\documentclass[preprint]{imsart}
\pdfoutput=1
\usepackage[top=2.8cm, bottom=2.8cm, left=2.8cm, right=2.8cm]{geometry}
\usepackage[utf8]{inputenc}
\usepackage{xr}

\usepackage{tikz}

\usepackage{amsthm,amsmath,amssymb, booktabs}
\RequirePackage{threeparttable}
\usepackage{parskip}
\usepackage[authoryear]{natbib}
\usepackage{hypernat}
\usepackage{marvosym}
\usepackage[hang,small,bf]{caption}
\usepackage{mathtools}

\usepackage{hyperref}
\hypersetup{
    colorlinks,%
    citecolor=blue,%
    filecolor=black,%
    linkcolor=blue,%
    urlcolor=blue
}
\usepackage{hypernat}
\usepackage[none]{hyphenat}
\usepackage{enumerate}
\usepackage{color}
\usepackage{bbm}
\usepackage{thmtools}
\usepackage{xcolor}
\usepackage{subcaption}
\usepackage{textcomp}
\usepackage{graphicx}
\usepackage{float}
\usepackage{algpseudocode}
\usepackage{algorithm}

\RequirePackage{adjustbox}
\RequirePackage{blindtext}
\newcommand{\indep}{\rotatebox[origin=c]{90}{$\models$}}
\RequirePackage{color}   
\numberwithin{equation}{section}

\newtheoremstyle{exampstyle}
{8pt} 
{8pt} 
{\it} 
{} 
{\bfseries} 
{.} 
{.5em} 
{} 

\theoremstyle{exampstyle}

\newtheorem{theorem}{Theorem}[section]

\newtheorem{lemma}{Lemma}
\newtheorem{corollary}[theorem]{Corollary}

\startlocaldefs

\newcommand{\eat}[1]{}

\renewcommand{\bar}[1]{\overline{#1}}
\renewcommand{\hat}[1]{\widehat{#1}}

\theoremstyle{plain}

\def\beq{\begin{equation}}
\def\eeq{\end{equation}}
\def\ba{\begin{enumerate}[(a)]}
\def\bei{\begin{enumerate}[(i)]}
\def\be{\begin{enumerate}[(1)]}
\def\ee{\end{enumerate}}
\def\bi{\begin{itemize}}
\def\ei{\end{itemize}}
\def\beg{\begin{eg}}
\def\eeg{\end{eg}}
\def\bd{\begin{defn}}
\def\ed{\end{defn}}
\def\bt{\begin{thm}}
\def\et{\end{thm}}
\def\bl{\begin{lemma}}
\def\el{\end{lemma}}
\def\bfac{\begin{fact}}
\def\efac{\end{fact}}

\def\bc{\begin{corollary}}
\def\ec{\end{corollary}}
\def\bp{\begin{prop}}
\def\ep{\end{prop}}
\def\bo{\begin{observe}}
\def\eo{\end{observe}}
\def\bas{\begin{assumption}}
\def\eas{\end{assumption}}

\endlocaldefs

\begin{document}

\begin{frontmatter}
\title{Robust Bayesian Inference for Big Data: Combining Sensor-based Records with Traditional Survey Data}


\runtitle{Robust Bayesian Inference for Big Data}

\begin{aug}
\author[A]{\fnms{Ali} \snm{Rafei}},   
\author[B]{\fnms{Carol A. C.} \snm{Flannagan}},   
\author[A]{\fnms{Brady T.} \snm{West}}   
\and
\author[A]{\fnms{Michael R.} \snm{Elliott}\thanks{\textit{Corresponding author; address: 426 Thompson St. Ann Arbor, MI 48109. Rm 4068 ISR, email: \href{mailto:mrelliot@umich.edu}{mrelliot@umich.edu}.}}}   
\address[A]{Survey Methodology Program, University of Michigan}
\vspace{-10pt}
\address[B]{University of Michigan Transportation Research Institute}
\vspace{10pt}
\end{aug}

\begin{abstract}
Big Data often presents as massive non-probability samples. Not only is the selection mechanism often unknown, but larger data volume amplifies the relative contribution of selection bias to total error. Existing bias adjustment approaches assume that the conditional mean structures have been correctly specified for the selection indicator or key substantive measures. In the presence of a reference probability sample, these methods rely on a pseudo-likelihood method to account for the sampling weights of the reference sample, which is parametric in nature. Under a Bayesian framework, handling the sampling weights is an even bigger hurdle. To further protect against model misspecification, we expand the idea of double robustness such that more flexible non-parametric methods as well as Bayesian models can be used for prediction. In particular, we employ Bayesian additive regression trees, which not only capture non-linear associations automatically but permit direct quantification of the uncertainty of point estimates through its posterior predictive draws. We apply our method to sensor-based naturalistic driving data from the second Strategic Highway Research Program using the 2017 National Household Travel Survey as a benchmark.
\end{abstract}


\end{frontmatter}


\section{Introduction}\label{S:1}
The 21\textsuperscript{st} century is witnessing a re-emergence of non-probability sampling in various domains \citep{murdoch2013inevitable, daas2015big, lane2016big, senthilkumar2018big}. Probability sampling is facing new challenges, mainly because of a steady decline in response rates \citep{groves2011three, johnson2017big, miller2017there}. At the same time, new modes of data collection via sensors, web portals, and smart devices have emerged that routinely capture a variety of human activities. These automated processes have led to an ever-accumulating massive volume of unstructured information, so-called ``Big Data'' \citep{couper2013sky, kreuter201412, japec2015big}. Being easy to access, inexpensive to collect, and highly detailed, this broad range of data is perceived to be valuable for producing official statistics as an alternative or supplement to probability surveys \citep{struijs2014official, kitchin2015opportunities}.
However, ``Big Data'' typically have a self-selecting data-generating process, which can lead to biased estimates. When this is the case, the larger data volume in the non-probability sample increases the relative contribution of selection bias to absolute or squared error. 
\cite{meng2018statistical} call this phenomenon a ``Big Data Paradox'', and these authors show both theoretically and empirically that the impact of selection bias on the effective sample size can be extremely large.\par

The motivating application in this article comes from naturalistic driving studies (NDS), which are one real-world example of Big Data for rare event investigations. Since traffic collisions are inherently rare events, measuring accurate pre-crash behaviors as well as exposure frequency in normal driving demands accurate long-term follow-up of the population of drivers. Thus, NDS are designed to continually monitor 

\pagebreak

\noindent
drivers' behavior via in-vehicle sensors, cameras, and advanced wireless technologies \citep{guo2009modeling}. The detailed information collected by NDS are considered a rich resource for assessing various aspects of 
transportation such as traffic safety, crash causality, and travel patterns \citep{huisingh2018distracted, tan2017development}.
In particular, we consider the sensor-based Big Data from the second phase of the Strategic Highway Research Program (SHRP2), which is the largest NDS conducted to date. This study recruited a convenience sample from geographically restricted regions (6 US states:  Florida, Indiana, New York, North Carolina, Pennsylvania, and Washington) and attempted to oversample both younger and older drivers,
leading to potential selection bias in the sample mean of some trip-related variables \citep{antin2015naturalistic}. 
To deal with this, we employ the 2017 National Household Travel Survey (NHTS) as a ``reference survey'', which can serve as a probability sample representing the population of American drivers \citep{santos2011summary}. While daily trip measures in SHRP2 are recorded via sensors, NHTS asks respondents to self-report their trip measures through an online travel log. By analyzing the aggregated data at the day level, we develop adjusted sensor-based estimates from SHRP2 for measures such as frequency of trips, trip duration, trip speed, and starting time of trip that can be compared with
self-reported weighted estimates in NHTS to assess the performance of our proposed methods in terms of bias and efficiency, as well as estimates of maximum speed, brake use per mile driven, and stop time that are available only in SHRP2. \par

Standard design-based approaches cannot be applied to non-probability samples for the simple reason that the probabilities of selection are unknown \citep{chen2019doubly}. Thus the American Association for Public Opinion Research (AAPOR) task force on non-probability samples recommends that adjustment methods should rely on models and external auxiliary information \citep{baker2013report}. In the presence of a relevant probability sample with a set of common auxiliary variables, which is often called a ``reference survey'', two general approaches can be taken: (1) \emph{prediction modeling} (PM)--fitting models on the non-probability sample to predict the response variable for units in the reference survey \citep{rivers2007sampling, kim2012combining, wang2015forecasting, kim2018combining}, and (2) \emph{quasi-randomization} (QR)--estimating the probabilities of being included in the non-probability sample, also known as propensity scores (PS), while treating the Big Data as a quasi-random sample \citep{lee2006propensity, lee2009estimation, valliant2011estimating}. Our focus is on the PM setting, since in our application the key measures of interest are not available in the probability (reference) survey, and our goal is to use prediction to impute them.\par

Correct specification of the model predicting the outcome is critical for imputation. To help relax this assumption, the PM approach can be combined with the QR method, in a way that the adjusted estimate of a population quantity is consistent if either the propensity or the outcome model holds. Augmented inverse propensity weighting (AIPW) was the first of these so-called ``doubly-robust'' (DR) methods \citep{robins1994estimation}, with applications to causal inference \citep{scharfstein1999adjusting, bang2005doubly, tan2006distributional, kang2007demystifying, tan2018robust} and adjustment for non-response bias \citep{kott1994note, kim2006imputation, kott2006using, haziza2006nonresponse, kott2010using}. Further extension to multiple robustness has been developed by \cite{han2013estimation}, where multiple models are specified and consistency is obtained as long as at least one of the models is correctly specified. \cite{chen2019doubly} offer further adjustments to adapt the AIPW estimator to a non-probability sampling setting where an external benchmark survey is available. While their method employs a modified pseudo-likelihood approach to estimate the selection probabilities for the non-probability sample, a parametric model is used to impute the outcome for units of the reference survey.

In a non-probability sample setting, \cite{rafei2020big} utilized BART in the QR approach outlined in \cite{Elliott2017Inference}. In this paper, we extend \cite{rafei2020big} in two major ways: first, by blending the QR and PS methods into a novel DR method that is made even more robust by using BART, which provides a strong non-parametric predictive tool by automatically capturing non-linear associations as well as high-order interactions \citep{chipman2007bayesian}. The proposed method separates the propensity model from the sampling weights in a two-step process, allowing for a broader range of models to be utilized for imputing the missing inclusion probabilities. 
This allows us to consider both parametric (linear and generalized linear models) and non-parametric (BART) models for both propensity and outcome. In addition, the posterior predictive distribution produced by BART makes it easier to quantify the uncertainty due to the imputation of pseudo-weights and the outcome variable \citep{tan2018robust, kaplan2012two}. Second, we derive asymptotic variance estimators for the previously proposed QR estimators in \cite{rafei2020big} as well as proofs of the double robustness of the proposed DR estimators.\par

The rest of the article is organized as follows. In Section~\ref{S:2}, we develop the theoretical background behind the proposed methods and associated variance estimators. A simulation study is designed in Section~\ref{S:3} to assess the repeated sampling properties of the proposed estimator, i.e. bias and efficiency. Section~\ref{S:4} 
uses the NHTS to develop adjusted estimates from the SHRP2 using the methods discussed and developed in the previous sections. All the statistical analyses in both the simulations and empirical studies have been performed using R version 3.6.1. The annotated R code is available for public use at \href{https://github.com/arafei/drinfNP}{https://github.com/arafei/drinfNP}. Finally, Section~\ref{S:6} reviews the strengths and weaknesses of the study in more detail and suggests some future research directions.\par


\section{Methods}\label{S:2}
\subsection{Notation}\label{S:2.1}
Denote by $U$ a finite population of size $N<\infty$. We consider the values of a scalar outcome variable, $y_i$, $i=1, 2, ... , N$ and $x_i=[1, x_{i1}, x_{i2}, ... , x_{ip}]$, the values of a $p$-dimensional vector of relevant auxiliary variables, $X$. Let $S_B$ be a non-probability sample of size $n_B$ selected from $U$. The goal is to estimate an unknown finite population quantity, e.g. $Q(Y)$. Here, the quantity of interest is considered to be the finite population mean that is a function of the outcome variable, i.e. $Q(Y)= \bar y_U=\sum_{i=1}^Ny_i/N$. Suppose $\delta^B_{i}=I(i\in S_B)$ $(i=1, ... , N)$ is the inclusion indicator variable of the ``big data'' survey $S_B$ of size $n_B$ in $U$. Further, we initially assume that given $X$, elements in $S_B$ are independent draws from $U$, but later, we relax this assumption by considering $S_B$ to have a single-stage clustering design as is the case in the real-data application of this article.\par

Suppose $S_R$ is a parallel reference survey of size $n_R$, for which the same set of covariates, $X$, has been measured. We also define $d_i=[1, d_{i1}, d_{i2}, ... , d_{iq}]$, the values of a $q$-dimensional vector of design variables for the reference survey. 
We assume $y_i$ is unobserved for $i\in S_R$; otherwise inference could be directly drawn based on $S_R$. Also, let $\delta^R_i=I(i\in S_R)$ denote the inclusion indicator variable associated with $S_R$ for $i\in U$. To avoid unnecessary complexity, we assume that units of $S_R$ are selected independently. Being a full probability sample implies that the selection mechanism in $S_R$ is ignorable given its design features, i.e. $f(\delta^R_i| y_i, d_i)=f(\delta^R_i|d_i)$ for $i\in U$, where $d_i$ denotes a $q$-dimensional vector of associated design variables. Thus, one can define the selection probabilities and sampling weights in $S_R$ as $\pi^R_i=p(\delta^R_i=1|d_i)$ and $w^R_i=1/\pi^R_i$, respectively, for $i\in U$, which we assume are known.\par

While $X$ and $D$ may overlap or correlate, we define $x^*_i=[x_i, d_i]^T$, the $(p+q)$-dimensional vector of all auxiliary variables associated with $S_B$ and $S_R$. To be able to make unbiased inference for $S_B$, we consider the following assumptions for $S_B$:
\begin{enumerate}
    \item[\textbf{C1.}] \textbf{Positivity}---$S_B$ actually does have a probabilistic sampling mechanism, albeit unknown. That means $p(\delta^B_i=1|x_i)>0$ for all possible values of $x_i$ in $U$.
    \item[\textbf{C2.}] \textbf{Ignorability}---the selection mechanism of $S_B$ is fully governed by $X$, which implies $Y\indep\delta^B|X$. Then, for $i\in U$, the unknown ``pseudo-inclusion'' probability associated with $S_B$ is defined as $\pi^A_i=p(\delta^B_i=1|x_i)$.
    \item[\textbf{C3.}] \textbf{Independence}---conditional on $X^*$, $S_R$ and $S_B$ are selected independently, i.e. $\delta^B\indep\delta^R|X^*$.
\end{enumerate}
Note that the first two assumptions are collectively called ``strong ignorability'' by \cite{rosenbaum1983central}. Considering \textbf{C1-C3}, the joint density of $y_i$, $\delta^B_i$ and $\delta^R_i$ can be factorized as below:
\begin{equation}\label{eq:2.1}
f(y_i, \delta^B_i, \delta^R_i| x^*_i; \theta, \beta, \gamma)=f(y_i|x_i^*; \theta)f(\delta^B_i|x_i; \beta)f(\delta^R_i|d_i; \gamma), \hspace{4mm} \forall i\in U
\end{equation}
where $\Psi=(\theta, \beta, \gamma)$ are some distributional parameters. While $\theta$ and $\beta$ are unknown, $\gamma$ may be known as $S_R$ is a probability sample. A QR approach involves modeling $f(\delta^B_i|x^*_i; \beta)$, whereas a PM approach deals with modeling $f(y_i|x^*_i; \theta)$. 


Now suppose $S_B$ and $S_R$ have trivial overlap, i.e. $p(\delta^B_i+\delta^R_i=2)\approx 0$. 
This assumption is reasonable when the sampling fraction in both samples is small. Note that under the ignorable assumption, the propensity model for $S_B$ depends on $X$ observed for the entire population. Thus, given the combined sample, $S=S_B\cup S_R$, with $n=n_B+n_R$ being the sample size, it is reasonable to expect that the pseudo-inclusion probabilities, $\pi^B_i$'s, are a function of both $x_i$ and $d_i$ for $i\in S$. Let $z_i=I(i\in S_B|\delta_i=1)$ be the indicator of subject $i$ belonging to the non-probability sample in the combined sample where $\delta_i =\delta^B_i+\delta^R_i$. Note that since $S_B\cap S_R=\emptyset$, $\delta_i$ can take values of either $0$ or $1$ as below:
\begin{equation*}
\delta_i = 
    \begin{cases}
        0, & \hspace{3mm}\text{if}\hspace{3mm}\delta^R_i=0\hspace{3mm} and \hspace{3mm}\delta^B_i=0\\
        1, & \hspace{3mm}\text{if}\hspace{3mm}\delta^R_i=1\hspace{6mm} or \hspace{3mm}\delta^B_i=1
    \end{cases}       
\end{equation*}

\begin{figure}[htp]
\centering\includegraphics[width=0.6\linewidth]{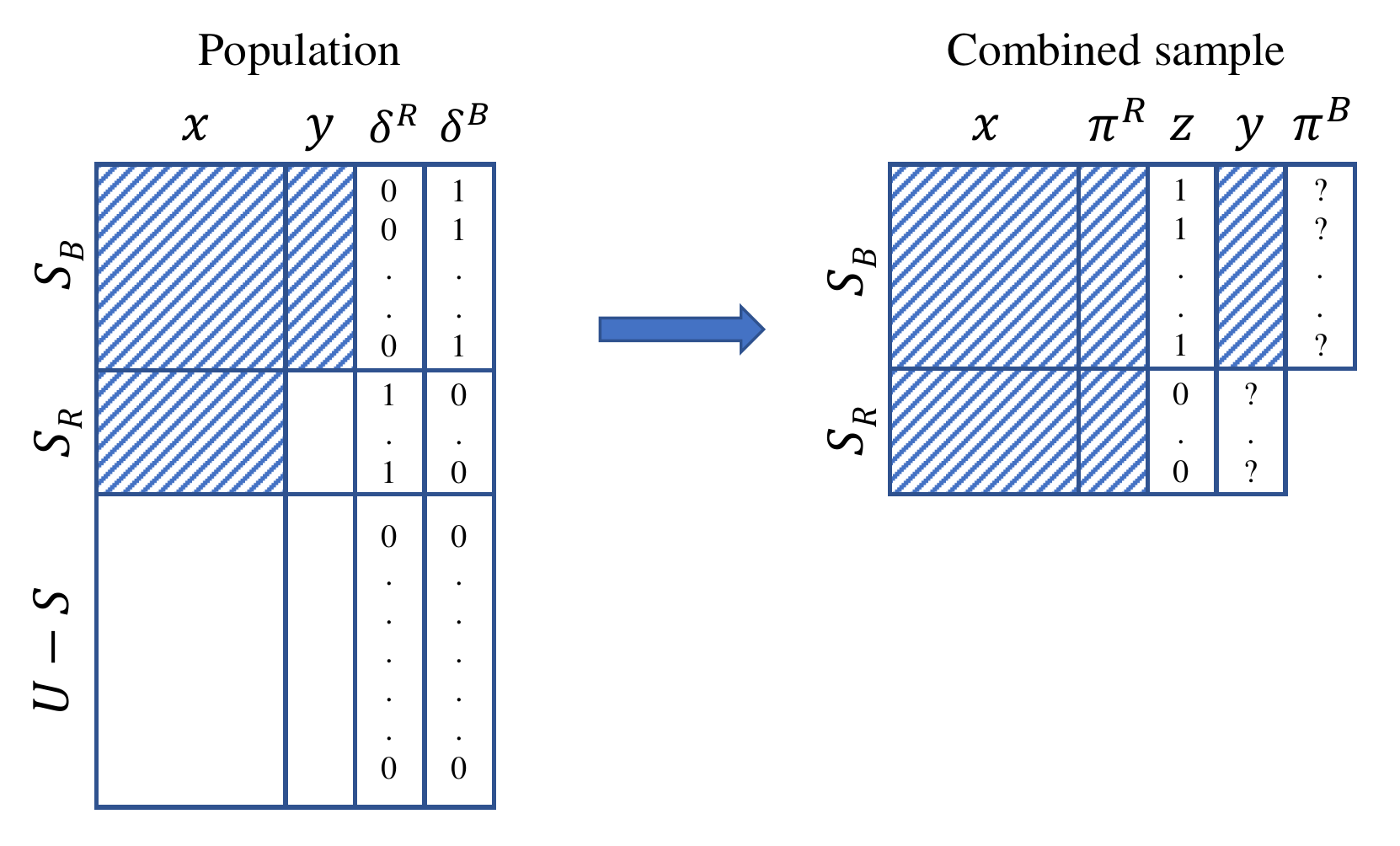}
\caption{Data structure in the population and the combined sample}\label{fig:1}
\end{figure}

Figure~\ref{fig:1} illustrates the data structure in both the finite population and the combined sample.

\subsection{Quasi-randomization}\label{S:2.2}
In QR, the non-probability sample is treated as if the self-selection mechanism of population units mimics a stochastic process, but with unknown selection probabilities. Then, attempts are made to estimate these missing quantities in $S_B$ based on external information. Conditional on $x^*_i$, suppose $\pi^B_i$ follows a logistic regression model in the finite population. We have
\begin{equation}\label{eq:2.2}
\pi^B(x_i; \beta)=p(\delta^B_i=1 | x_i; \beta)=\frac{\exp\{\beta^T_0x_i\}}{1+\exp\{\beta^T_0x_i\}}, \hspace{5mm} \forall i\in U
\end{equation}
where $\beta$ is a vector of model parameters in $U$. Using a modified pseudo-maximum likelihood approach (PMLE), \cite{chen2019doubly} demonstrate that, given $S$, a consistent estimate of $\beta$ can be obtained by solving the following estimating equation with respect to $\beta$:
\begin{equation}\label{eq:2.3}
U(\beta)=\sum_{i=1}^{n_B}x_i - \sum_{i=1}^{n_R} \pi^B(x_i;\beta)x_i/\pi^R_i=0
\end{equation}
The estimates of the $\pi^B_i$'s, which we also call propensity scores (PS), are obtained by plugging the solution of Eq.~\ref{eq:2.3}, i.e. $\hat\beta$, into Eq.~\ref{eq:2.2}. It is important to note that the proposed PS estimator by \cite{chen2019doubly} depends implicitly on $d_i$ in addition to $x_i$, because we know that $\pi^R_i$ is a determinstic function of $d_i$ for $i\in U$. Under certain regularity conditions, the authors show that the inverse PS weighted (IPSW) mean from $S_B$ yields a consistent and asymptotically unbiased estimate for the population mean.\par 

Obviously, the possible solutions of Eq.~\ref{eq:2.3} are not a typical output of logistic regression procedures in the existing standard software. With one additional assumption, which is mutual exclusiveness of the two samples, i.e. $S_B\cap S_R=\emptyset$, we show that estimating $\pi^B_i$'s can be reduced to modeling $Z_i$ for $i\in S$ instead of modeling $\delta^B_i$ for $i\in U$. Intuitively, one can view the selection process of the $i$-th population unit in $S_B$ as being initially selected in the joint sample ($\delta_i=1$) and then being selected in $S_B$ given the combined sample ($Z_i=1$). By conditioning on $x^*_i$, the selection probabilities in $S_B$ are factorized as
\begin{equation}
\begin{aligned}\label{eq:2.4}
p(\delta^B_i=1 | x^*_i)&=p(\delta^B_i=1, \delta_i=1 | x^*_i)\\
&=p(\delta^B_i=1 | \delta_i=1, x^*_i)p(\delta_i=1 | x^*_i)\\
& = p(Z_i=1|x^*_i)p(\delta_i=1 | x^*_i)\hspace{5mm}i\in S
\end{aligned}
\end{equation}
Note that the last expression in Eq.~\ref{eq:2.4} results from the definition of $Z_i$ given $S$. The same factorization can be derived for the selection probabilities in $S_R$. Thus, we have
\begin{align}\label{eq:2.5}
p(\delta^R_i=1 | x^*_i)=p(Z_i=0 | x^*_i)p(\delta_i=1 | x^*_i)
\end{align}
By dividing the two sides of the equations~\ref{eq:2.4} and~\ref{eq:2.5}, one can get rid of $p(\delta_i=1 | x^*_i)$ and obtain the pseudo-selection probabilities in $S_B$ as below:
\begin{equation}\label{eq:2.6}
p(\delta^B_i=1 | x^*_i)=p(\delta^R_i=1 | x^*_i)\frac{p(Z_i=1|x^*_i)}{p(Z_i=0|x^*_i)}
\end{equation}
It is clear that $p(\delta^R_i=1 | x^*_i)=\pi^R_i$ as $x^*_i$ contains $d_i$ and the sampling design of $S_R$ is known given $d_i$. As will be seen in Section~\ref{S:2.4}, conditioning on $d_i$ is vital for the DR estimator, as Chen's method is limited to situations where the dimension of the auxiliary variables must be the same in QR and PM.\par

Note that Eq.~\ref{eq:2.6} is identical to the pseudo-weighting formula \cite{Elliott2017Inference} derive for a non-probability sample. Unlike the PMLE approach, modeling $Z_i$ in $S$ can be performed using the standard binary logistic regression or any alternative classification methods, such as supervised machine learning algorithms. Under a logistic regression model, we have
\begin{equation}\label{eq:2.7}
    p(Z_i=1|x^*_i)=\frac{\exp\{\beta^T_1x_i^*\}}{1+\exp\{\beta^T_1x_i^*\}}
\end{equation}
where $\beta_1$ denotes the vector of model parameters being estimated via maximum likelihood estimation (MLE). Hence, in situations where $\pi^R_i$ is known or can be calculated for $i\in S_B$, the estimate of $\pi^B_i$ for $i\in S_B$ is given by
\begin{equation}\label{eq:2.8}
    \hat\pi^B_i=\pi^R_i\exp\{\hat\beta^T_1x_i^*\}=\pi^R_i\frac{p_i(\hat\beta_1)}{1-p_i(\hat\beta_1)}
\end{equation}
where $\hat\beta_1$ denotes the MLE estimate of the logistic regression model parameters, and $p_i(\hat\beta_1)$ is a shorthand of $p(Z_i=1|x^*_i;\hat\beta_1)$. Intuitively, one can envision that the first factor in~\ref{eq:2.8} treats $S_B$ as if it is selected under the design of $S_R$, and the second factor attempts to balance the distribution of $x$ in $S_B$ with respect to that in $S_R$.\par

Having $\pi^B_i$ estimated based on~\ref{eq:2.8} for all $i\in S_B$, one can construct the H{\'a}jek-type pseudo-weighted estimator for the finite population mean as below:
\begin{equation}\label{eq:2.9}
    \hat{\bar{y}}_{PAPW}=\frac{1}{\hat N_B}\sum_{i=1}^{n_B}\frac{y_i}{\hat\pi^B_i}
\end{equation}
where $\hat N_B=\sum_{i=1}^{n_B}1/\pi^B_i$. Hereafter, we refer to the estimator in~\ref{eq:2.9} as propensity-adjusted probability weighting (PAPW). Under mild regularity conditions, the ignorable assumption in $S_B$ given $x$, the logistic regression model and the additional assumption of $S_B\cap S_R=\emptyset$, Appendix~\ref{S:7.1} shows that this estimator is consistent and asymptotically unbiased for $\bar y_U$.  Further, when $\pi_i^R$ is known, the sandwich-type variance estimator for $\hat{\bar{y}}_{PAPW}$ is given by
\begin{equation}\label{eq:2.27}
\begin{aligned}
\widehat{Var}\left(\hat{\bar y}_{PAPW}\right)=
\frac{1}{N^2}\sum_{i=1}^{n_B} \big\{1-\hat\pi^B_i\big\}\left(\frac{y_i - \hat{\bar y}_{PAPW}}{\hat\pi^B_i}\right)^2 &-2\frac{\hat b^T}{N^2}\sum_{i=1}^{n_B}\big\{1-p_i(\hat\beta_1)\big\}\left(\frac{y_i-\hat{\bar y}_{PAPW}}{\hat\pi^B_i}\right)x^*_i\\
&+\hat b^T\left[\frac{1}{N^2}\sum_{i=1}^np_i(\hat\beta_1)x^*_ix_i^{*T}\right]\hat b
\end{aligned}
\end{equation}
where
\begin{equation}\label{eq:2.28}
    \hat b^T=\bigg\{\frac{1}{N}\sum_{i=1}^{n_B}\left(\frac{y_i-\hat{\bar y}_{PAPW}}{\hat\pi^B_i}\right)x^{*T}_i\bigg\}\bigg\{\frac{1}{N}\sum_{i=1}^n p_i(\hat\beta_1)x^*_ix_i^{*T}\bigg\}^{-1}
\end{equation}
where $\hat\pi^B_i$ is the estimated pseudo-selection probability based on~\ref{eq:2.9} for $i\in S_B$.
See Appendix~\ref{S:7.1} for the derivation.

In situations where $\pi^R_i$ is unknown for $i\in S_B$, \cite{Elliott2017Inference} suggest predicting this quantity for units of the non-probability sample. Note that, in this situation, it is no longer required to condition on $d_i$ in addition to $x_i$. Treating $\pi^R_i$ as a random variable for $i\in S_B$ conditional on $x_i$, one can obtain this quantity by regressing the $\pi^R_i$'s on the $x_i$'s in the reference survey. We have
\begin{equation}\label{eq:2.11}
\begin{aligned}
p(\delta^R_i=1|x_i) &=\int_0^1 p(\delta^R_i=1| \pi^R_i, x_i)p(\pi^R_i|x_i)d\pi^R_i\\
&=\int_0^1\pi^R_ip(\pi^R_i|x_i)d\pi^R_i\\
&=E(\pi^R_i|x_i) \hspace{5mm}i\in S_R.
\end{aligned}
\end{equation}
However, since the outcome is continuous bounded taking values within $(0, 1)$, fitting a $Beta$ regression model is recommended \citep{ferrari2004beta}. Note that, $\pi^R_i$ is fixed given $d_i$ as $S_R$ is a probability sample, but conditional on $x_i$, $\pi^R_i$ can be regarded as a random variable. \par

\cite{rafei2020big} call this approach propensity-adjusted probability prediction (PAPP). This two-step derivation of pseudo-inclusion probabilities is especially useful, as it separates sampling weights in $S_R$ from the propensity model computationally. When the true model is unknown, this feature enables us to fit a broader and more flexible range of models, such as algorithmic tree-based methods. It is worth noting that modeling $E(\pi^R_i|x_i)$ does not impose an additional ignorable assumption in $S_R$ given $x$, because in the extreme case if $\delta^R_i\indep x_i$, that means weighted and unweighted distributions of $x$ are identical in $S_R$, and therefore the $\pi^R_i$'s can be safely ignored in propensity modeling.\par


\subsection{Prediction modeling approach}\label{S:2.3}
An alternative approach to deal with selectivity in Big Data is modeling $f(y|x^*)$ \citep{smith1983validity}. In a fully model-based fashion, this essentially involves imputing $y$ for the population non-sampled units, $U-S_B$. When $x^*$ is unobserved for non-sampled units, it is recommended that a synthetic population is generated by undoing the selection mechanism of $S_R$ through a non-parametric Bayesian bootstrap method using the design variables in $S_R$ \citep{dong2014nonparametric, zangeneh2015bayesian}. In the non-probability sample context, \cite{Elliott2017Inference} propose an extension of the General Regression Estimator (GREG) when only summary information about $x^*$, such as totals, is known regarding $U$. In situations where an external probability sample is available with $x^*$ measured, an alternative is to limit the outcome prediction to the units in $S_R$, and then, use design-based approaches to estimate the population quantity \citep{rivers2007sampling, kim2018combining}.\par

However, to the best of our knowledge, none of the prior literature distinguish the role of $D$ from $X$ in the conditional mean structure of the outcome, while the likelihood factorization in Eq.~\ref{eq:2.1} indicates that predicting $y$ requires conditioning not only on $x$ but also on $d$. Suppose $U$ is a realization of a repeated random sampling process from a super-population under the following model:
\begin{equation}\label{eq:2.12}
\begin{aligned}
E(y_i|x^*_i; \theta)=m(x^*_i; \theta) \hspace{5mm} \forall i\in U
\end{aligned}
\end{equation}
where $m(x^*_i; \theta)$ can be either a parametric model with $m$ being a continuous differentiable function or an unspecified non-parametric form. Under the \emph{ignorable} condition where
\begin{equation}\label{eq:2.13}
\begin{aligned}
f(y_i|x^*_i, z_i=1;\theta)=f(y_i|x^*_i, z_i=0;\theta)=f(y_i|x_i, d_i;\theta)
\end{aligned}
\end{equation}
an MLE estimate of $\theta$ can be obtained by regressing $Y$ on $X^*$ given $S_B$. The predictions for units in $S_R$ are then given by
\begin{equation}\label{eq:2.14}
\begin{aligned}
\hat y_i=E(y_i|x^*_i, z_i=0; \hat\theta)=m(x^*_i; \hat\theta) \hspace{5mm} \forall i\in S_R
\end{aligned}
\end{equation}
where $m(x^*_i; \hat\theta)=\hat\theta^Tx^*_i$. Once $y$ is imputed for all units in the reference survey, the population mean can be estimated through the H{\'a}jek formula as below:
\begin{equation}\label{eq:2.15}
\begin{aligned}
\hat{\bar y}_{PM}=\frac{1}{\hat N_R}\sum_{i=1}^{n_R}\frac{\hat y_i}{\pi^R_i}
\end{aligned}
\end{equation}
where $\hat y_i=m(x^*_i; \hat\theta)$ for $i\in S_R$, $\hat N_R=\sum_{i=1}^{n_R}w^R_i$ and $\pi^R_i$ is the selection probability for subject $i\in S$.\par 
The asymptotic properties of the estimator in~\ref{eq:2.15}, including consistency and unbiasedness, have been investigated by \cite{kim2018combining}. Note that in situations where $\pi^R_i$ is available for $i\in S_B$, one can use $w^R_i$ instead of the high-dimensional $d_i$ as a predictor in $m(.)$. This method is known as linear-in-the-weight prediction (LWP) \citep{scharfstein1999adjusting, bang2005doubly, zhang2011comparative}. However, since outcome imputation relies fully on extrapolation, even minor misspecification of the underlying model can be seriously detrimental to bias correction.\par

\subsection{Doubly robust adjustment approach}\label{S:2.4}
To reduce the sensitivity to model misspecification, \cite{chen2019doubly} reconcile the two aforementioned approaches, i.e. QR and PM, in a way that estimates remain consistent even if one of the two models is incorrectly specified. Their method involves an extension of the augmented inverse propensity weighting (AIPW) proposed by \cite{robins1994estimation}. When $N$ is known, the expanded AIPW estimator takes the following form:
\begin{equation}\label{eq:2.16}
\bar y_{DR}=\frac{1}{N}\sum_{i=1}^{n_B}\frac{\{y_i-m( x^*_i;\theta)\}}{\pi^B(x^*_i;\beta)}+\frac{1}{N}\sum_{j=1}^{n_R}\frac{m(x^*_i;\theta)}{\pi^R_j}
\end{equation}
where given $x^*$, $\beta$ and $\theta$ are estimated using the modified PMLE and MLE methods mentioned in sections~\ref{S:2.2} and~\ref{S:2.3}, respectively. The theoretical proof of the asymptotic unbiasedness of $\bar y_{DR}$ under the correct modeling of $\pi^B(x^*_i;\beta)$ or $m(x^*_i;\theta)$ is reviewed in Appendix~\ref{S:7.1}. \par

To avoid using $\pi^R$ in modeling $\delta^B_i$ because of the PMLE restrictions we discussed in Section~\ref{S:2.2}, in this study, we suggest estimating $\pi^B_i$ for $i\in S_B$ in Eq.~\ref{eq:2.16} based on the PAPW/PAPP method depending on whether $\pi^R_i$ is available for $i\in S_B$ or not. As a result, in situations where $\pi^R_i$ is known for $i\in S_B$, our proposed DR estimator is given by
\begin{equation}\label{eq:2.17}
\hat{\bar y}_{DR}=\frac{1}{N}\sum_{i=1}^{n_B}\frac{1}{\pi^R_i}\left[\frac{1-p_i(\beta_1)}{p_i(\beta_1)}\right]\{y_i-m( x^*_i;\theta)\}+\frac{1}{N}\sum_{j=1}^{n_R}\frac{m(x^*_i;\theta)}{\pi^R_j}
\end{equation}
wher $\pi^B(x^*_i;\beta)$ is substituted using Eq.~\ref{eq:2.8}. We demonstrate that this form of the AIPW estimator is identical to that defined by \cite{kim2014doubly} in the non-response adjustment context under probability surveys. Assuming that $y_i$ is fully observed for $i\in S_R$, let us define the following HT-estimator for the population mean:
\begin{equation}\label{eq:2.18}
\hat{\bar y}_{U}=\frac{1}{N}\sum_{i=1}^{n_R}\frac{y_i}{\pi^R_i}
\end{equation}
Now, one can easily conclude that
\begin{equation}\label{eq:2.19}
\begin{aligned}
\hat{\bar y}_{DR}&=\frac{1}{N}\sum_{i=1}^{n}\frac{1}{\pi^R_i}\left[Z_i\left(\frac{1-p_i(\beta_1)}{p_i(\beta_1)} \right)\{y_i-m( x^*_i;\theta)\}+(1-Z_i)m(x^*_i;\theta)\right]\\
&=\hat{\bar y}_{U}+\frac{1}{N}\sum_{i=1}^{n}\frac{1}{\pi^R_i}\left[\frac{Z_i}{p_i(\beta_1)}-1\right]\big\{y_i-m(x^*_i;\theta)\big\}
\end{aligned}
\end{equation}
where $p_i(\beta_1)=p(Z_i=1|x_i^*;\beta_1)$. The formula in~\ref{eq:2.19} is similar to what is derived by \cite{kim2014doubly}. Therefore, the rest of the theoretical proof of asymptotic unbiasedness, i.e. $\hat{\bar y}_{DR} - \bar{\hat y}_U = O_p(n_B^{-1/2})$, in \cite{kim2014doubly} should hold for our modified AIPW estimator in~\ref{eq:2.17} as well.\par

To preserve the DR property for both the point and variance estimator of $\bar y_{DR}$, as suggested by \cite{kim2014doubly}, one can solve the following estimating equations simultaneously given $S$ to obtain the estimate of $(\beta_1, \theta)$. The aim is to cancel the first-order derivative terms in the Taylor-series expansion of $\hat{\bar y}_{DR}-\hat{\bar y}_U$ under QR and PM. These estimating equations are given by
\begin{equation}\label{eq:2.20}
\begin{aligned}
\frac{\partial}{\partial \beta_1}\left[\hat{\bar y}_{DR}-\hat{\bar y}_U\right]&=\frac{1}{N}\sum_{i=1}^{n}\frac{Z_i}{\pi^R_i}\left[\frac{1}{p_i(\beta_1)}-1\right]\{y_i-m( x^*_i;\theta)\}x^*_i=0\\
\frac{\partial}{\partial \theta}\left[\hat{\bar y}_{DR}-\hat{\bar y}_U\right]&=\frac{1}{N}\sum_{i=1}^{n}\frac{1}{\pi^R_i}\left[\frac{Z_i}{p_i(\beta_1)}-1\right]\dot m( x^*_i;\theta)=0
\end{aligned}
\end{equation}
where $\dot m$ is the derivative of $m$ with respect to $\beta_1$. Under a linear regression model, $\dot m(x^*_i)=x^*_i$. Therefore, given the same regularity conditions, ignorability in $S_B$, the logistic regression model as well as the additional imposed assumption of $S_B\cap S_R=\emptyset$, one can show that the proposed DR estimator is consistent and approximately unbiased given that either the QR or PM model holds.\par

It is important to note that the system of equations in~\ref{eq:2.20} may not have unique solutions unless the dimension of covariates in QR and PM is identical. Therefore, the AIPW estimator by \cite{chen2019doubly} may not be applicable here, as our likelihood factorization suggests that conditioning on $d_i$ is necessary at least for the PM. Furthermore, when $\pi^R_i$ is known for $i\in S_B$, one can replace the $q$-dimensional $d_i$ with the $1$-dimensional $w^R_i$ in modeling both QR and PM. \cite{bang2005doubly} shows that estimators based on a linear-in-weight prediction model remains consistent.\par

\subsection{Extension of the proposed method under a two-step Bayesian framework}\label{S:2.5}
A fully Bayesian approach specifies a model for the joint distribution of selection indicator, $\delta^B_i$, and the outcome variable, $y_i$, for $i\in U$ \citep{mccandless2009bayesian, an20104}. This requires multiply generating synthetic populations and fitting the QR and PM models on each of them repeatedly \citep{little2007bayesian, zangeneh2015bayesian}, which can be computationally expensive under a Big Data setting. While joint modeling may result in good frequentist properties \citep{little2004model}, feedback occurs between the two models \citep{zigler2013model}. This can be controversial in the sense that PS estimates should not be informed by the outcome model \citep{rubin2007design}. Here, we are interested in modeling the PS and the outcome separately through the two-step framework proposed by \cite{kaplan2012two}. The first step involves fitting Bayesian models to multiply impute the PS and the outcome by randomly subsampling the posterior predictive draws, and Rubin's combining rules are utilized as the second step to obtain the final point and interval estimates. This method not only is computationally efficient as it suffices to fit the models once and on the combined sample but also cuts the undesirable feedback between the models as they are fitted separately. Bayesian modeling can be performed either parametrically or non-parametrically.

\subsubsection{Parametric Bayes} As the first step, we employ Bayesian Generalized Linear Models to handle multiple imputations of $\pi^B_i$ and $y_i$ for $i\in S$, and $\pi^R_i$ if it is unknown for $i\in S_R$. Under a standard Bayesian framework, a set of independent prior distributions are assigned to the model parameters, and conditional on the observed data, the associated posterior distributions are simulated through an appropriate MCMC method, such as Metropolis–Hastings algorithm. We propose the following steps:
\begin{equation*}\label{eq:2.220}
\begin{aligned}
Step 1: \hspace{15mm} (\gamma^T, \phi, \beta^T, \theta^T, \sigma) &\sim p(\gamma)p(\phi)p(\beta)p(\theta)p(\sigma)\\
Step 2: \hspace{25mm} \pi^R_i|x_i,\gamma, \phi &\sim Beta(\phi[logit^{-1}(\gamma^Tx_i)], \phi[1-logit^{-1}(\gamma^Tx_i)])\\
Step 3: \hspace{30mm} Z_i|x_i,\beta &\sim Bernoulli(logit^{-1}\{\beta^Tx_i\})\\
Step 4: \hspace{27mm} Y_i|x^*_i,\theta, \sigma &\sim Normal(\theta^Tx^*_i, \sigma^2)
\end{aligned}
\end{equation*}
where $(\gamma^T, \phi)$, $\beta^T$ and $(\theta^T, \sigma)$ are the parameters associated with modeling $\pi^R_i$ in a Beta regression ($Step2$), $Z_i$ in a binary logistic regression ($Step3$) and $Y_i$ is a linear regression ($Step4$), respectively, and $p(.)$ denotes a prior density function. Note that in situations where $\pi^R_i$ is calculable for $i\in S_B$, $Step2$ should be skipped, and $x_i$ should be replaced by $x^*_i$ for $Step3$. Standard weak or non-informative priors for the regression parameter models can be used \citep{kaplan2012two}. We also note that $Step2$, which will be required for the estimation of $\pi^R_i$ when not provided directly or through the availability of $d_i$ in $S_B$, relies on a reasonably strong association between the available $x_i$ and $\pi^R_i$ in order to accurately estimate  $\pi^R_i$.  We explore the effect of differing degrees of this association via simulation in Sections~\ref{S:3.2} and \ref{S:3.3}. \par

Suppose $\Theta^{(m)T}=\left[(\gamma^{(m)T}, \phi^{(m)}, \beta^{(m)T}, \theta^{T(m)}, \sigma^{(m)}\right]$ is the $m$-th unit of an $M$-sized random sample from the MCMC draws associated with the posterior distribution of the models parameters. Then, given that $\pi^R_i$ is known for $i\in S_B$, one can obtain the m-th draw of the proposed AIPW estimator as below:
\begin{equation}\label{eq:2.230}
{\bar y}^{(m)}_{DR}=\frac{1}{\hat N_B}\sum_{i=1}^{n_B}\frac{y_i-\theta^{(m)T}x^*_i}{\pi_i^R \exp[\beta^{(m)T}x^*_i]} +\frac{1}{\hat N_R}\sum_{j=1}^{n_R}\frac{\theta^{(m)T}x^*_i}{\pi^R_j}
\end{equation}
where $\theta^{(m)T}x^*_i$ corresponds to an imputation of $y_i$ for the unobserved values in the probability sample. In situations where $\pi^R_i$ is unknown for $i\in S_B$, the $m$-th draw of the AIPW estimator is given by
\begin{equation}\label{eq:2.240}
{\bar y}^{(m)}_{DR}=\frac{1}{\hat N_B}\sum_{i=1}^{n_B}\bigg\{\frac{1+\exp[ \gamma^{(m)T}x_i]}{\exp[ \gamma^{(m)T}x_i]} \bigg\}\bigg\{\frac{y_i-\theta^{(m)T}x^*_i}{\exp[\beta^{(m)T}x_i]}\bigg\} +\frac{1}{\hat N_R}\sum_{j=1}^{n_R}\frac{\theta^{(m)T}x^*_i}{\pi^R_j}.
\end{equation}
Having ${\bar y}^{(m)}_{DR}$ for all $m=1, 2, ... , M$, then, Rubin's combining rule for the point estimate \citep{rubin2004multiple} can be employed to obtain the final AIPW estimator as below:
\begin{equation}\label{eq:2.26}
{\bar y}_{DR}=\frac{1}{M}\sum_{m=1}^M{\bar y}^{(m)}_{DR}.
\end{equation}
If at least one of the underlying models is correctly specified, we would expect that this estimator is approximately unbiased. The variance estimation under the two-step Bayesian method is discussed in Section~\ref{S:2.6}.
\subsubsection{Non-parametric Bayes} Despite the prominent feature of double robustness, for a given non-probability sample, neither QR nor PM have a known modeling structure in practice. When both working models are invalid, the AIPW estimator will be biased and a non-robust estimator based on PM may produce a more efficient estimate than the AIPW \citep{kang2007demystifying}. To show the advantage of our modified estimator in~\ref{eq:2.17} over that proposed by \cite{chen2019doubly}, we employ Bayesian Additive Regression Trees (BART) as a predictive tool for multiply imputing the $\pi^B_i$'s as well as the $y_i$'s in $S$. A brief introduction to BART has been provided in Appendix~\ref{S:7.2}. 
\par

Suppose BART approximates a continuous outcome variable through an implicit function $f$ as below: 
\begin{equation}\label{eq:2.21}
y_i=f(x^*_i)+\epsilon_i \hspace{10mm} \forall i\in S_B
\end{equation}
where $\epsilon_i\sim N(0, \sigma^2)$. Accordingly, one can train BART in $S_B$ and multiply impute $y_i$ for $i\in S_R$ using the simulated posterior predictive distribution. Regarding QR, we consider two situations; first, $\pi^R_i$ is known for $i\in S_B$. Under this circumstance, it suffices to model $z_i$ on $x_i^*$ in $S$ to estimate $\pi^B_i$ for $i \in S_B$. For a binary outcome variable, BART utilizes a data augmentation technique with respect to a given \emph{link} function, to map $\{0, 1\}$ values to $\mathbb{R}$
via a \emph{probit} link. 
Suppose
\begin{equation}\label{eq:2.22}
\Phi^{-1}[p(Z_i=1|x^*_i)]=h(x^*_i) \hspace{10mm} \forall i\in S
\end{equation}
where $\Phi^{-1}$ is the inverse CDF of the standard normal distribution. Hence, using the posterior predictive draws generated by BART in~\ref{eq:2.22}, $p(Z_i=1|x^*_i)$ and consequently $\pi^B_i$ can be imputed multiply for $i\in S_B$. For a given imputation $m$ $(m=1, 2, ... , M)$, one can expand the DR estimator in~\ref{eq:2.17} as below:
\begin{equation}\label{eq:2.23}
{\bar y}^{(m)}_{DR}=\frac{1}{\hat N_B}\sum_{i=1}^{n_B}\frac{1}{\pi_i^R}\bigg\{\frac{1-\Phi[ h^{(m)}(x^*_i)]}{\Phi[ h^{(m)}(x^*_i)]} \bigg\}\big\{y_i- f^{(m)}( x^*_i)\big\} +\frac{1}{\hat N_R}\sum_{j=1}^{n_R}\frac{ f^{(m)}(x^*_j)}{\pi^R_j}
\end{equation}
where $ f^{(m)}(.)$ and $ h^{(m)}(.)$ are the constructed sum-of-trees associated with the $m$-th MCMC draw in~\ref{eq:2.21} and \ref{eq:2.22}, respectively, after training BART on the combined sample.

Secondly, in situations where $\pi^R_i$ is not available for $i\in S_B$, we suggest applying BART to multiply impute the missing $\pi^R_i$'s in $S_B$. Since the outcome is continuous bounded within $(0, 1)$, a \emph{logit} transformation of the $\pi^R_i$'s can be used as the outcome variable in BART to map the values to $\mathbb{R}$. Such a model is presented by
\begin{equation}\label{eq:2.24}
\log\left(\frac{\pi^R_i}{1-\pi^R_i}\right)=k(x_i)+\epsilon_i \hspace{10mm} \forall i\in S_R
\end{equation}
where $k$ is a sum-of-trees function approximated by BART. Under this circumstance, ${\bar y}_{DR}$ based on the $m$-th draw from the posterior distribution is given by
\begin{equation}\label{eq:2.25}
{\bar y}^{(m)}_{DR}=\frac{1}{\hat N_B}\sum_{i=1}^{n_B}\bigg\{\frac{1+\exp[ k^{(m)}(x_i)]}{\exp[ k^{(m)}(x_i)]} \bigg\}\bigg\{\frac{1-\Phi[ h^{(m)}(x_i)]}{\Phi[ h^{(m)}(x_i)]} \bigg\}\big\{y_i- f^{(m)}( x^*_i)\big\} +\frac{1}{\hat N_R}\sum_{j=1}^{n_R}\frac{ f^{(m)}(x^*_j)}{\pi^R_j}.
\end{equation}
Having ${\bar y}^{(m)}_{DR}$ estimated for $m=1,2, ... , M$, one can eventually use Rubin's combining rule \citep{rubin2004multiple} to obtain the ultimate point estimate as in~\ref{eq:2.26}.


\subsection{Variance estimation}\label{S:2.6}
To obtain an unbiased variance estimate for the the proposed DR estimator, one needs to account for three sources of uncertainty: (i) the uncertainty due to estimated pseudo-weights in $S_B$, (ii) the uncertainty due to the predicted outcome in both $S_B$ and $S_R$, and (iii) the uncertainty due to sampling itself. In the following, we consider two scenarios:

\subsubsection{Scenario I: $\pi^R_i$ is known for $i\in S_B$} 
In this scenario the derivation of the asymptotic variance estimator for $\hat{\bar y}_{DR}$ is straightforward and follows \cite{chen2019doubly}. It is given by
\begin{equation}\label{eq:2.29}
\widehat{Var}(\hat{\bar y}_{DR})=\hat V_1 + \hat V_2 - \hat B(\hat V_2)
\end{equation}
where $V_1=Var(\hat{\bar y}_{PM})$ under the sampling design of $S_R$. $V_2$ is the variance of $\hat{\bar y}_{DR}-\hat{\bar y}_U$ under the joint sampling design of $S_R$ and the PS model. This quantity can be estimated from $S_R$ as below:
\begin{equation}\label{eq:2.30}
\hat V_2=\frac{1}{N^2}\sum_{i=1}^{n_B}\left[\frac{1-\hat\pi^B_i}{(\hat\pi^B_i)^2}\right]\{y_i-m(x^*_i;\hat\theta)\}^2.
\end{equation}
Finally, $B(\hat V_2)$ corrects for the bias in $V_2$ under the PM, and is given by
\begin{equation}\label{eq:2.31}
\hat B(\hat V_2)=\frac{1}{N^2}\sum_{i=1}^{n}\left[\frac{Z_i}{\hat\pi^B_i}-\frac{1-Z_i}{\pi^R_i} \right]\hat\sigma^2_i
\end{equation}
where $\hat\sigma^2_i=\widehat{Var}(y_i|x_i)$. Since the quantity in~\ref{eq:2.31} tends to zero asymptotically under the QR model, the derived variance estimator in~\ref{eq:2.29} is DR. 

\subsubsection{Scenario II: $\pi_i^R$ is unknown for $i\in S_B$} To estimate the variance of $\hat{\bar y}_{DR}$ in~\ref{eq:2.17} under the GLM, we employ the bootstrap repeated replication method proposed by \cite{rao1988resampling}. For a given replication $b$ $(b=1, 2, ... , B)$, we draw replicated bootstrap subsamples, $S_R^{(b)}$ and $S_B^{(b)}$, of sizes $n_R-1$ and $n_B-1$ from $S_R$ and $S_B$, respectively. The sampling weights in $S_R^{(b)}$ are updated as below:
\begin{equation}\label{eq:2.32}
w^{(b)}_i=w_i\frac{n_R}{n_R-1}h_i \hspace{10mm} \forall i\in S^{(b)}_R
\end{equation}
where $h_i$ is the number of times the $i$-th unit has been repeated in $S_B^{(b)}$. Let's assume $\hat{\bar y}^{(b)}_{DR}$ is the DR estimate based on the $b$-th combined bootstrap sample, $S^{(b)}$, using Eq.~\ref{eq:2.7}. The variance estimator is given by
\begin{equation}\label{eq:2.33}
\widehat{Var}(\hat{\bar y}_{DR})=\frac{1}{B}\sum_{b=1}^B\left[\hat{\bar y}^{(b)}_{DR}-\bar{\bar y}_{DR}\right]^2
\end{equation}
where $\bar{\bar y}_{DR}=\sum_{b=1}^B\hat{\bar y}^{(b)}_{DR}/B$. Note that when $S_R$ and $S_B$ are clustered, which is the case in our application, bootstrap subsamples are selected from the primary sampling units (PSU), and $n_R$ and $n_B$ are replaced by their respective PSU sizes.

To estimate the variance of $\hat{\bar y}_{DR}$ under a Bayesian framework, whether parametric or non-parametric, we treat $y_i$ for $i\in S_R$, and $\pi^R_i$ and $e_i$ for $i \in S_B$, as missing values in Eq.~\ref{eq:2.17} and multiply impute these quantities using the Monte Carlo Markov Chain (MCMC) sequence of the posterior predictive distribution generated by BART. For $M$ randomly selected MCMC draws, one can estimate $\hat{\bar y}^{(m)}_{DR}$ for $m=1, 2, ... , M$ based on Eq.~\ref{eq:2.17}. Following Rubin's combining rule for variance estimation under multiple imputation, the final variance estimate of $\hat{\bar y}_{DR}$ is given as below:
\begin{equation}\label{eq:2.34}
\widehat{Var}(\hat{\bar y}_{DR})=\bar V_W+\left(1+\frac{1}{M}\right)V_B
\end{equation}
where $\bar V_W=\sum_{m=1}^M \widehat{Var}(\hat{\bar y}^{(m)}_{DR})/M$, $V_B=\sum_{m=1}^M(\hat{\bar y}^{(m)}_{DR}-\bar{\bar y}_{DR})^2/(M-1)$ and $\bar{\bar y}_{DR}=\sum_{m=1}^M\hat{\bar y}^{(m)}_{DR}/M$. We have shown in the Appendix~\ref{S:7.1} that the within-imputation component can be approximated by
\begin{equation}\label{eq:2.340}
\widehat{Var}(\hat{\bar y}^{(m)}_{DR})\approx\frac{1}{\hat N^2_B}\sum_{i=1}^{n_B}\frac{var(y_i)}{\left(\hat\pi^B_i\right)^2}+\frac{1}{\hat N^2_R}var\left(\frac{1}{\pi^R_i}\right)\bigg\{\sum_{i=1}^{n_R}\left(\hat y^{(m)}_i\right)^2 + n_R\left(\frac{\hat t_R}{\hat N_R}\right)^2-2\sum_{i=1}^{n_R}\hat y^{(m)}_i \bigg\}
\end{equation}
where $t_R=\sum_{i=1}^{n_R}\hat y^{(m)}_i/\pi^R_i$. Note that when $S_R$ or $S_B$ is clustered, under a Bayesian framework, it is important to fit multilevel models to obtain unbiased variance \citep{zhou2020bayesian}. In addition, one needs to account for the intraclass correlation across the sample units in $\widehat{Var}(\hat{\bar y}^{(m)}_{DR})$ for $m=1, 2, ... , M$. Further, one may use the extension of BART with random intercept to properly specify the working models under a cluster sampling design \citep{tan2016predicting}.


\section{Simulation Study}\label{S:3}
Three simulations are studied in this section to assess the performance of our proposed methods and associated variance estimators in terms of bias magnitude and other repeated sampling properties. 
In Simulation 1, we mimic the simulation design in \cite{chen2019doubly} to compare the proposed methods. Here the probability of selection in the probability sample is a fixed linear combination of a subset of the covariates that govern selection into the non-probability sample.  In Simulation 2 we separate the design variable for the probability sample and the selection covariate for the non-probability sample in order to consider different associations between these values, as well as misspecification of the functional form of the means to consider the advantages of BART in modeling.  Simulation 3 extends Simulation 2 to allow for cluster sampling in the probability sample to better match the design of the National Household Transportation Survey in the application.

\subsection{Simulation I}\label{S:3.1} The design of our first simulation is inspired by the one implemented in \cite{chen2019doubly}. For all three studies, the non-probability samples are given a random selection mechanism with unequal probabilities, but it is later assumed that these selection probabilities are unknown at the stage of analysis, and the goal is to adjust for the selection bias using a parallel probability sample whose sampling mechanism is known.
We conduct the simulation under both asymptotic frequentist and two-step Bayesian frameworks. Consider a finite population of size $N=10^6$ with $z=\{z_1, z_2, z_3, z_4\}$ being a set of auxiliary variables generated as follows:
\begin{equation}\label{eq:3.1}
z_1\sim Ber(p=0.5) \hspace{15mm} z_2\sim U(0, 2) \hspace{15mm} z_3\sim Exp(\mu=1) \hspace{15mm} z_4\sim \chi^2_{(4)}
\end{equation}
and $x=\{x_1, x_2, x_3, x_4\}$ is defined as a function of $z$ as below:
\begin{equation}\label{eq:3.2}
x_1=z_1 \hspace{10mm} x_2=z_2+0.3z_1 \hspace{10mm} x_3=z_3+0.2(x_1 + x_2) \hspace{10mm} x_4=z_4+0.1(x_1+x_2+x_3)
\end{equation}
Given $x$, a continuous outcome variable $y$ is defined by
\begin{equation}\label{eq:3.3}
y_i=2+x_{1i}+x_{2i}+x_{3i}+x_{4i}+\sigma\epsilon_i
\end{equation}
where $\epsilon_i\sim N(0, 1)$, and $\sigma$ is defined such that the correlation between $y_i$ and $\sum_{k=1}^4x_{ki}$ equals $\rho$, which takes one of the values $\{0.2, 0.5, 0.8\}$. Further, associated with the design of $S_B$, a set of selection probabilities are assigned to the population units through the following logistic model:
\begin{equation*}\label{eq:3.4}
\log\left(\frac{\pi^B_i}{1-\pi^B_i}\right)=\gamma_0+0.1x_{1i}+0.2x_{2i}+0.1x_{3i}+0.2x_{4i}
\end{equation*}
where $\gamma_0$ is determined such that $\sum_{i=1}^{N}\pi^B_i=n_B$. For $S_R$, we assume $\pi^R_i\propto \gamma_1 + z_{3i}$ where $\gamma_1$ is obtained such that $\max\{\pi^R_i\}/\min\{\pi^R_i\}=50$. Hence, $\pi^R_i$ is assumed to be known for $i\in S_B$ as long as $z_3$ is observed in $S_B$. Using these measures of size, we repeatedly draw pairs of samples of sizes $n_R=100$ and $n_B=1,000$ associated with $S_R$ and $S_B$ from $U$ through a Poisson sampling method. Note that units in both $S_R$ and $S_B$ are independently selected, and $n_R<<n_B$, which might be the case in a Big Data setting. Extensions with $n_B=100$ and $n_B=10,000$ for both frequentist and Bayesian methods are provided in Appendix~\ref{S:7.3}.\par 

Once $S_B$ and $S_R$ are drawn from $U$, we assume that the $\pi^B_i$'s for $i\in S_B$ and $y_i$'s for $i\in S_R$ are unobserved.
The simulation is then iterated $K=5,000$ times, where the bias-adjusted mean, standad error (SE), and associated 95\% confidence interval (CI) for the mean of $y$ are estimated in each iteration. Under the frequentist approach, the AIPW point estimates are obtained by simultaneously solving the estimating equations in \ref{eq:2.20}. In addition, the proposed two-step method is used to derive the AIPW point estimates under the parametric Bayes. Also, to estimate the variance, we use the DR asymptotic method proposed by \cite{chen2019doubly}, and the conditional variance formula in Eq.~\ref{eq:2.34} under the frequentist and Bayesian approaches, respectively. For the latter, we set flat priors to the model parameters, and simulate the posteriors using $1,000$ MCMC draws after omitting an additional $1,000$ draws as the burn-in period. We then get a random sample of size $M=200$ from the posterior draws to obtain the point and variance estimates.\par

To evaluate the repeated sampling properties of the competing method, relative bias (rBias), relative root mean square error (rMSE), the nominal coverage rate of 95\% CIs (crCI) and SE ratio (rSE) are calculated as below:
\begin{align}
rbias(\hat{\bar y}_{DR}) &=100 \times\frac{1}{K}\sum_{k=1}^K \left(\hat{\bar y}^{(k)}_{DR}-\bar y_U\right) /\bar y_U\\
rMSE(\hat{\bar y}_{DR}) &=100 \times\sqrt{\frac{1}{K}\sum_{k=1}^K\left(\hat{\bar y}^{(k)}_{DR}-\bar y_U\right)^2} /\bar y_U\\
crCI(\hat{\bar y}_{DR}) &=100 \times \frac{1}{K}\sum_{k=1}^K I\left(\big|\hat{\bar y}^{(k)}_{DR} - \bar y_U\big| <z_{0.975}\sqrt{var(\hat{\bar y}^{(k)}_{DR})}\right)\\
rSE(\hat{\bar y}_{DR}) &= \frac{1}{K}\sum_{k=1}^K \sqrt{var(\hat{\bar y}^{(k)}_{DR})}/\sqrt{\frac{1}{K-1}\sum_{k=1}^K \left(\hat{\bar y}^{(k)}_{DR}-\bar{\bar y}_{DR}\right)^2}
\end{align}
where $\hat{\bar y}^{(k)}_{DR}$ denotes the DR adjusted sample mean from iteration $k$, $\bar{\bar y}_{DR}=\sum_{k=1}^K \hat{\bar y}^{(k)}_{DR}/K$, $\bar y_U$ is the finite population true mean, and $var(.)$ represents the variance estimate of the adjusted mean based on the sample. Finally, we consider model misspecification to test the DR property by removing $x_4$ from the predictors of the working model. Here $K=5000$. \par

\begin{table}[htp]
\caption{Comparing the performance of the bias adjustment methods and associated asymptotic variance estimator under the frequentist approach in the first simulation study for $\rho=\{0.2, 0.5, 0.8\}$}\label{tab:1}
\begin{threeparttable}
\scriptsize{\begin{tabular}{r r r r r r r r r r r r r r r r}
\toprule
 & \multicolumn{4}{c}{\textbf{$\rho=0.2$}}  &  &  \multicolumn{4}{c}{\textbf{$\rho=0.5$}}  &  &  \multicolumn{4}{c}{\textbf{$\rho=0.8$}}\\\cline{2-5}\cline{7-10}\cline{12-15}
\textbf{Method} & rBias & rMSE & crCI   & rSE &  & rBias & rMSE & crCI   & rSE &  & rBias & rMSE & crCI   & rSE \\
\midrule
\multicolumn{10}{l}{\textbf{Probability sample ($S_R$)}}\\
\hline
\hspace{2mm} Unweighted   & 8.528 & 19.248 & 92.6 & 1.009 &  & 8.647 & 11.065 & 77.4 & 1.018 &  & 8.682 & 9.719 & 50.9 & 1.020\\
\hspace{2mm} Fully weighted   & -0.029 & 20.276 & 94.7 & 1.001 &  & 0.006 & 8.035 & 95.1 & 1.010 &  & 0.015 & 5.008 & 94.9 & 1.008\\
\hline
\multicolumn{10}{l}{\textbf{Non-probability sample ($S_B$)}}  &    &    &    &  \\
\hline
\hspace{2mm} Unweighted   & 31.742 & 32.230 & 0.0 & 1.009 &  & 31.937 & 32.035 & 0.0 & 1.012 &  & 31.996 & 32.049 & 0.0 & 1.013\\
\hspace{2mm} Fully weighted   & 0.127 & 6.587 & 95.4 & 1.013 &  & 0.078 & 2.583 & 95.7 & 1.014 &  & 0.061 & 1.554 & 95.4 & 1.012\\
\hline
\multicolumn{10}{l}{\textbf{Non-robust adjustment}}   &    &    &    &  \\
\hline
\multicolumn{10}{l}{Model specification: True}   &    &    &    &  \\
\hline
\hspace{2mm} PAPW   & -1.780 & 8.088 & 97.0 & 1.107 &  & -1.906 & 4.734 & 95.7 & 1.103 &  & -1.947 & 4.186 & 94.0 & 1.100\\
\hspace{2mm} IPSW   & -3.054 & 10.934 & 97.2 & 1.305 &  & -3.134 & 8.145 & 95.2 & 1.173 &  & -3.160 & 7.778 & 92.4 & 1.067\\
\hline
\hspace{2mm} PM   & 0.490 & 7.577 & 95.2 & 1.007 &  & 0.190 & 4.668 & 94.6 & 0.991 &  & 0.095 & 4.204 & 94.6 & 0.985\\
\hline
\multicolumn{10}{l}{Model specification: False}   &    &    &    &  \\
\hline
\hspace{2mm} PAPW   & 26.338 & 27.089 & 3.1 & 1.112 &  & 26.434 & 26.618 & 0.0 & 1.123 &  & 26.461 & 26.580 & 0.0 & 1.128\\
\hspace{2mm} IPSW   & 28.269 & 28.917 & 0.6 & 1.021 &  & 28.474 & 28.648 & 0.0 & 1.018 &  & 28.536 & 28.654 & 0.0 & 1.014\\
\hline
\hspace{2mm} PM   & 28.093 & 28.750 & 0.6 & 1.022 &  & 28.315 & 28.494 & 0.0 & 1.022 &  & 28.382 & 28.505 & 0.0 & 1.021\\
\hline
\multicolumn{10}{l}{\textbf{Doubly robust adjustment}}   &    &    &    &  \\
\hline
\multicolumn{10}{l}{Model specification: QR--True, PM--True}   &    &    &    &  \\
\hline
\hspace{2mm} AIPW--PAPW   & 0.238 & 8.070 & 95.2 & 1.017 &  & 0.100 & 4.787 & 95.0 & 0.996 &  & 0.056 & 4.235 & 94.6 & 0.987\\
\hspace{2mm} AIPW--IPSW   & 0.105 & 7.861 & 95.1 & 1.019 &  & 0.053 & 4.737 & 94.8 & 0.996 &  & 0.036 & 4.222 & 94.6 & 0.987\\
\hline
\multicolumn{10}{l}{Model specification: QR--True, PM--False}   &    &    &    &  \\
\hline
\hspace{2mm} AIPW--PAPW   & 0.311 & 8.197 & 95.4 & 1.021 &  & 0.172 & 4.988 & 95.0 & 1.013 &  & 0.127 & 4.460 & 95.2 & 1.011\\
\hspace{2mm} AIPW--IPSW   & 0.222 & 7.962 & 95.5 & 1.024 &  & 0.170 & 4.901 & 95.4 & 1.019 &  & 0.152 & 4.405 & 95.3 & 1.018\\
\hline
\multicolumn{10}{l}{Model specification: QR--False, PM--True}   &    &    &    &  \\
\hline
\hspace{2mm} AIPW--PAPW   & 0.877 & 13.362 & 96.9 & 1.028 &  & 0.327 & 6.089 & 95.8 & 1.027 &  & 0.154 & 4.523 & 95.2 & 1.006\\
\hspace{2mm} AIPW--IPSW   & 0.609 & 12.532 & 96.6 & 1.025 &  & 0.232 & 5.842 & 95.5 & 1.022 &  & 0.113 & 4.464 & 95.3 & 1.003\\
\hline
\multicolumn{10}{l}{Model specification: QR--False, PM--False}   &    &    &    &  \\
\hline
\hspace{2mm} AIPW--PAPW   & 28.301 & 28.995 & 1.0 & 1.024 &  & 28.392 & 28.579 & 0.0 & 1.021 &  & 28.419 & 28.546 & 0.0 & 1.018\\
\hspace{2mm} AIPW--IPSW   & 28.104 & 28.762 & 0.7 & 1.024 &  & 28.313 & 28.493 & 0.0 & 1.023 &  & 28.376 & 28.500 & 0.0 & 1.022\\
\bottomrule
\end{tabular}}
    \begin{tablenotes}
      \footnotesize
      \item PAPW: propensity adjusted probability weighting; IPSW: Inverse propensity score weighting; QR: quasi-randomization; PM: prediction model; AIPW: augmented inverse propensity weighting.\
    \end{tablenotes}
  \end{threeparttable}
\end{table}

\begin{table}[htp]
\caption{Comparing the performance of the bias adjustment methods and associated variance estimator under the two-step parametric Bayesian approach in the first simulation study for $\rho=\{0.2, 0.5, 0.8\}$}\label{tab:2}
\begin{threeparttable}
\scriptsize{\begin{tabular}{r r r r r r r r r r r r r r r r}
\toprule
 & \multicolumn{4}{c}{\textbf{$\rho=0.2$}}  &  &  \multicolumn{4}{c}{\textbf{$\rho=0.5$}}  &  &  \multicolumn{4}{c}{\textbf{$\rho=0.8$}}\\\cline{2-5}\cline{7-10}\cline{12-15}
\textbf{Method} & rBias & rMSE & crCI   & rSE &  & rBias & rMSE & crCI   & rSE &  & rBias & rMSE & crCI   & rSE \\
\midrule
\multicolumn{10}{l}{\textbf{Non-robust adjustment}} &  &  &  & \\
\hline
\multicolumn{10}{l}{Model specification: True} &  &  &  & \\
\hline
\hspace{2mm} Unweighted   & 8.528 & 19.248 & 92.6 & 1.009 &  & 8.647 & 11.065 & 77.4 & 1.018 &  & 8.682 & 9.719 & 50.9 & 1.020\\
\hspace{2mm} Fully weighted   & -0.029 & 20.276 & 94.7 & 1.001 &  & 0.006 & 8.035 & 95.1 & 1.010 &  & 0.015 & 5.008 & 94.9 & 1.008\\
\hline
\multicolumn{10}{l}{\textbf{Non-probability sample ($S_B$)}}   &    &    &    &  \\ 
\hline
\hspace{2mm} Unweighted   & 31.620 & 32.106 & 0.0 & 1.014 &  & 31.906 & 32.003 & 0.0 & 1.015 &  & 31.993 & 32.045 & 0.0 & 1.017\\
\hspace{2mm} Fully weighted   & 0.026 & 6.615 & 95.3 & 1.010 &  & 0.052 & 2.604 & 95.2 & 1.007 &  & 0.059 & 1.564 & 95.2 & 1.006\\
\hline
\multicolumn{10}{l}{\textbf{Non-robust adjustment}}   &    &    &    &  \\ 
\hline
\multicolumn{10}{l}{Model specification: True}   &    &    &    &  \\ 
\hline
\hspace{2mm} PAPW   & -1.846 & 8.148 & 96.3 & 1.081 &  & -1.890 & 4.749 & 96.9 & 1.163 &  & -1.906 & 4.195 & 96.6 & 1.200\\
\hspace{2mm} IPSW   & 0.113 & 7.566 & 96.5 & 1.076 &  & 0.117 & 4.302 & 97.7 & 1.140 &  & 0.117 & 3.759 & 97.9 & 1.164\\
\hline
\hspace{2mm} PM   & 0.385 & 7.534 & 95.2 & 1.027 &  & 0.151 & 4.644 & 95.1 & 1.001 &  & 0.078 & 4.190 & 95.0 & 0.989\\
\hline
\multicolumn{10}{l}{Model specification: False}   &    &    &    &  \\ 
\hline
\hspace{2mm} PAPW   & 26.290 & 27.041 & 2.3 & 1.051 &  & 26.499 & 26.687 & 0.0 & 1.071 &  & 26.562 & 26.684 & 0.0 & 1.083\\
\hspace{2mm} IPSW   & 28.151 & 28.784 & 0.5 & 1.038 &  & 28.446 & 28.612 & 0.0 & 1.025 &  & 28.535 & 28.647 & 0.0 & 1.015\\
\hline
\hspace{2mm} PM   & 27.981 & 28.641 & 0.8 & 1.040 &  & 28.291 & 28.472 & 0.0 & 1.025 &  & 28.384 & 28.510 & 0.0 & 1.015\\
\hline
\multicolumn{10}{l}{\textbf{Doubly robust adjustment}}   &    &    &    &  \\ 
\hline
\multicolumn{10}{l}{Model specification: QR--True, PM--True}   &    &    &    &  \\ 
\hline
\hspace{2mm} AIPW--PAPW   & 0.115 & 8.093 & 96.9 & 1.097 &  & 0.057 & 4.764 & 97.1 & 1.121 &  & 0.037 & 4.219 & 97.2 & 1.130\\
\hspace{2mm} AIPW--IPSW   & 0.009 & 7.803 & 96.6 & 1.083 &  & 0.019 & 4.704 & 96.7 & 1.106 &  & 0.020 & 4.206 & 97.0 & 1.114\\
\hline
\multicolumn{10}{l}{Model specification: QR--True, PM--False}   &    &    &    &  \\ 
\hline
\hspace{2mm} AIPW--PAPW   & -0.016 & 7.930 & 97.2 & 1.108 &  & -0.080 & 4.444 & 97.9 & 1.166 &  & -0.098 & 3.842 & 98.1 & 1.193\\
\hspace{2mm} AIPW--IPSW   & -0.079 & 7.648 & 96.8 & 1.095 &  & -0.074 & 4.411 & 97.7 & 1.151 &  & -0.069 & 3.867 & 97.9 & 1.175\\
\hline
\multicolumn{10}{l}{Model specification: QR--False, PM--True}   &    &    &    &  \\ 
\hline
\hspace{2mm} AIPW--PAPW   & 0.557 & 7.693 & 96.4 & 1.086 &  & 0.214 & 4.669 & 96.8 & 1.092 &  & 0.105 & 4.195 & 96.6 & 1.090\\
\hspace{2mm} AIPW--IPSW   & 0.392 & 7.526 & 96.0 & 1.067 &  & 0.155 & 4.637 & 96.3 & 1.077 &  & 0.080 & 4.189 & 96.4 & 1.078\\
\hline
\multicolumn{10}{l}{Model specification: QR--False, PM--False}   &    &    &    &  \\ 
\hline
\hspace{2mm} AIPW--PAPW   & 28.167 & 28.864 & 1.4 & 1.096 &  & 28.359 & 28.549 & 0.0 & 1.082 &  & 28.416 & 28.548 & 0.0 & 1.068\\
\hspace{2mm} AIPW--IPSW   & 27.990 & 28.647 & 1.0 & 1.069 &  & 28.289 & 28.471 & 0.0 & 1.059 &  & 28.379 & 28.506 & 0.0 & 1.049\\
\bottomrule
\end{tabular}}
    \begin{tablenotes}
      \footnotesize
      \item PAPW: propensity adjusted probability weighting; IPSW: Inverse propensity score weighting; QR: quasi-randomization; PM: prediction model; AIPW: augmented inverse propensity weighting.\\
    \end{tablenotes}
  \end{threeparttable}
\end{table}

Table~\ref{tab:1} summarizes the results of the first simulation study under the frequentist approach. As illustrated, unweighted estimates of the population mean are biased in both $S_R$ and $S_B$. For the non-robust estimators, when the working model is valid, it seems that PM outperforms QR consistently in terms of bias correction across different $\rho$ values. While PAPW works slightly better than IPSW with respect to bias, when the QR model is true, the latter tends to overestimate the variance slightly according to the values of rSE. In addition, the smaller value of rMSE indicates that PAPW is more efficient than IPSW. For the PM, both crCI and rSE reflect accurate estimates of the variance for all values of $\rho$. When the working model is incorrect, point estimates associated with both QR and PM are biased, but the variance remains unbiased. These findings hold across all three values of $\rho$.\par

For the DR methods, it is evident that estimates based on both PAPW and IPSW remain unbiased when at least one of the PM or QR models holds. Also, the values of crCI and rSE reveal that the asymptotic variance estimator is DR for both methods. Comparing the rMSE values, the AIPW estimate based on IPSW is slightly more efficient than the one based on PAPW. While the variance estimates remain unbiased under the false-false model specification status, point estimates are severely biased. Finally, the performance of both AIPW estimators improves with respect to bias reduction especially when the QR model is misspecified.\par

For the Bayesian approach, the simulation results are displayed in Table~\ref{tab:2}. Note that we no longer are able to use the PMLE approach. Instead, we apply the PAPP method assuming that $\pi^R_i$ is unknown for $i\in S_B$. As illustrated, PAPP outperforms all the non-robust methods with respect to bias. Surprisingly, the magnitude of bias is even smaller in the Bayesian PAPP than the QR methods examined under the frequentist framework. In addition, estimates under the Bayesian approach are slightly more efficient than those obtained under the frequentist methods. While variance is approximately unbiased for $\rho=0.2$, there is evidence that PM and QR increasingly underestimate and overestimate the true variance, respectively, as the value of $\rho$ increases. Regarding the DR methods, the Bayesian and frequentist methods yield similar results. The DR property holds for all values of $\rho$, when at least one of the working models are correctly specified.\par

\subsection{Simulation II}\label{S:3.2} In the previous simulation study, we violated the ignorable assumption in order to misspecify the working model by dropping a key auxiliary variable. Now, we focus on a situation where models misspecified with respect to the functional form of their conditional means. To this end, we consider non-linear associations and two-way interactions in constructing of the outcome variables as well as the selection probabilities. This also allows us to examine the flexibility of BART as a non-parametric method when the true functional form of the underlying models is unknown. In addition, to simulate a more realistic situation, this time, two separate sets of auxiliary variables are generated, $D$ associated with the design of $S_B$, and $X$ associated with the design of $S_R$. However, we allow the two variables to be correlated through a bivariate Gaussian distribution as below:
\begin{equation}\label{eq:3.7}
\begin{pmatrix}
	d\\
	x
\end{pmatrix} \sim MVN \left(
\begin{pmatrix}
	0\\
	0
\end{pmatrix} , 
\begin{pmatrix}
	1   &  & \rho \\
	\rho & &   1
\end{pmatrix}
\right)
\end{equation}
Note that $\rho$ controls how strongly the sampling design of $S_R$  is associated with that of $S_B$, which in turn controls the quality of our assumption that $\pi^R_i$ can be well estimated by $x_i$ ($Step2$ in Section 2.5.1). In addition, the values of $d_i$ can be either observed or unobserved for $i\in S_B$. In this simulation, we set $\rho=0.2$, but later we check other values ranging from $0$ to $0.9$ as well.\par

To generate the outcome variable in $U$, we consider the following non-linear model:
\begin{equation}\label{eq:3.8}
y_i = 2f_k(x_i) - d^2_i + 0.5x_id_i + \sigma\epsilon_i
\end{equation}
where $\epsilon_i\sim N(0, 1)$, and $\sigma$ is determined such that the correlation between $y_i$ and $2f_k(x_i) - d_i^2 + 0.5x_id_i$ equals $0.5$ for $i\in U$. The function $f_k(.)$ is assumed to take one of the following forms:
\begin{equation}\label{eq:3.8}
SIN: f_1(x)=sin(x) \hspace{15mm} EXP: f_2(x)=\exp(x/2) \hspace{15mm} SQR: f_3(x)=x^2/3
\end{equation}
We then consider an informative sampling strategy with unequal probabilities of inclusion, where the selection mechanism of $S_B$ and $S_R$ depends on $x$ and $d$, respectively. Thus, each $i\in U$ is assigned two values corresponding to the probabilities of selection in $S_R$ and $S_B$ through a $logistic$ function as below:
\begin{equation}\label{eq:3.90}
\begin{aligned}
\pi^R(x_i)=P(\delta^R_i=1|d_i) & = \frac{\exp\{\gamma_0+0.2d^2_i\}}{1+\exp\{\gamma_0+0.2d^2_i\}}\\
\pi_k^A(x_i)=P_k(\delta^A_i=1|x_i) & = \frac{\exp\{\gamma_1+f_k(x_i)\}}{1+\exp\{\gamma_1+f_k(x_i)\}}
\end{aligned}
\end{equation}
where $\delta^R_i$ and $\delta^A_i$ are the indicators of being selected in $S_R$ and $S_B$, respectively.\par 

Associated with $S_R$ and $S_B$, independent samples of size $n_R=100$ and $n_A=1,000$ were selected randomly from $U$ with Poisson sampling at the first stage and simple random sampling at the second stage. The sample size per cluster, $n_\alpha$, was $1$ and $50$ for $S_R$ and $S_B$, respectively. The model intercepts, $\gamma_0$ and $\gamma_1$ in Eq.~\ref{eq:3.90}, are obtained such that $\sum_{i=1}^{N}\pi^R_i=n_R$ and $\sum_{i=1}^{N}\pi^R_i=n_A$. We restrict this simulation to Bayesian analysis based on the proposed PAPW and PAPP methods but focus on how well the non-parametric Bayes performs over the parametric Bayes in situations when the true structure of both underlying models are supposed to be unknown. The rest of the simulation design is similar to that defined in Simulation I, except for the way we specify a working model. This is done by including only the main and linear effects of $X$ and $D$ in the PM model, and the main and linear effect of $X$ in the QR model. BART's performance is examined under the assumption that the true functional form of both QR and PM models is unknown, and thus, only main effects are included in BART. 

The findings of this simulation for the two-step Bayesian approach with $1,000$ MCMC draws and $M=200$ are exhibited numerically in Table~\ref{tab:301}. Regarding the non-robust methods, both QR and PM estimators show unbiased results across the three defined functions, i.e. SIN, EXP and SQR, as long the working GLM is valid, with the minimum value of rBias associated with the PAPP method. According to the rSE values, there is evidence that PAPW and PAPP overestimate the variance, and PM underestimate the variance to some degrees, especially under the EXP and SQR scenarios. When the specified GLM is wrong, as seen, point estimates are biased for both QR and PM methods across all three functions. However, BART produces approximately unbiased results with smaller values of rMSE than GLM. In general, the PM method outperforms the QR methods under BART with respect to bias, but results based on the PAPP method are more efficient. In addition, BART tends to overestimate the variance under both QR and PM methods.\par

When it comes to the DR adjustment, Bayesian GLM produces unbiased results across all the three defined functions if the working model of either QR or PM holds. However, the variance is slightly underestimated for the SIN function when the PM specified model is wrong, and it is overestimated for the EXP function under all model-specification scenarios. As expected, point estimates are biased when the GLM is misspecified for both QR and PM. However, BART tends to produce unbiased estimates consistently across all three functions, and the magnitude of both rBias and rMSE are smaller in the AIPW estimator based on PAPP compared to the AIPW estimator based on PAPW. Finally, as in the non-robust method, variance under BART is overestimated compared to the GLM.\par

\begin{table}[htp]
\caption{Comparing the performance of the bias adjustment methods and associated variance estimator under the two-step parametric Bayesian approach in the second simulation study for $\rho=0.2$}\label{tab:301}
\begin{threeparttable}
\setlength{\tabcolsep}{3pt}
\scriptsize{\begin{tabular}{r r r r r r r r r r r r r r r r}
\toprule
 & \multicolumn{4}{c}{\textbf{SIN}}  &  &  \multicolumn{4}{c}{\textbf{EXP}}  &  &  \multicolumn{4}{c}{\textbf{SQR}}\\\cline{2-5}\cline{7-10}\cline{12-15}
\textbf{Model-method} & rBias & rMSE & crCI   & rSE &  & rBias & rMSE & crCI   & rSE &  & rBias & rMSE & crCI   & rSE \\
\midrule
\multicolumn{15}{l}{\textbf{Probability sample ($S_R$)}}\\
\hline
\hspace{2mm} Unweighted   & -17.210 & 23.109 & 80.0 & 0.999 &    & -8.406 & 11.126 & 78.3 & 1.000 &  & -17.302 & 20.563 & 65.8 & 1.002\\
\hspace{2mm} Fully weighted   & -0.623 & 17.027 & 94.4 & 0.987 &    & -0.303 & 7.947 & 94.6 & 0.987 &  & -0.675 & 13.219 & 94.0 & 0.975\\
\hline
\multicolumn{15}{l}{\textbf{Non-probability sample ($S_B$)}}\\
\hline
\hspace{2mm} Unweighted   & 33.063 & 33.379 & 0.0 & 1.003 &    & 40.307 & 40.409 & 0.0 & 1.079 &  & 49.356 & 49.570 & 0.0 & 1.016\\
\hspace{2mm} Fully weighted   & 0.019 & 6.010 & 95.1 & 1.006 &    & 0.005 & 2.755 & 94.9 & 1.005 &  & 0.009 & 3.948 & 95.0 & 0.992\\
\hline
\multicolumn{15}{l}{\textbf{Non-robust adjustment}}\\
\hline
\multicolumn{15}{l}{Model specification: True}\\
\hline
\hspace{2mm} GLM--PAPW   & -0.425 & 9.257 & 96.3 & 1.072 &    & -0.185 & 4.262 & 98.7 & 1.257 &  & -0.325 & 6.649 & 98.4 & 1.213\\
\hspace{2mm} GLM--PAPP   & 0.018 & 8.460 & 95.7 & 1.018 &    & 0.040 & 3.870 & 98.6 & 1.238 &  & -0.037 & 5.914 & 98.8 & 1.222\\
\hline
\hspace{2mm} GLM--PM   & -0.411 & 9.899 & 94.7 & 0.982 &    & -0.371 & 4.504 & 94.4 & 0.988 &  & -0.762 & 8.115 & 92.5 & 0.947\\
\hline
\multicolumn{15}{l}{Model specification: False}\\
\hline
\hspace{2mm}GLM--PAPW   & 7.180 & 11.635 & 86.4 & 1.027 &    & 2.511 & 5.299 & 97.2 & 1.316 &  & 52.170 & 52.559 & 0.0 & 1.102\\
\hspace{2mm}GLM--PAPP   & 7.647 & 11.265 & 78.0 & 0.954 &    & 3.025 & 5.425 & 96.2 & 1.277 &  & 53.095 & 53.397 & 0.0 & 1.122\\
\hspace{2mm} BART--PAPW   & 4.035 & 10.078 & 97.0 & 1.217 &    & 2.811 & 5.129 & 98.4 & 1.472 &  & 8.356 & 11.082 & 97.2 & 1.468\\
\hspace{2mm} BART--PAPP   & 1.098 & 8.530 & 96.7 & 1.121 &    & 1.108 & 4.120 & 98.9 & 1.391 &  & 4.482 & 7.479 & 98.0 & 1.401\\

\hline
\hspace{2mm} GLM--PM   & 5.870 & 10.542 & 87.9 & 0.972 &    & -6.589 & 9.264 & 82.5 & 0.976 &  & 48.993 & 49.409 & 0.0 & 0.994\\
\hspace{2mm} BART--PM   & 0.577 & 9.635 & 97.0 & 1.115 &    & 0.087 & 4.501 & 97.5 & 1.155 &  & 0.249 & 8.276 & 96.1 & 1.062\\
\hline
\multicolumn{15}{l}{\textbf{Doubly robust adjustment}}\\
\hline
\multicolumn{15}{l}{Model specification: QR--True, PM--True}\\
\hline
\hspace{2mm} GLM--AIPW--PAPW   & -0.450 & 9.930 & 95.8 & 1.023 &    & -0.165 & 4.593 & 98.2 & 1.200 &  & -0.458 & 8.116 & 96.5 & 1.089\\
\hspace{2mm} GLM--AIPW--PAPP   & -0.452 & 9.925 & 95.8 & 1.020 &    & -0.162 & 4.592 & 98.1 & 1.193 &  & -0.453 & 8.106 & 96.5 & 1.086\\
\hline
\multicolumn{15}{l}{Model specification: QR--True, PM--False}\\
\hline
\hspace{2mm} GLM--AIPW--PAPW   & -0.279 & 9.996 & 93.2 & 0.926 &    & 0.310 & 5.697 & 98.8 & 1.303 &  & -0.338 & 7.128 & 97.5 & 1.154\\
\hspace{2mm} GLM--AIPW--PAPP   & -0.134 & 9.418 & 94.1 & 0.961 &    & 0.508 & 4.977 & 99.5 & 1.475 &  & -0.275 & 7.376 & 97.6 & 1.152\\
\hline
\multicolumn{15}{l}{Model specification: QR--False, PM--True}\\
\hline
\hspace{2mm} GLM--AIPW--PAPW   & -0.411 & 10.098 & 96.1 & 1.024 &    & -0.176 & 4.715 & 98.5 & 1.234 &  & -0.771 & 8.122 & 95.5 & 1.057\\
\hspace{2mm} GLM--AIPW--PAPP   & -0.417 & 10.101 & 96.0 & 1.021 &    & -0.173 & 4.705 & 98.4 & 1.229 &  & -0.778 & 8.119 & 95.4 & 1.057\\
\hline
\multicolumn{15}{l}{Model specification: QR--False, PM--False}\\
\hline
\hspace{2mm}GLM--AIPW--PAPW   & 9.015 & 13.176 & 84.1 & 1.000 &    & 6.735 & 8.693 & 94.1 & 1.456 &  & 50.835 & 51.288 & 0.0 & 1.019\\
\hspace{2mm}GLM--AIPW--PAPP   & 9.191 & 12.717 & 84.9 & 1.082 &    & 6.787 & 8.181 & 96.7 & 1.761 &  & 51.667 & 52.131 & 0.0 & 1.047\\
\hspace{2mm}BART--AIPW--PAPW   & 0.425 & 10.071 & 97.9 & 1.184 &    & 0.122 & 4.689 & 99.3 & 1.407 &  & -0.259 & 8.349 & 98.0 & 1.231\\
\hspace{2mm}BART--AIPW--PAPP   & -0.144 & 9.794 & 97.8 & 1.184 &    & -0.100 & 4.541 & 99.3 & 1.405 &  & -0.245 & 8.329 & 97.7 & 1.203\\
\bottomrule
\end{tabular}}
    \begin{tablenotes}
      \footnotesize
      \item PAPW: propensity adjusted probability weighting; PAPP: propensity adjusted probability Prediction; QR: quasi-randomization; PM: prediction model; AIPW: augmented inverse propensity weighting.
    \end{tablenotes}
  \end{threeparttable}
\end{table}

\subsection{Simulation III}\label{S:3.3} Since the non-probability sample in the application of this study is clustered, we performed a third simulation study. To this end, the hypothetical population is assumed to be clustered with $A=10^3$ clusters, each of size $n_\alpha=10^3$ ($N=10^6$). Then, three cluster-level covariates, $\{x_1, x_2, d\}$, are defined with the following distributions:
\begin{equation}\label{eq:3.7}
\begin{pmatrix}
	d_\alpha\\
	x_{0\alpha}\\
	x_{1\alpha}
\end{pmatrix} \sim MVN \left(
\begin{pmatrix}
	0\\
	0\\
	1
\end{pmatrix} , 
\begin{pmatrix}
	1     &     -\rho/2   &   \rho \\
	-\rho/2  &  1         &   -\rho/2\\
	\rho  &  -\rho/2      &    1
\end{pmatrix}
\right)
\hspace{10mm} x_{2\alpha} = I(x_{0\alpha}>0)
\end{equation}
where $d$ denotes a design variable in $S_R$, and $\{x_1, x_2\}$ describes the selection mechanism in $S_B$. Primarily, we set $\rho=0.8$, but later we check other values ranging from $0$ to $0.9$ as well. Note that $\rho$ controls how strongly the sampling design of $S_R$ is associated with that of $S_B$. Furthermore, we assume that both $d$ and $x$ are observed for units of $S$.\par

Again, to be able to assess BART's performance, we consider non-linear associations with polynomial terms and two-way interactions in construction of the outcome variables as well as the selection probabilities. Two outcome variables are studied, one continuous ($y_c$) and one binary ($y_b$), which both depend on $\{x, d\}$ as below:

\begin{align}\label{eq:3.8}
y^c_{\alpha i}|x_\alpha, d_\alpha & \sim N(\mu=1+0.5x_{1\alpha}^2+0.4x_{1\alpha}^3-0.3x_{2\alpha}-0.2x_{1\alpha}x_{2\alpha}-0.1d_\alpha+u_\alpha, \sigma^2=1)\\
y^b_{\alpha i}|x_\alpha, d_\alpha & \sim Ber\left(p=\frac{\exp\{-1+0.1x_{1\alpha}^2+0.2x_{1\alpha}^3-0.3x_{2\alpha}-0.4x_{1\alpha}x_{2}-0.5d_\alpha+u_\alpha\}}{1+\exp\{-1+0.1x_{1\alpha}^2+0.2x_{1\alpha}^3-0.3x_{2\alpha}-0.4x_{1\alpha}x_{2}-0.5d_\alpha+u_\alpha\}}\right)
\end{align}
where $u_\alpha\sim N(0, \sigma^2_u)$, and $\sigma^2_u$ is determined such that the intraclass correlation equals $0.2$ \citep{oman2001modelling, hunsberger2008testing}. For each $i\in U$, we then consider the following set of selection probabilities associated with the design of the $S_R$ and $S_B$:
\begin{equation}\label{eq:3.91}
\begin{aligned}
\pi^R(x_\alpha)=P(\delta^R_\alpha=1|d_\alpha) & = \frac{\exp\{\gamma_0+0.5d_\alpha\}}{1+\exp\{\gamma_0+0.5d_\alpha\}}\\
\pi^B(x_\alpha)=P(\delta^B_\alpha=1|x_\alpha) & = \frac{\exp\{\gamma_1-0.1x_{1\alpha}+0.2x_{1\alpha}^2+0.3x_{2\alpha}-0.4x_{1\alpha}x_{2\alpha}\}}{1+\exp\{\gamma_1-0.1x_{1\alpha}+0.2x_{1\alpha}^2+0.3x_{2\alpha}-0.4x_{1\alpha}x_{2\alpha}\}}
\end{aligned}
\end{equation}
where $\delta^R_i$ and $\delta^B_i$ are the indicators of being selected in $S_R$ and $S_B$, respectively. Associated with $S_R$ and $S_B$, two-stage cluster samples of size $n_R=100$ and $n_B=10,000$ were selected randomly from $U$ with Poisson sampling at the first stage and simple random sampling at the second stage. The sample size per cluster, $n_\alpha$, was $1$ and $50$ for $S_R$ and $S_B$, respectively. The model intercepts, $\gamma_0$ and $\gamma_1$ in~\ref{eq:3.91}, are obtained such that $\sum_{i=1}^{N}\pi^R_i=n_R$ and $\sum_{i=1}^{N}\pi^R_i=n_B/n_\alpha$.\par

The rest of the simulation design is similar to that defined in Simulation II, except for the methods we use for point and variance estimation. In addition to the situation where $\pi^R_i$ is known for $i\in S_B$, we consider a situation where $\pi^R$ is unobserved for $i\in S_B$ and draw the estimates based on PAPP. Furthermore, unlike the simulation I, DR estimates are achieved by separately fitting the QR and PM models, and to get the variance estimates, a bootstrap technique is applied with $B=200$ based on \cite{rao1988resampling}. Finally, under BART, Rubin's combining rules are employed to derive the point and variance estimates based on the random draws of the posterior predictive distribution. As in Simulations II, we consider different scenarios of model specification. To misspecify a model, we only include the main effects in the working model.  Also, under BART, no interaction or polynomial is included as input. 

\begin{figure}[htp]
\centering\includegraphics[scale=0.30]{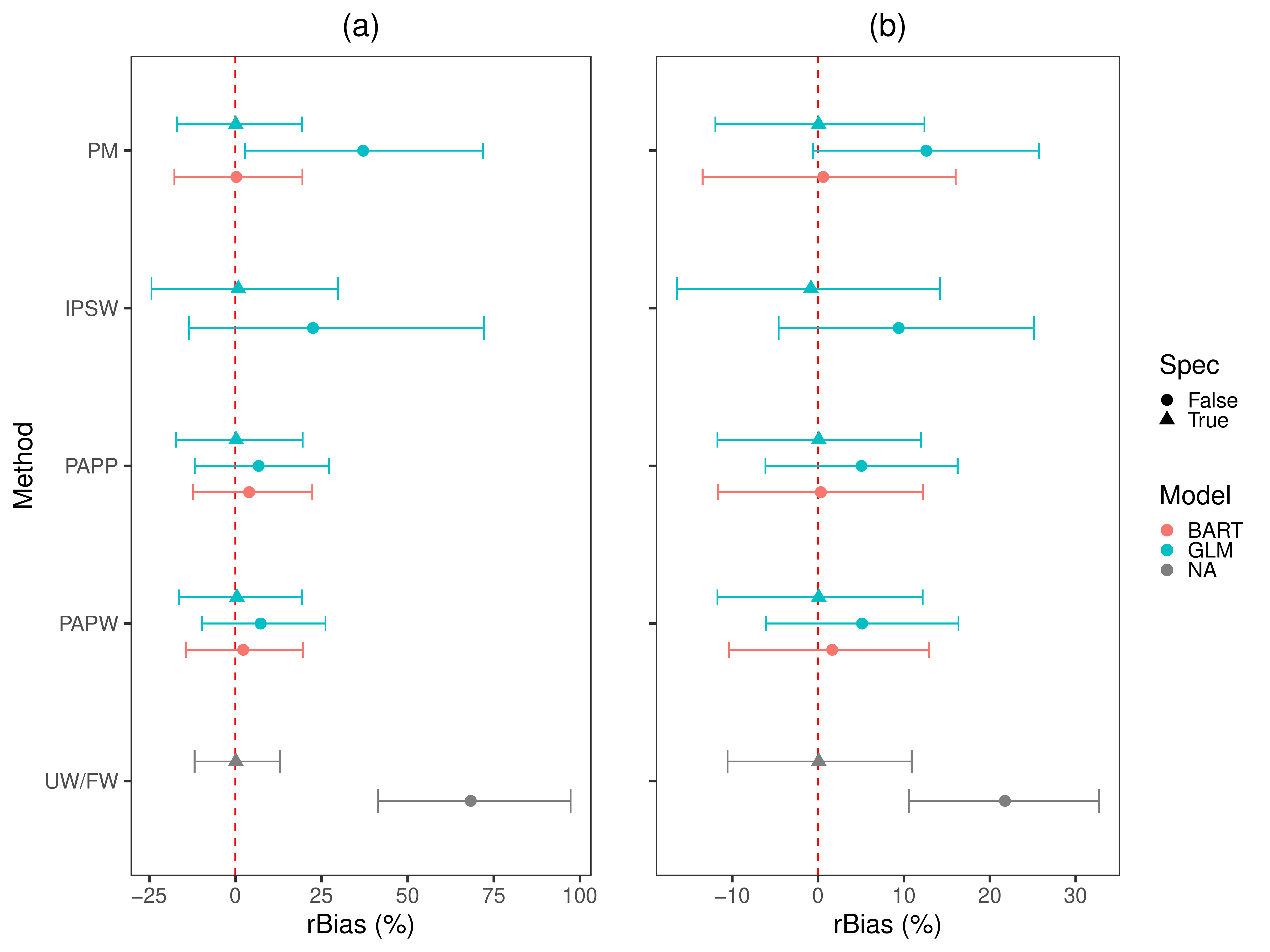}
\caption{Comparing the performance of the non-robust approaches for (a) the continuous outcome ($Y_c$) and (b) the binary outcome ($Y_b$) when the model is correctly specified. Error bars represent the 2.5\% and 97.5\% percentiles of the empirical distribution of bias over the simulation iterations. UW: unweighted; FW: fully weighted; PM: prediction model; PAPP: propensity adjusted probability prediction; IPSW: inverse propensity score weighting.}\label{fig:2}
\end{figure}

The means of the synthesized $U$ for the outcome variables were $\bar y^c_U=3.39$ and $\bar y^b_U=0.40$. Figure~\ref{fig:2} compares the bias magnitude and efficiency across the non-robust methods. As illustrated, point estimates from both $S_R$ and $S_B$ are biased if the true sampling weights are ignored. After adjusting, for both continuous and binary outcomes, the bias is close to zero under both QR and PM methods when the working model is correct. However, the lengths of the error bars reveal that the proposed PAPW/PAPP method is more efficient than the IPSW. When only main effects are included in the model, all adjusted estimates are biased except for those based on BART. Note that BART cannot be applied under IPSW. Further details about the simulation results for the non-robust methods are displayed in Appendix~\ref{S:7.3}. We see that IPSW tends to have slightly larger magnitudes of rBias and rMSE for both $y^c$ and $y^b$. Also, the values of rSE close to $1$ indicate that the Rao \& Wu's bootstrap method of variance estimation performs well under both QR and PM approaches. However, $95\%$ coverage is achieved only when the working model is correct.\par

\begin{figure}[htp]
\centering\includegraphics[scale=0.30]{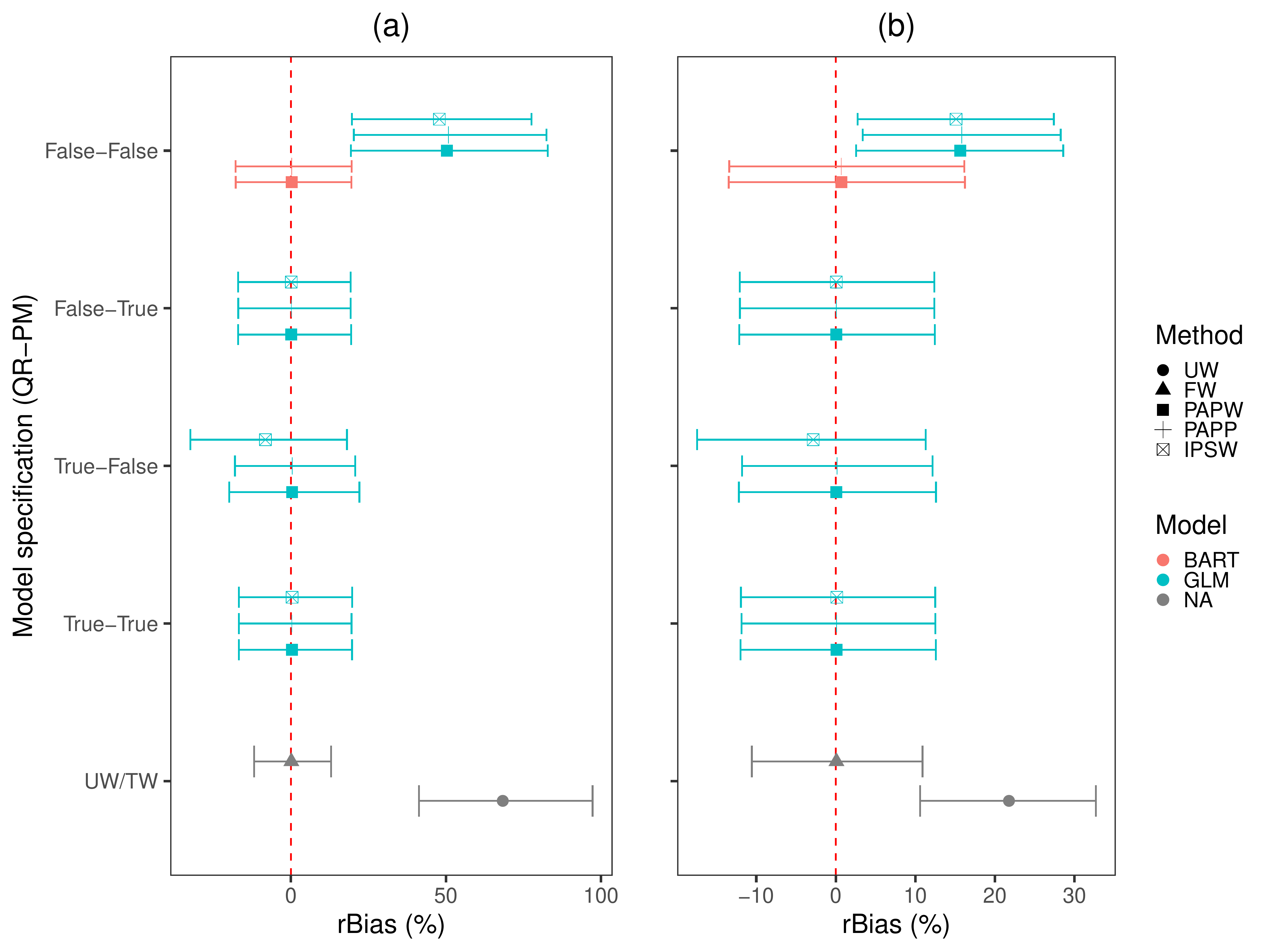}
\caption{Comparing the performance of the doubly robust estimators under different model-specification scenarios for (a) the continuous outcome ($Y_c$) and (b) the binary outcome ($Y_b$). 95\% CIs have been generated based on the 2.5\% and 97.5\% percentiles of the empirical distribution of bias over the simulation iterations. UW: unweighted; FW: fully weighted; PAPP: propensity adjusted probability prediction; IPSW: inverse propensity score weighting.}\label{fig:3}
\end{figure}

\begin{figure}[htp]
\centering\includegraphics[scale=0.30]{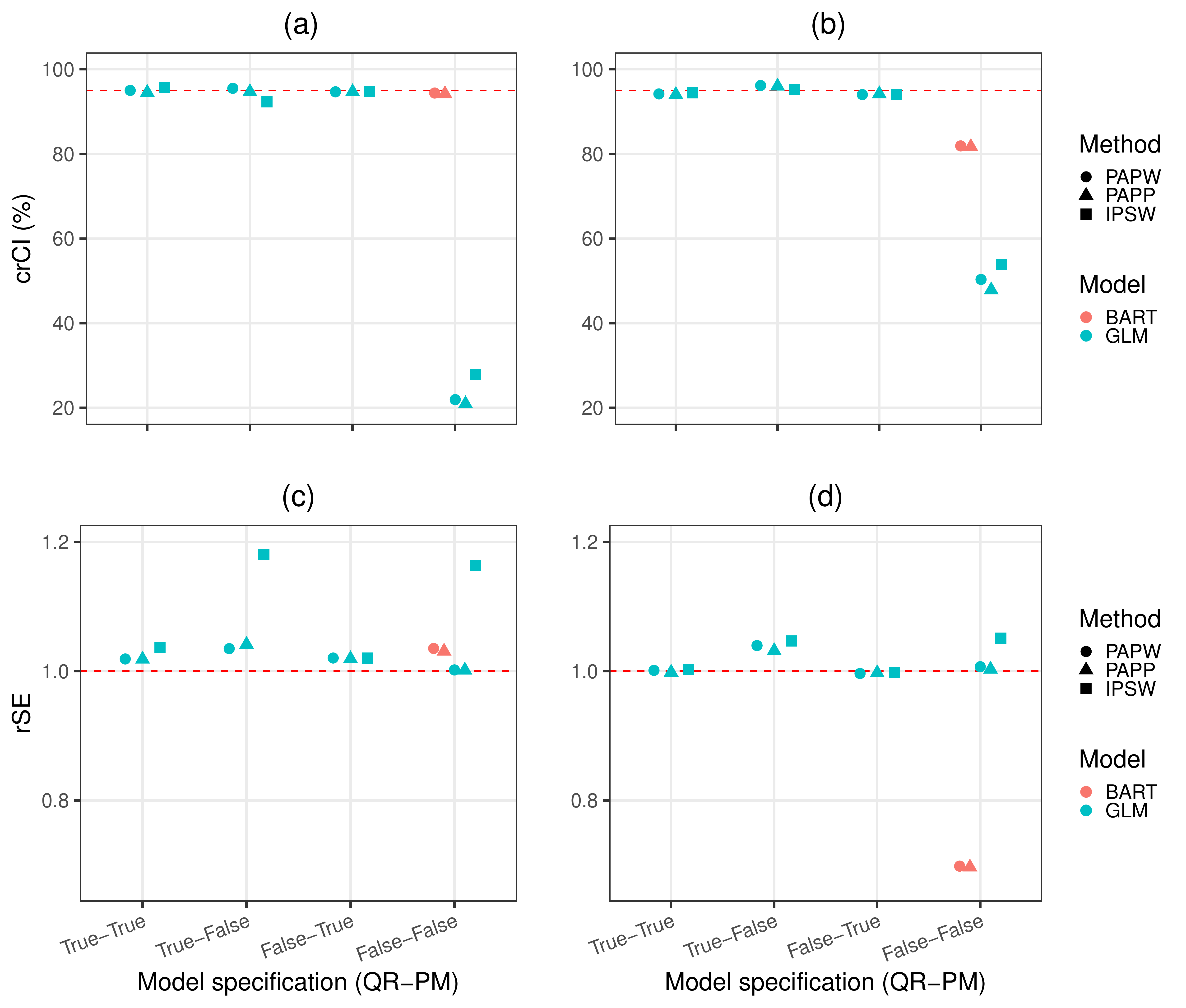}
\caption{Comparing the 95\% CI coverage rates for the means of (a) continuous outcome and (b) binary outcome and SE ratios for (c) continuous outcome and (d) binary outcome across different DR methods under different model specification scenarios. UW: unweighted; FW: fully weighted; PAPP: propensity adjusted probability prediction; IPSW: inverse propensity score weighting.}\label{fig:4}
\end{figure}

In Figure~\ref{fig:3}, we depict the results of the DR estimators under different permutations of model specification. One can immediately infer that AIPW produces unbiased results when either the PM or QR model holds. However, in situations where the true underlying models for both PM and QR are unknown, the point estimates based on BART remains unbiased under both the PAPW and PAPP approaches. Furthermore, under the GLM, it is evident that AIPW estimates based on PAPW/PAPP are slightly less biased and more efficient than those based on IPSW when the PM is incorrect. Details of the numerical results can be found in Appendix~\ref{S:7.3}. The latter compares BART with GLM under a situation where both working models are wrong. Results showing the performance of the bootstrap variance estimator are provided in Figure~\ref{fig:4}. The crCI values are all close to the correct value unless both working models are incorrectly specified. When the models are incorrrectly specified, the BART approach yields correct variance estimation for the continuous outcome; variance is underestimated and anticonservative for the binary outcome, although closer to nominal coverage than competing methods. To conclude, we observe that when neither the PM nor QR model are known, BART based on PAPP produces unbiased and efficient estimates with accurate variance.\par

\begin{figure}[htp]
\centering\includegraphics[scale=0.53]{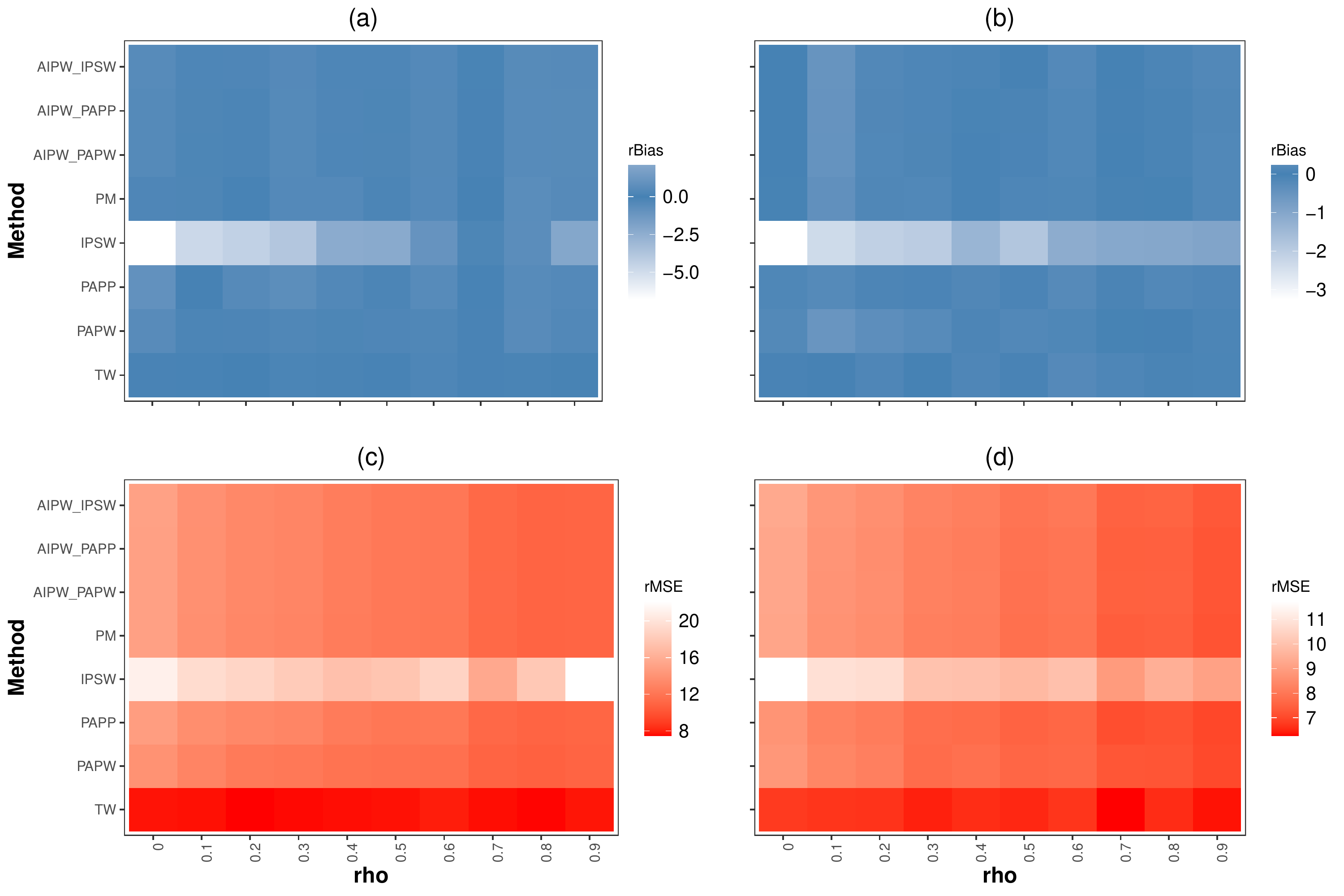}
\caption{Comparing the rBias for the means of (a) continuous outcome and (b) binary outcome and rMSE for the means of (c) continuous outcome and (d) binary outcome across different adjustment methods and different values of $\rho$. UW: unweighted; FW: fully weighted; PAPP: propensity adjusted probability prediction; IPSW: inverse propensity score weighting.}\label{fig:5b}
\end{figure}

As the final step, we replicate the simulation for different values of $\rho$ ranging from $0$ to $0.9$ to show how stable the competing methods perform in terms of rbias and rMSE. Figure~\ref{fig:5b} depicts changes in the values of rBias and rMSE for different adjustment methods as the value of $\rho$ increases. Generally, it seems that the value of rMSE decreases for all competing methods as $\rho$ increases, but for all values of $\rho$, PAPW are PAPP are less biased than IPSW. It is only when $\rho=0$ for the continuous variable that IPSW outperforms the PPAW/PAPP in bias reduction. However, when $d$ is highly correlated with $x$, there is also evidence of better performance by PAPP than IPSW in terms of bias reduction. We believe this is mainly because the stronger association between $x$ and $d$ implies that the additional ignorable assumption under PAPP is better met, while this correlation causes a sort of collinearity in IPSW leading to a loss of efficiency. The rest of the methods did not show significant changes as the value of $\rho$ increases. Numerical values associated with Figure~\ref{fig:5b} have been provided in Appendix~\ref{S:7.3}.\par

\section{Strategic Highway Research Program 2 Analysis}\label{S:4}
We briefly describe SHRP2, the non-probability sample, and the NHTS, the probability sample, as well as the variables used for statistical adjustment.

\subsection{Strategic Highway Research Program 2}\label{S:4.1}
SHRP2 is the largest naturalistic driving study conducted to date, with the primary aim to assess how people interact with their vehicle and traffic conditions while driving \citep{SHRP2013}. About $A=3,140$ drivers aged $16-95$ years were recruited from six geographically dispersed sites across the United States (Florida, Indiana, New York, North Carolina, Pennsylvania, and Washington), and over five million trips and $50$ million driven miles have been recorded. The average follow-up time per person was $\bar n_\alpha=440$ days. A quasi-random approach was initially employed to select samples by random cold calling from a pool of $17,000$ pre-registered volunteers. However, because of the low success rate along with budgetary constraints, the investigators later chose to pursue voluntary recruitment. Sites were assigned one of three pre-determined sample sizes according to their population density \citep{campbell2012shrp}. The youngest and oldest age groups were oversampled because of the higher crash risk among those subgroups. Thus, one can conclude that the selection mechanism in SHRP2 is a combination of convenience and quota sampling methods. Further description of the study design and recruitment process can be found in \cite{antin2015naturalistic}. \par

SHRP2 data are collected in multiple stages. Selected participants are initially asked to complete multiple assessment tests, including executive function and cognition, visual perception, visual-cognitive, physical and psychomotor capabilities, personality factors, sleep-related factors, general medical condition, driving knowledge, etc. In addition, demographic information such as age, gender, household income, education level, and marital status as well as vehicle characteristics such as vehicle type, model year, manufacturer, and annual mileage are gathered at the screening stage. A trip in SHRP2 is defined as the time interval during which the vehicle is operating. The in-vehicle sensors start recording kinematic information, the driver's behaviors, and traffic events continuously as soon as the vehicle is switched on. 
Trip-related information such as average speed, duration, distance, and GPS trajectory coordinates are obtained by aggregating the sensor records at the trip level \citep{campbell2012shrp}. 

\subsection{National Household Travel Survey data}\label{S:4.2}
In the present study, we use data from the eighth round of the NHTS conducted from March 2016 through May 2017 as the reference survey. The NHTS is a nationally representative survey, repeated cross-sectionally approximately every seven years. It is aimed at characterizing personal travel behaviors among the civilian, non-institutionalized population of the United States. The 2017 NHTS was a mixed-mode survey, in which households were initially recruited by mailing through an address-based sampling (ABS) technique. Within the selected households, all eligible individuals aged $\geq5$ years were requested to report the trips they made on a randomly assigned weekday through a web-based travel log. Proxy interviews were requested for younger household members who were $\leq15$ years old. \par 

The overall sample size was 129,696, of which roughly 20\% was used for national representativity and the remaining 80\% was regarded as add-ons for the state-level analysis. The recruitment response rate was 30.4\%, of which 51.4\% reported their trips via the travel logs \citep{santos2011summary}. In NHTS, a travel day is defined from 4:00 AM of the assigned day to 3:59 AM of the following day on a typical weekday. A trip is defined as that made by one person using any mode of transportation. While trip distance was measured by online geocoding, the rest of the trip-related information was based on self-reporting. A total of 264,234 eligible individuals aged $\geq$5 took part in the study, for which 923,572 trips were recorded \citep{mcguckin2018summary}.\par


\subsection{Auxiliary variables and analysis plan}\label{S:4.3}
Because of the critical role of auxiliary variables in maintaining the ignorable assumption for the selection mechanism of the SHRP2 sample, particular attention was paid to identify and build as many common variables as possible in the combined sample that are expected to govern both selection mechanism and outcome variables in SHRP2. However, since the SHRP2 sample is gathered from a limited geographical area, in order to be able to generalize the findings to the American population of drivers, we had to assume that no other auxiliary variable apart from those investigated in this study will define the distribution of the outcome variables. This assumption is in fact embedded in the ignorable condition in the SHRP2 given the set of common observed covariates. Three distinct sets of variables were considered: (i) demographic information of the drivers, (ii) vehicle characteristics, and (iii) day-level information. These variables and associated levels/ranges are listed in Table~\ref{tab:4}. 


Our focus was on inference at the day level, so SHRP2 data were aggregated. We constructed several trip-related outcome variables such as daily frequency of trips, daily total trip duration, daily total distance driven, mean daily trip average speed, and mean daily start time of trips that were available in both datasets as well as daily maximum speed, daily frequency of brakes per mile, and daily percentage of trip with a full stop, which was available in SHRP2 only. The final sample sizes of the complete day-level datasets were $n_B=837,061$ and $n_R=133,582$ in SHRP2 and NHTS, respectively.\par


In order to make the two datasets more comparable, we filtered out all the subjects in NHTS who were not drivers or were younger than $16$ years old or used public transportation or transportation modes other than cars, SUVs, vans, or light pickup trucks. One major structural difference between NHTS and SHRP2 was that in the NHTS, participants' trips were recorded for only one randomly assigned weekday, while in SHRP2, individuals were followed up for several months or years. Therefore, to properly account for the potential intraclass correlation across sample units in SHRP2, we treated SHRP2 participants as clusters for variance estimation. For BART, we fitted random intercept BART \citep{tan2016predicting}. In addition, since the $\pi^R_i$ were not observed for units of SHRP2, we employed the PAPP and IPSW methods to estimate pseudo-weights, so variance estimation under the GLM was based on the Rao \& Wu bootstrap method throughout the application section.\par

\begin{table}[htp]
\caption{List of auxiliary variables and associated levels/ranges that are used to adjust for selection bias in SHRP2}\label{tab:4}
\begin{tabular}{l l l}
\toprule
\textbf{Auxiliary variables (scale)}\hspace{10mm} & \textbf{Levels/range}\\
\midrule
\textbf{Demographic information} & \\
\hline
gender & (female, male)\\
age (yrs) & (16-24, 25-34, 35-44, 45-54, 55-64, 65-74, 75+)\\
race & (White, Black, other)\\
ethnicity & (Hispanic, non-Hispanic)\\
birth country & (citizen, alien)\\
education level & ($\leq$HS, HS completed, associate, graduate, post-graduate)\\
household income ($\times$\$1,000) & (0-49k, 50-99k, 100-149k, 150k+)\\
household size & (1, 2, 3-5, 6-10, 10+)\\
job status & (part-time, full time)\\
home ownership & (owner, renter) \\
pop. size of resid. area ($\times1,000$) & (0-49, 50-200, 200-500, 500+)\\
\hline
\textbf{Vehicle characteristics} & \\
\hline
age (yrs) & (0-4, 5-9, 10-14, 15-19, 20+)\\
type & (passenger car, Van, SUV, truck)\\  
make & (American, European, Asian)\\
mileage ($\times$1000km) & (0-4, 5-9, 10, 10-19, 20-49, 50+)\\
fuel type &  (gas, other)\\
\hline
\textbf{Day-level information} & \\
\hline
weekend indicator of trip day & \{0,1\}\\
season of trip day & (winter, spring, summer, fall)\\
\bottomrule
\end{tabular}
\end{table}


\subsection{Results}\label{S:5}
According to Figure~\ref{fig:12} of Appendix~\ref{S:7.4}, one can visually infer that the largest discrepancies between the sample distribution of auxiliary variables in SHRP2 and that in the population stem from participants' age, race and population size of residential area as well as vehicles' age and vehicles' type. The youngest and eldest age groups have been oversampled as are Whites and non-Hispanics. In addition, we found that the proportion of urban dwellers is higher in SHRP2 than that in the NHTS. In terms of vehicle characteristics, SHRP2 participants tend to own passenger cars more than the population average, whereas individuals with other vehicle types were underrepresented in SHRP2.\par

\begin{figure}[htp]
\centering\includegraphics[scale=0.38]{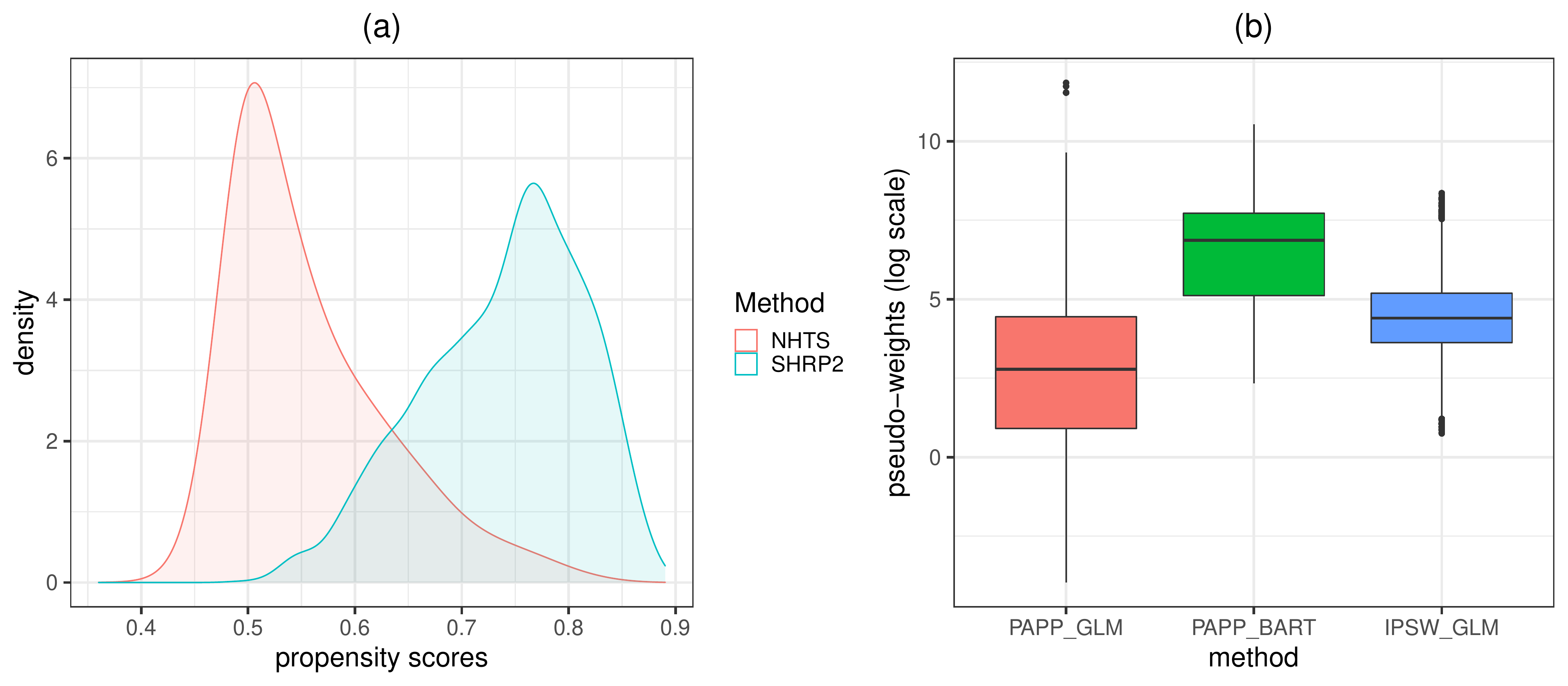}
\caption{Comparing the distribution of (a) estimated propensity scores between SHRP2 and NHTS using BART and (b) estimated pseudo-weights in SHRP2 across the applied quasi-randomization methods}\label{fig:5}
\end{figure}

As the first step of QR, we checked if there is any evidence of a lack of common distributional support between the two studies for the auxiliary variables. Figure~\ref{fig:5}a compares the kernel density of the estimated PS using BART across the two samples. As illustrated, a notable lack of overlap appears on the left tail of the PS distribution in SHRP2. 
However, owing to the huge sample size in SHRP2, we believe this does not jeopardize the positivity assumption. 
The available auxiliary variables are strong predictors of the NHTS selection probabilities for SRHP2: the average pseudo-$R^2$ was for BART $73\%$ in a $10$-fold cross validation.\par

In Figure~\ref{fig:5}b, we compare the distribution of estimated pseudo-weights across the QR methods. It seems that PAPP based on BART is the only method that does not produce influential weights. 
Also, the highest variability in the estimated pseudo-weights belonged to the PAPP method under GLM. 
As can be seen in Figure~\ref{fig:6} in Appendix \ref{S:7.4}, the largest values of area under the ROC curve (AUC) and the largest values of (pseudo)-$R^2$ in the radar across different trip-related outcome variables are associated with BART, versus classification and regression trees (CART) or GLM when  modeling $Z$ on $X$ and $Y$ on $X$, respectively. 
Additionally, Figure~\ref{fig:11} in Appendix \ref{S:7.4} exhibits how pseudo-weighting based on PAPP-BART improves the imbalance in the distribution of $X$ in SHRP2 with respect to the weighted distribution of NHTS.\par


\begin{figure}[htp]
\centering\includegraphics[scale=0.33]{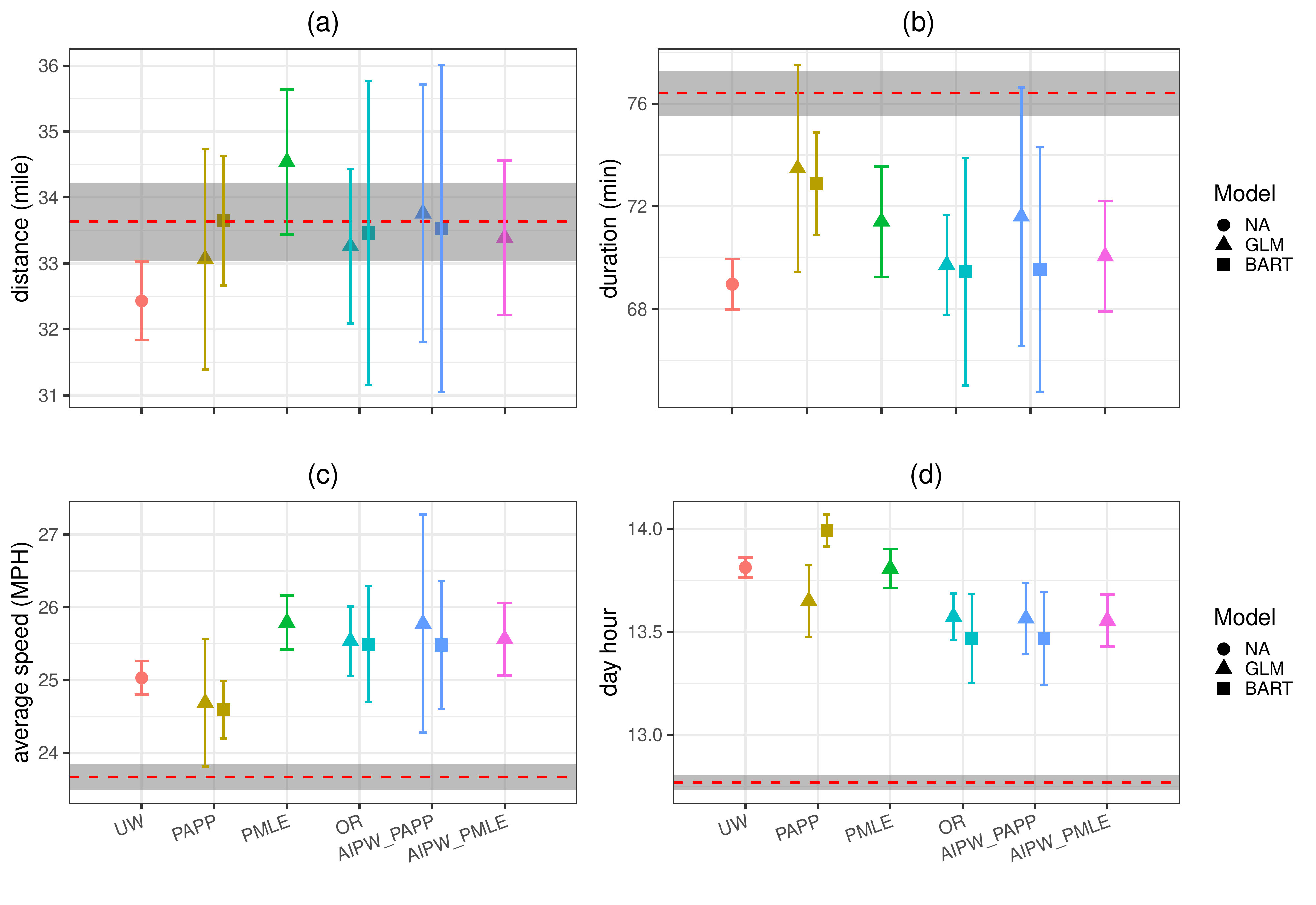}
\caption{Evaluation of pseudo-weights by comparing weighted estimates of daily frequency of trips between NHTS and SHRP2: (a) Mean daily total trip duration, (b) Mean daily total distance driven, (c) Mean trip average speed, and (d) Mean daily start hour of trips. The dashed line and surrounding shadowed area represent weighted estimates and 95\% CIs in NHTS, respectively. UW: unweighted; PAPP: propensity adjusted probability prediction; IPSW: inverse propensity score weighting; NA: not applicable.}\label{fig:7}
\end{figure}

Figure~\ref{fig:7} depicts the adjusted sample means for some trip-related measures that were available in both SHRP2 and NHTS. The methods we compare here encompass PAPP, IPSW and PM as the non-robust approaches, and AIPW with PAPP and AIPW with IPSW as the DR approaches. Also, a comparison is made between GLM and BART for all the methods except those involving IPSW. Our results suggest that, as expected, the oversampling of younger and older drivers lead to underestimating miles driven and length of trips, and overestimating the time of the first trip of the day; other factors may impact these variables, as well as the average speed of a given drive. For three of these four variables (total trip duration, total distance driven, and start hour of daily trip), there appeared to be improvement with respect to bias considering the NHTS weighted estimates as the benchmark, although only trip duration appears to be fully corrected. 
In Figure~\ref{fig:12}, we display the posterior predictive density of mean daily total distance driven under PAPP, PM and AIPW-PAPP. Note that the narrow variance associated with the PAPP approach is due to the fact that the posterior predictive distribution under pseudo-weighting does not account for the clustering effects in SHRP2. It is in fact $\bar V_W$ in~\ref{eq:2.34} that is capturing this source of uncertainty in variance estimation.\par

\begin{figure}[htp]
\centering\includegraphics[scale=0.4]{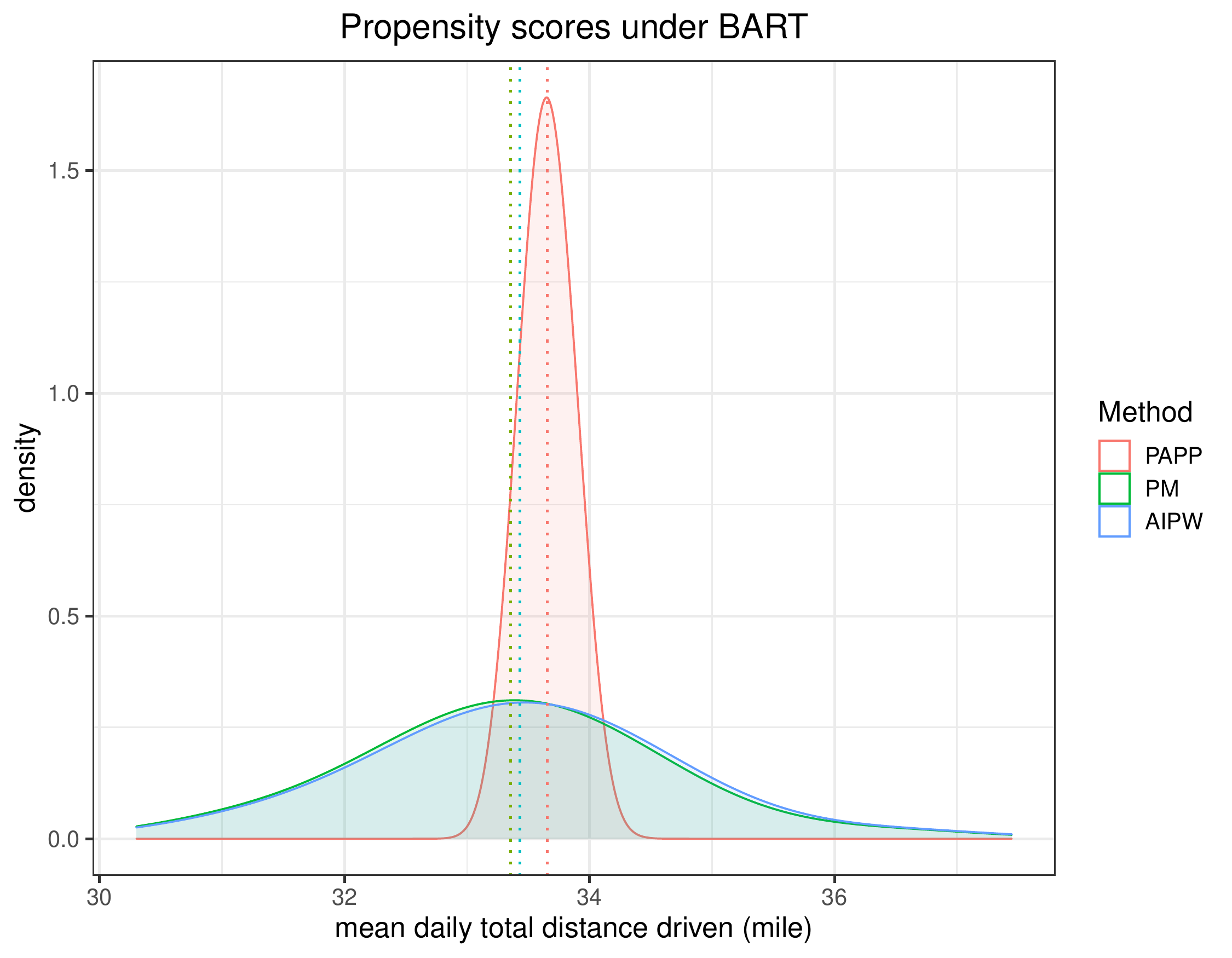}
\caption{The posterior predictive distributions of the adjusted sample mean of daily total distance driven based on BART}\label{fig:12}
\end{figure}

Among the QR methods, we observed that the PAPP based on BART gives the most accurate estimate with respect to bias for this variable. However, the relatively narrow 95\% CI associated with BART may indicate that BART does not properly propagate the uncertainty in pseudo-weighting. Regarding the PM, it seems BART performs as well as GLM, but with wider uncertainty. As a consequence, the AIPW estimator performs the same in terms of bias across different QR methods. The AIPW estimator based on IPSW, on the other hand, seems to be is more efficient than the ones based on PAPP. However, these findings are not consistent across the outcome variables. 

\begin{figure}[htp]
\centering\includegraphics[scale=0.33]{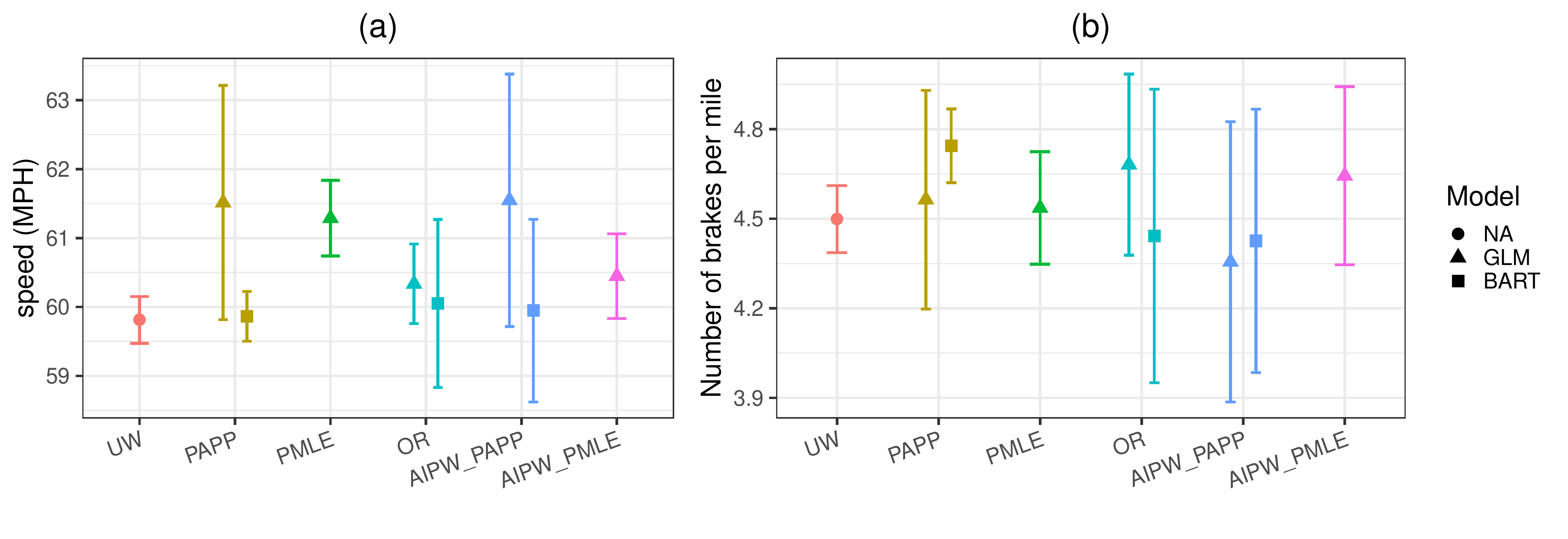}\\\includegraphics[scale=0.33]{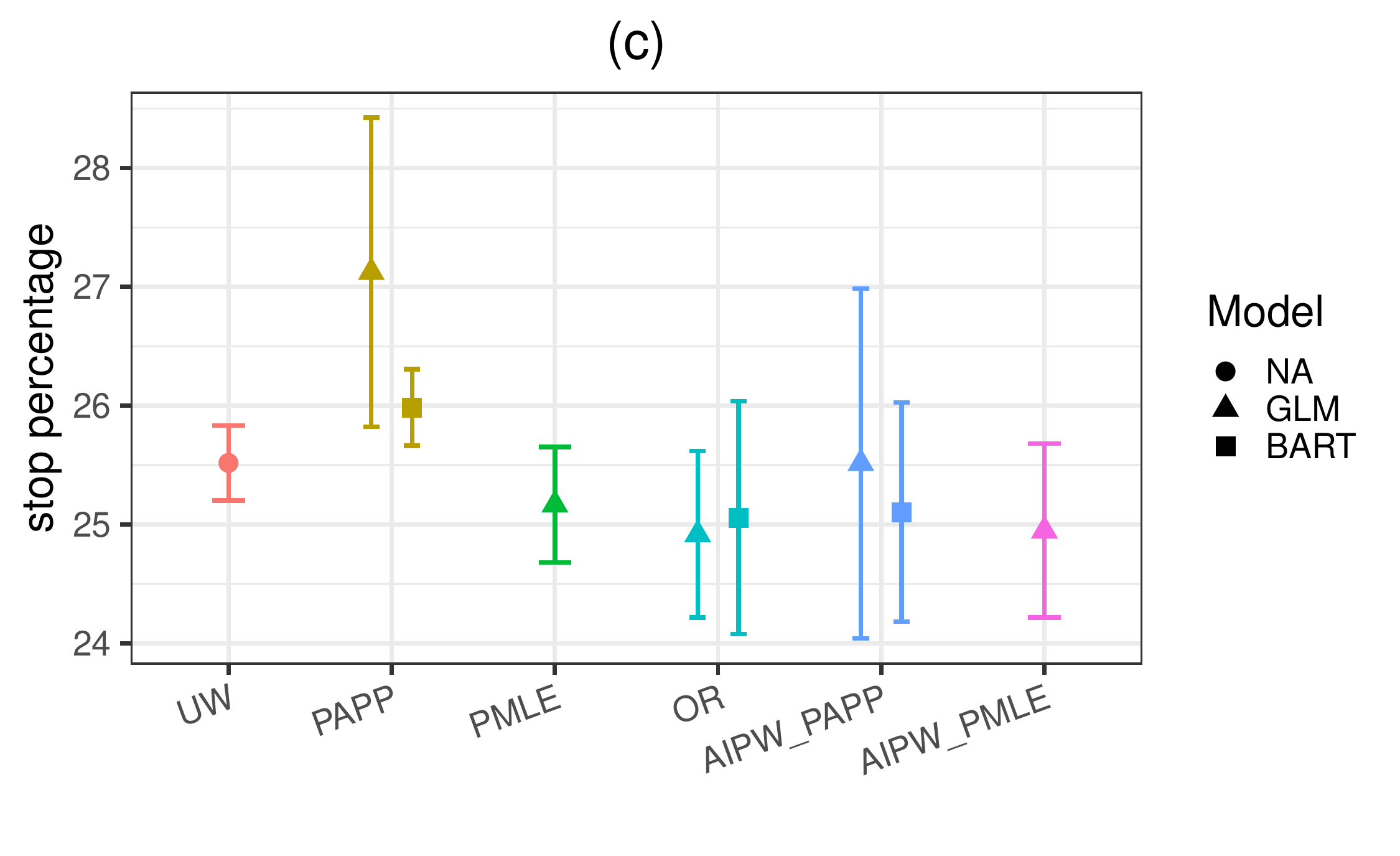}
\caption{Adjusted estimates of some SHRP2-specific outcomes: (a) Mean daily maximum speed, (b) daily frequency of brakes per mile driven, and (c) daily percentage of stop time. UW: unweighted; PAPP: propensity adjusted probability prediction; IPSW: inverse propensity score weighting.}\label{fig:8}
\end{figure}

\begin{figure}[htp]
\centering\includegraphics[scale=0.3]{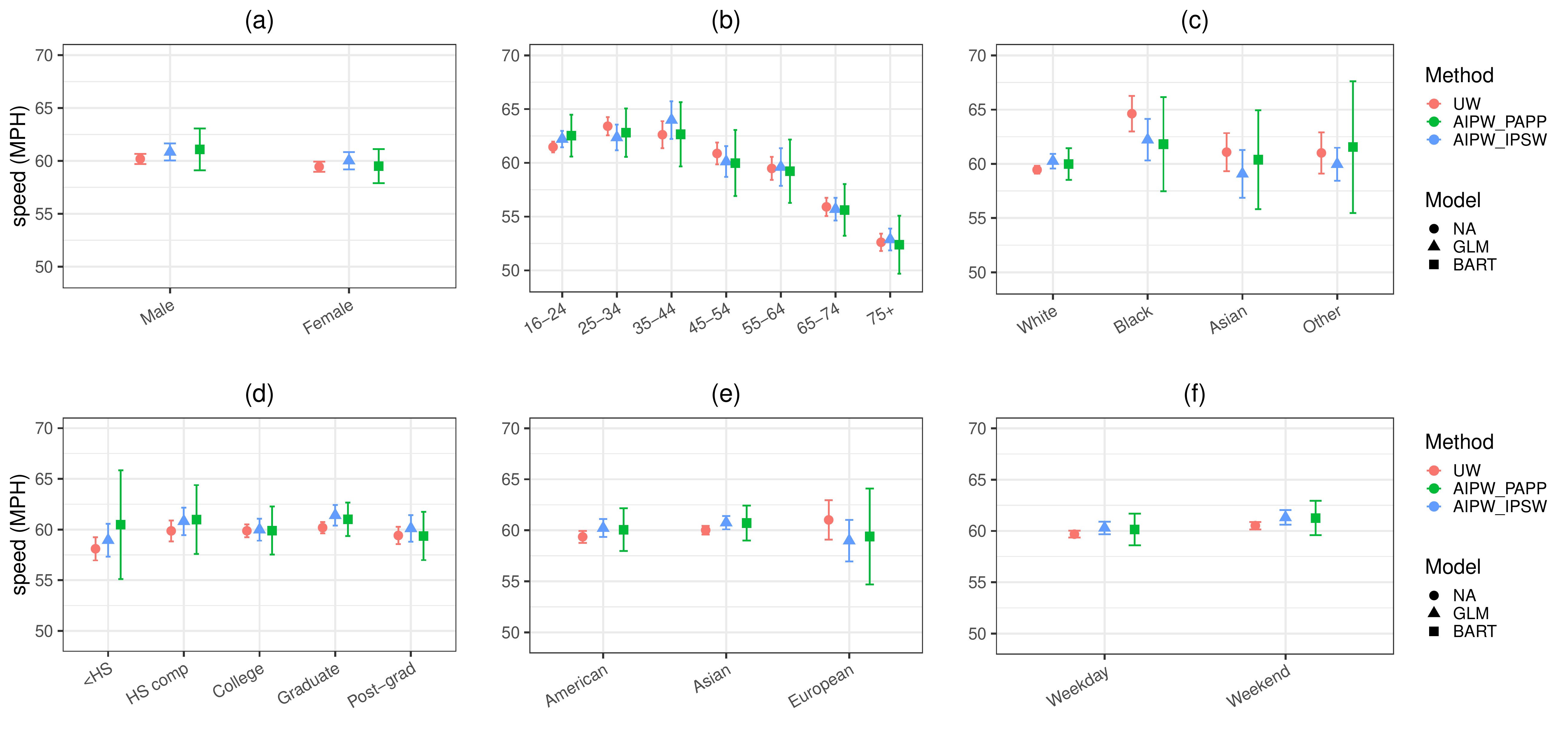}
\caption{Bias adjusted estimates of mean daily maximum speed (MPH) driven by (a) gender, (b) age groups, (c) race, (d) education, (e) vehicle manufacturer, and (f) weekend indicator. UW: unweighted; PAPP: propensity adjusted probability prediction; IPSW: inverse propensity score weighting; NA: not applicable.}\label{fig:9}
\end{figure}

Results related to the adjusted means for some SHRP2-specific outcome variables are summarized in Figure~\ref{fig:8}. These variables consist of (a) daily maximum speed, (b) frequency of brakes per mile, and (c) percentage of trip duration when the vehicle is fully stopped. For the daily maximum speed, we take one further step and present the DR adjusted mean based on the IPSW-GLM and PAPP-BART by some auxiliary variables in Figure~\ref{fig:9}. As illustrated, higher levels of mean daily maximum speed are associated with males, age group 35-44 years, Blacks, high school graduates, Asian cars, and weekends. According to the lengths of 95\% CIs, one can see that the AIPW-PAPP-BART consistently produces more efficient estimates than the AIPW-IPSW-GLM. Further numerical details of these findings by the auxiliary variables have been provided in Tables~\ref{tab:5}-\ref{tab:11} in Appendix~\ref{S:7.4}.

\section{Discussion}\label{S:6}
In this study, we proposed a doubly robust (DR) adjustment method for finite population inference in non-probability samples when a well-designed probability sample is available as a benchmark. Combining the ideas of pseudo-weighting with prediction modeling, our method involved a modified version of AIPW, which is DR in the sense that estimates are consistent if either underlying model holds. More importantly, the proposed method permitted us to apply a wider class of predictive tools, especially supervised algorithmic methods. To better address model misspecification, our study employed BART to multiply impute both pseudo-inclusion probabilities and the outcome variable. We also proposed a method to estimate the variance of the DR estimator based on the posterior predictive draws simulated by BART. In a simulation study, we then assessed the repeated sampling properties of our proposed estimator. Finally, we apply it to real Big Data from naturalistic driving studies with the aim to improve the potential selection bias in the estimates of finite population means.\par

Generally, the simulation findings revealed that our modified AIPW method produces less biased estimates than its competitors, especially when $n_R<<n_B$. When at least one of the models, i.e. QR or PM, is correctly specified, all the DR methods generated unbiased results, though our estimator was substantially more efficient with narrower 95\% CIs. However, when both working models are invalid, our findings suggest that DR estimates based on the GLM can be severely biased. However, under BART, it seems that estimates remain approximately unbiased if the true model structure associated with both QR and PM is unknown to the researcher. In contrast to the conventional IPSW estimator, we found that the new proposed estimator produces more stable results in terms of bias and efficiency across different sampling fractions and various degrees of association between the sampling designs of $S_R$ and $S_B$.\par

Generally, the results of the application suggest near total removal of bias for only one of the four variables that can be estimated from the reference survey (daily total distance driven). We believe this failure originates from several sources. First and foremost, the bias observed in the final estimates is very likely to be mixed with measurement error because we compared the results of sensor data with self-reported data as a benchmark. Second, there was evidence of departure from the positivity assumption in SHRP2. Studies show that even a slight lack of common support in the distribution of auxiliary variables may lead to inflated variance and aggravated bias \citep{hill2013assessing}. Part of this can be due to the fact that we attempted to generalize the results to the general population of American drivers, while SHRP2 data was restricted to six states. 
Another reason might be deviation from the ignorable assumptions: The associations between the auxiliary variables and the outcome variables were relatively weak and varying across the variables.\par


Our study was not without weaknesses. 
First, our approach assumes the ideal situation where the $d_i$ are available in the non-probability sample, since that is demanded by the general theory linking together the probability and non-probability samples.  In practice it can be difficult to fully meet this requirement, and indeed in many practical settings it might be that only the available subset of $x_i^*$ is required to fully model selection into the non-probability sample and the outcome variable, or alternatively, that the available components of  $x_i^*$  will provide a much better approximation to the true estimates than simply using the non-probability sample without correction. Second, our adjustment method assumes that the two samples are mutually exclusive. However, in many Big Data scenarios (though not the one we consider), the sampling fraction may be non-trivial, so the two samples may overlap substantially. In such a situation, it is important to check how sensitive our proposed pseudo-weighting approach is to this assumption. Extensions may be plausible to account for the duplicate units of the population in the pooled sample. Third, our multiple imputation variance estimator (Eq.~\ref{eq:2.34}) ignores covariance between $\overline{V}_W$ and $V_B$ induced by the weights.  This covariance is typically negative and leads to conservative inference, as seen in the modest overestimation of variance in the BART estimations in Simulations 2 and 3.  Use of a bootstrap procedure such as that described in the simulation study of \cite{chen2019doubly} may be an alternative, although impractical in our setting given the computational demands of fitting the BART models to each bootstrap sample.  Another drawback is that the combined dataset may be subject to differential measurement error in the variables. This issue is particularly acute in our SHPR2 analysis, because the definition of a \textit{trip} may not be identical between the two studies: although trip measures in the SHRP2 are recorded by sensors, in the NHTS trip measures are memory and human estimation based, as they are self-reported. Having such error-prone information either as the outcome or as an auxiliary variable may lead to biased results. Finally, we failed to use the two-step Bayesian method under GLM for the application part, because SHRP2 data were clustered demanding for Bayesian generalized linear mixed effect models to properly estimate the variance of the DR estimators required computational resources beyond our reach. This prompted us to apply resampling techniques to the actual data instead of a full Bayesian method.\par

There are a number of potential future directions for our research. First, we would like to expand the asymptotic variance estimator under PAPP when $\pi_i^R$ cannot be computed for $i\in S_B$. Alternatively, one may be interested in developing a fully model-based approach, in which a synthetic population is created by undoing the sampling stages via a Bayesian bootstrap method, and attempts are made to impute the outcome for non-sampled units of the population \citep{dong2014nonparametric, zangeneh2015bayesian, an2008robust}. The synthetic population idea makes it easier to incorporate the design features of the reference survey into adjustments, especially when Bayesian inference is of interest. While correcting for selection bias, one can adjust for the potential measurement error in the outcome variables as well if there exists a validation dataset where both mismeasured and error-free values of the variables are observed \citep{kim2020data}. When combining data from multiple sources, it is also likely that auxiliary variables are subject to differential measurement error. \cite{hong2017bayesian} propose a Bayesian approach to adjust for a different type of measurement error in a causal inference context. Also, in a Big Data setting, fitting models can be computationally demanding. To address this issue, it might be worth expanding the divide-and-recombine techniques for the proposed DR methods.  Finally, as noted by a reviewer, the basic structure of our problem (see Figure~\ref{fig:1}) approximates that tackled by ``data fusion'' methods, developed primarily in the computer science literature.  While this literature does not appear to have directly addressed issues around sample design, it may be a useful vein of research to mine for future connections to non-probability sampling research.


\section{Acknowledgement}
The present study was part of the doctoral research of the first author of this article at the Michigan Program in Survey and Data Science. The authors would like to thank all the researchers and staff who have been involved in collecting the data of both NHTS 2017 and SHRP2. Our gratitude also goes to professors Katharine Abraham, Stanley Presser and Joseph Sedarski at the University of Maryland who have improved the quality of this article with their valuable comments. The findings and conclusions of this paper are those of the authors and do not necessarily represent the views of the 
Virginia Tech Transportation Institute (VTTI), SHRP2, the Transportation Research Board, or the National Academy of Sciences.

\section{Conflict of Interest} The authors declare that there was no conflict of interest in the current research.

\clearpage
\newpage

\bibliographystyle{chicago}
\bibliography{arXiv-Rafei-DR-paper}


\setcounter{page}{1}

\section{Appendix}\label{S:7}

\subsection{Theoretical proofs}\label{S:7.1}
Suppose there exists an infinite sequence of finite populations $U_\nu$ of sizes $N_\nu$ with $\nu=1, 2, ... , \infty$. Corresponding to $U_\nu$ are a non-probability sample $S_{B,\nu}$ and a probability sample $S_{R,\nu}$ with $n_{B,\nu}$ and $n_{R,\nu}$ being the respective sample sizes. Also, let us assume that $N_\nu{\to}\infty$, $n_{B,\nu}{\to}\infty$ and $n_{R,\nu}{\to}\infty$ as $\nu{\to}\infty$, while $n_{B,\nu}/N_\nu{\to}f_B$, and $n_{R,\nu}/N_\nu{\to}f_R$ with $0<f_R<1$ and $0<f_B<1$. However, from now on, we suppress the subscript $\nu$ for rotational simplicity. In order to be able to make unbiased inference based on $S_B$, we consider the following conditions:
\begin{enumerate}
\item The set of observed auxiliary variables, $X$, fully governs the selection mechanism in $S_B$. This is called an \emph{ignorable} condition, implying $p(\delta^B_i=1|y_i, x_i)=p(\delta_i^B=1|x_i)$ for $i\in U$.
\item The $S_B$ actually does have a probability sampling mechanism, albeit unknown. This means $p(\delta^B_i=1|x_i)>0$ for all $i\in U$.
\item Units of $S_R$ and $S_B$ are selected independently from $U$ given the observed auxiliary variables, $X^*$, i.e. $\delta^R_i\indep \delta^B_j|X^*$ for $i\neq j$.
\item The sampling fractions, $f_R$ and $f_B$, are small enough such that the possible overlap between $S_R$ and $S_B$ is negligible, i.e. $S_R\cap S_B=\emptyset$.
\item The true underling models for $Y|X^*$ and $\delta_B|X$ and $\delta^R|X$ are known.
\end{enumerate}
In addition, to be able to drive the asymptotic properties of the proposed estimators, we consider the following regularity conditions according to \cite{chen2019doubly}:
\begin{enumerate}
\item For any given $x$, $\partial m(x;\theta)/\partial\theta$ exists and is continuous with respect to $\theta$, and $|\partial m(x;\theta)/\partial\theta|\leq h(x;\theta)$ for $\theta$ in the neighborhood of $\theta$, and $\sum_{i=1}^Nh(x_i;\theta)=O(1)$.
\item For any given $x$, $\partial^2 m(x;\theta)/\partial\theta^T$ exists and is continuous with respect to $\theta$, and $max_{j,l} | \partial^2 m(x;\theta)/\partial\theta_j\partial\theta_l | \leq k(x;\theta)$ for $\theta$ in the neighborhood of $\theta$, and $\sum_{i=1}^N k(x_i;\theta)=O(1)$.
\item For $u_i=\{x_i,y_i, m(x_i;\theta)\}$, the finite population and the sampling design in $S_R$ satisfy $N^{-1}\sum_{i=1}^{n_R}u_i/\pi^R_i-N^{-1}\sum_{i=1}^Nu_i=O_p(n_R^{-1/2})$.
\item There exist $c_1$ and $c_2$ such that $0<c_1\leq N\pi^B_i/n_B\leq c_2$ and $0<c_1\leq N\pi^R_i/n_R\leq c_2$ for all $i\in U$.
\item The finite population and the propensity scores satisfy $N^{-1}\sum_{i=1}^N y_i^2=O(1)$, $N^{-1}\sum_{i=1}^N||x_i||^3=O(1)$, and $N^{-1}\sum_{i=1}^N\pi^B_i(1-\pi^B_i)x_ix_i^T$ is a positive definite matrix.
\end{enumerate}
Note that while we assume $\pi^R_i$ is calculable for $i\in S_B$ throughout the proofs, extensions can be provided for situations where $\pi^R_i$ need to be predicted for $i\in S_B$.
\subsubsection{Asymptotic properties of PAPW estimator}\label{S:7.1.1}
Since $\hat\beta_1$ is the MLE estimate of $\beta_1$ in the logistic regression of $Z_i$ on $x^*_i$, it is clear that $\hat\beta_1\overset{p}{\to}\beta_1$. Two immediate result of this are that $\hat\pi^B_i\overset{p}{\to}\pi^B_i$ and $E(\hat\pi^B_i|x^*_i)=\pi^B_i$ where $\hat\pi^B_i$ is defined as in~\ref{eq:2.8}. Now, we prove the consistency and asymptotic unbiasedness of the PAPW estimator in~\ref{eq:2.15}. To this end, we show that $\hat{\bar y}_{PAPW}-{\bar y}_{U}=O_p(n_B^{-1/2})$. Consider the following set of estimating equations:

\begin{align}
    \Phi_n(\eta) &= \begin{bmatrix}
    n^{-1}\sum_{i=1}^n Z_i(y_i-\bar y_U)/\pi^B_i\\\\
    n^{-1}\sum_{i=1}^n\{Z_i-p_i(\beta_1)\}x^*_i
    \end{bmatrix}
    = \begin{bmatrix}
    N^{-1}\sum_{i=1}^N \delta^B_i(y_i-\bar y_U)/\pi^B_i\\\\
    N^{-1}\sum_{i=1}^N\delta_i\{Z_i-p_i(\beta_1)\}x^*_i
    \end{bmatrix}=0
\end{align}
where $\eta=(\bar y_U, \beta_1)$.\par

In the following, we show that $E_{\delta^B}[\Phi_n(\hat\eta)|x^*_i]=0$. We start with the first component of $\Phi_n(\hat\eta)$
\begin{equation*}
\begin{aligned}
    E_{\delta^B}\left[\frac{1}{N}\sum_{i=1}^N\frac{E_{\delta^B}(\delta^B_i)(y_i-\bar y_U)}{\pi^B_i}\big|x^*_i\right]
    &=\frac{1}{N}\sum_{i=1}^N\frac{E_{\delta^B}(\delta^B_i|x^*_i)(y_i-\bar y_U)}{\pi^B_i}\\
    &=\frac{1}{N}\sum_{i=1}^N\frac{\pi^B_i(y_i-\bar y_U)}{\pi^B_i}\\
    &=0
\end{aligned}
\end{equation*}
Noting that $E_{\delta^B}[\Phi_n(\hat\eta)]=E_{\delta}[E_Z\{\Phi_n(\hat\eta)|\delta_i=1\}]$, for the second component, we have
\begin{equation}
\begin{aligned}
    E_{\delta^B}\left[
    \frac{1}{N}\sum_{i=1}^N\delta_i\{Z_i-p_i(\beta_1)\}x_i\bigg|x_i\right]&=E_{\delta}\left[E_Z\bigg\{\frac{1}{N}\sum_{i=1}^N\delta_i\{Z_i-p_i(\beta_1)\}x_i\big|\delta_i=1, x_i\bigg\}\right]\\
    &=E_{\delta}\left[\frac{1}{N}\sum_{i=1}^N\delta_i\{E_Z(Z_i|\delta_i=1, x_i)-p_i(\beta_1)\}x^*_i\right]\\
    &=E_{\delta}\left[\frac{1}{N}\sum_{i=1}^N\delta_i\{p_i(\beta_1)-p_i(\beta_1)\}x^*_i\right]\\
    &=0
\end{aligned}
\end{equation}

Now, we apply the first-order Taylor approximation to $\Phi_n(\hat\eta)$ around $\eta_1$ as below:
\begin{equation}
    \hat\eta-\eta_1=[E\{\phi_n(\eta_1)\}]^{-1}\Phi_n(\eta_1)+O_p(n_B^{-1/2})
\end{equation}
where $\phi_n(\eta)=\partial\Phi_n(\eta)/\partial\eta$.
\begin{equation}
    \frac{\partial}{\partial\bar y_U}\left[\frac{1}{N}\sum_{i=1}^N\delta^B_i\frac{(y_i-\bar y_U)}{\pi^B_i}\right]=-\frac{1}{N}\sum_{i=1}^N\frac{\delta^B_i}{\pi^B_i}
\end{equation}

\begin{equation}
\begin{aligned}
    \frac{\partial}{\partial\beta_1}\left[\frac{1}{N}\sum_{i=1}^N\delta^B_i\frac{(y_i-\bar y_U)}{\pi^B_i}\right]&=\frac{\partial}{\partial\beta_1}\left[\frac{1}{N}\sum_{i=1}^N\frac{\delta^B_i}{\pi^B_i}\bigg\{\frac{p_i(\beta_1)}{1-p_i(\beta_1)}\bigg\}(y_i-\bar y_U)\right]\\
    &=-\frac{1}{N}\sum_{i=1}^N\frac{\delta^B_i}{\pi^B_i}(y_i-\bar y_U)x_i^{*T}
\end{aligned}
\end{equation}

\begin{equation}
    \frac{\partial}{\partial\beta_1}\left[\frac{1}{N}\sum_{i=1}^N\delta_i\big\{Z_i-p_i(\beta_1)\big\}\right]=-\frac{1}{N}\sum_{i=1}^N\delta_ip_i(\beta_1)\left[1-p_i(\beta_1)\right]x^*_ix_i^{*T}
\end{equation}
Therefore, we have
\begin{equation}
\phi_n(\eta_1)=
\begin{pmatrix}
-\frac{1}{N}\sum_{i=1}^N\frac{\delta^B_i}{\pi^B_i} & -\frac{1}{N}\sum_{i=1}^N\frac{\delta^B_i}{\pi^B_i}(y_i-\bar y_U)x_i^{*T}\\
0 & -\frac{1}{N}\sum_{i=1}^N\delta_ip_i(\beta_1)\left[1-p_i(\beta_1)\right]x^*_ix_i^{*T}
\end{pmatrix}
\end{equation}
Thus, it follows that $\hat{\bar y}_{PM}=\bar y_U + O_p(n_B^{-1/2})$.

Now, we turn to deriving the asymptotic variance estimator for $\hat{\bar y}_{PM}$. According to the sandwich formula, we have
\begin{equation}
    Var(\hat\eta_1)=\left[E\{\phi_n(\eta_1)\}\right]^{-1}Var\big\{\phi_n(\eta_1)\big\}\left[E\{\phi_n(\eta_1)\}^T\right]^{-1}+O_p(n_B^{-1})
\end{equation}
Given the fact that 
\begin{equation}
    E(\delta_i=1|x^*_i)=\frac{p(\delta^B_i=1|x^*_i)}{p(Z_i=1|x^*_i)}=\frac{\pi^R_i}{1-p_i(\beta_1)}
\end{equation}
It can be shown that
\begin{equation}
E\big\{\phi_n(\eta_1)\big\}=
\begin{pmatrix}
-1 & -\frac{1}{N}\sum_{i=1}^N(y_i-\bar y_U)x_i^{*T}\\
0 & -\frac{1}{N}\sum_{i=1}^N\pi^B_i\left[1-p_i(\beta_1)\right]x^*_ix_i^{*T}
\end{pmatrix}
\end{equation}
And
\begin{equation}
\left[E\big\{\phi_n(\eta_1)\big\}\right]^{-1}=
\begin{pmatrix}
-1 & b^T\\
0 & -\left[\frac{1}{N}\sum_{i=1}^N\pi^B_i\left[1-p_i(\beta_1)\right]x^*_ix_i^{*T}\right]^{-1}
\end{pmatrix}
\end{equation}
where
\begin{equation}
    b^T=\bigg\{\frac{1}{N}\sum_{i=1}^N(y_i-\bar y_U)x_i^{*T}\bigg\}\bigg\{\frac{1}{N}\sum_{i=1}^N\pi^B_i\big\{1-p_i(\beta_1)\big\}x^*_ix_i^{*T}\bigg\}^{-1}
\end{equation}
Now, the goal is to calculate $Var\big\{\phi_n(\eta_1)\big\}$. We know that
\begin{equation}
\begin{aligned}
    Var_{\delta^B}\left(\frac{1}{N}\sum_{i=1}^N\frac{\delta^B_i(y_i-\bar  y_U)}{\pi^B_i}\bigg|x_i\right)&=\frac{1}{N}\sum_{i=1}^N\frac{(y_i-\bar  y_U)^2}{(\pi^B_i)^2}\pi^B_i(1-\pi^B_i)\\
    &=\frac{1}{N}\sum_{i=1}^N\bigg\{\frac{1-\pi^B_i}{\pi^B_i}\bigg\}(y_i-\bar  y_U)^2
\end{aligned}
\end{equation}

\begin{equation}
\begin{aligned}
    Var_{\delta^B}\left(\frac{1}{N}\sum_{i=1}^N\delta_i\big\{Z_i-p_i(\beta_1)\big\}\bigg|x^*_i\right)&=E_\delta\left[Var_Z\left(\frac{1}{N}\sum_{i=1}^N\delta_i\big\{Z_i-p_i(\beta_1)\big\}\big|\delta_i=1, x^*_i\right)\right]\\
    &+Var_\delta\left[\delta_iE_Z\left(\frac{1}{N}\sum_{i=1}^N\delta_i\big\{Z_i-p_i(\beta_1)\big\}\big|\delta_i=1, x^*_i\right)\right]\\
    &=\frac{1}{N^2}E_\delta\left(\sum_{i=1}^N\delta^2_iVar_Z(Z_i)x^*_ix_i^{*T}\bigg|x^*_i\right)+0\\
    &=\frac{1}{N^2}\sum_{i=1}^N\pi^R_ip_i(\beta_1)x^*_ix_i^{*T}
\end{aligned}
\end{equation}

\begin{equation}
\begin{aligned}
    Cov\left(\frac{1}{N}\sum_{i=1}^N\frac{\delta^B_i(y_i-\bar y_U)}{\pi^B_i}, \frac{1}{N}\sum_{i=1}^N\delta_i\big\{Z_i-p_i(\beta_1)\big\}\bigg|x^*_i\right)&=E_\delta\left[E_Z\left(\frac{1}{N}\sum_{i=1}^N\delta_i\frac{Z_i(y_i-\bar y_U)}{\pi^B_i}\bigg|\delta_i=1,x^*_i\right)\right]\\
    &=\frac{1}{N^2}\sum_{i=1}^N\big\{1-p_i(\beta_1)\big\}(y_i-\bar y_U)x^*_i
\end{aligned}
\end{equation}
Therefore, we have
\begin{equation}
Var\big\{\Phi_n(\eta_1)\big\}=
\begin{pmatrix}
\frac{1}{N^2}\sum_{i=1}^N\{(1-\pi^B_i)/\pi^B_i\}(y_i - \bar y_U) & \frac{1}{N^2}\sum_{i=1}^N\big\{1-p_i(\beta_1\big)\}(y_i-\bar y_U)x_i^{*T}\\
\frac{1}{N^2}\sum_{i=1}^N\big\{1-p_i(\beta_1\big)\}(y_i-\bar y_U)x^*_i & \frac{1}{N^2}\sum_{i=1}^N\pi^B_i\big\{1-p_i(\beta_1)\big\}x^*_ix_i^{*T}
\end{pmatrix}
\end{equation}
The final asymptotic variance estimator of $\hat{\bar y}_{PAPW}$ is given by
\begin{equation}
Var\big\{\hat{\bar y}_{PAPW}\big\}=
\frac{1}{N^2}\sum_{i=1}^N \bigg\{\frac{1-\pi^B_i}{\pi^B_i}\bigg\}(y_i - \bar y_U)^2-2\frac{b^T}{N^2}\sum_{i=1}^N\big\{1-p_i(\beta_1)\big\}(y_i-\bar y_U)x^*_i+b^T\left[\frac{1}{N^2}\sum_{i=1}^N\pi^B_i\big\{1-p_i(\beta_1)\big\}x^*_ix_i^{*T}\right]b
\end{equation}
To obtain the variance estimate based on the observed samples of $S_B$ and $S_R$, we substitute the population components with their estimates from the samples.
\begin{equation}
\widehat{Var}\big\{\hat{\bar y}_{PAPW}\big\}=
\frac{1}{N^2}\sum_{i=1}^{n_B} \big\{1-\hat\pi^B_i\big\}\left(\frac{y_i - \bar y_U}{\hat\pi^B_i}\right)^2-2\frac{\hat b^T}{N^2}\sum_{i=1}^{n_B}\big\{1-p_i(\hat\beta_1)\big\}\left(\frac{y_i-\bar y_U}{\hat\pi^B_i}\right)x^*_i+\hat b^T\left[\frac{1}{N^2}\sum_{i=1}^np_i(\hat\beta_1)x^*_ix_i^{*T}\right]\hat b
\end{equation}
where
\begin{equation}
    \hat b^T=\bigg\{\frac{1}{N}\sum_{i=1}^{n_B}\left(\frac{y_i-\bar y_U}{\hat\pi^B_i}\right)x_i^{*T}\bigg\}\bigg\{\frac{1}{N}\sum_{i=1}^n p_i(\hat\beta_1)x^*_ix_i^{*T}\bigg\}^{-1}
\end{equation}

\subsubsection{Proof of doubly robustness}\label{S:7.1.5}
As discussed in the section~\ref{S:2.4}, a doubly robust estimator should be consistent even if either model is misspecified. To prove the doubly robustness property of the AIPW estimator proposed here, let initially assume that $\hat\theta\overset{p}{\to}\theta$ if the prediction model (PM) is correctly specified, and  $\hat\phi\overset{p}{\to}\phi$ and $\hat\beta\overset{p}{\to}\beta$ if the pseudo-weighting model is correctly specified. Given the true probabilities of selection in $S_B$, we know that HT estimator is design-unbiased for the population total, i.e.
\begin{equation}\label{eq:19}
\begin{aligned}
E\left(\sum_{i=1}^{n_B}y_i/\pi^B_i\right)&=E\left(\sum_{i=1}^{N}\delta^B_i y_i/\pi^B_i\right)\\&=\sum_{i=1}^{N}E(\delta^B_i) y_i/\pi^B_i\\&=\sum_{i=1}^{N}\pi^B_i y_i/\pi^B_i\\&=\sum_{i=1}^{N}y_i\\&=\hat y_U
\end{aligned}
\end{equation}
And the same result will be obtained for $S_R$. Therefore
\begin{equation}\label{eq:19}
\begin{aligned}
E\left(\sum_{i=1}^{n_B}y_i/\pi^B_i\right)&=E\left(\sum_{i=1}^{n_R}y_i/\pi^R_i\right)\\&=\hat y_U
\end{aligned}
\end{equation}
Now we have
\begin{equation}\label{eq:19}
\begin{aligned}
\hat y_{DR}\overset{p}{\to}E(\hat y_{DR})&=E\bigg\{\sum_{i=1}^{n_B}\frac{(y_i-\hat y_i)}{\hat\pi^B_i}+\sum_{i=1}^{n_R}\frac{\hat y_i}{\pi^R_i}\bigg\}\\
&=E\bigg\{\sum_{i=1}^{n_B}\frac{(y_i-\hat y_i)}{\hat\pi^B_i}+\sum_{i=1}^{n_B}\frac{\hat y_i}{\pi^B_i}\bigg\}\\
&=E\bigg\{\sum_{i=1}^{n_B}\frac{(y_i-\hat y_i)}{\hat\pi^B_i}+\frac{\hat y_i}{\pi^B_i}\bigg\}\\
&=E\bigg\{\sum_{i=1}^{n_B}\frac{\hat y_i}{\pi^B_i}+\frac{(y_i-\hat y_i)}{\hat\pi^B_i}-\frac{(\hat y_i-\hat y_i)}{\pi^B_i}\bigg\}
\end{aligned}
\end{equation}
\begin{equation}\label{eq:19}
\begin{aligned}
\hat y_{DR}\overset{p}{\to}E(\hat y_{DR})&=y_U + E\bigg\{\sum_{i=1}^{n_B}(y_i-\hat y_i)(\frac{1}{\hat\pi^B_i}-\frac{1}{\pi^B_i})\bigg\}\\
&=y_U + E\bigg\{\sum_{i=1}^{n_B}(y_i-\hat y_i)(\frac{\pi^B_i}{\hat\pi^B_i}-1)\bigg\}
\end{aligned}
\end{equation}
Under the ignorable assumption in $S_B$, we have $Y\indep\pi^B|X, \pi^R$. Hence
\begin{equation}\label{eq:19}
\begin{aligned}
\hat y_{DR}\overset{p}{\to}E(\hat y_{DR})&=y_U + E\bigg\{\sum_{i=1}^{n_B}(y_i-\hat y_i)(\frac{\pi^B_i}{\hat\pi^B_i}-1)\bigg\}\\
&=y_U + E\bigg\{E\bigg\{\sum_{i=1}^{n_B}(y_i-\hat y_i)(\frac{\pi^B_i}{\hat\pi^B_i}-1)|x_i, \pi^R_i\bigg\}\bigg\}\\
&=y_U + E\bigg\{\sum_{i=1}^{n_B}E(y_i-\hat y_i|x_i, \pi^R_i)E(\frac{\pi^B_i}{\hat\pi^B_i}-1|x_i, \pi^R_i)\bigg\}
\end{aligned}
\end{equation}
If we assume the pseudo-weighting model is correctly specified, then we expect $\hat\pi^B_i \overset{p}{\to}\pi^B_i$ and
\begin{equation}\label{eq:19}
\begin{aligned}
E\left(\frac{\pi^B_i}{\hat\pi^B_i}-1|x_i, \pi^R_i\right)\overset{p}{\to}\frac{\pi^B_i}{\pi^B_i}-1=0
\end{aligned}
\end{equation}
which implies that $\hat y_{DR} \overset{p}{\to} y_U$ regardless of whether the PM is correctly specified or not. In situations where the mean model is correctly specified, then we expect that $\hat y_i \overset{p}{\to} y_i$. Hence
\begin{equation}\label{eq:19}
\begin{aligned}
E\left(y_i-\hat y_i|x_i, \pi^R_i\right)\overset{p}{\to}E\left(y_i-y_i|x_i, \pi^R_i\right)=0
\end{aligned}
\end{equation}
which means that $\hat y_{DR} \overset{p}{\to} y_U$ even if the PW model is incorrectly specified.

\subsubsection{Variance estimation under the Bayesian approach}\label{S:7.1.5}
As discussed in Section~\ref{S:2.6}, in this study, we use Rubin's combining rule to estimate the variance of the AIPW estimator under the two-step Bayesian approach. The idea stems from the conditional variance formula, which involves two parts: (1) within-imputation variance and between-imputation variance. The latter is straightforward and achieves by taking the variance of the $\hat{\bar y}^{(m)}_{DR}$ across the $M$ MCMC draws. The within-imputation variance requires more attention as one needs to account for the intraclass correlations due to clustering and use linearization techniques when dealing with a ratio estimator.\par 

It is clear that this component is calculated conditional on the observed $\hat y^{(m)}_i$ for $i\in S$, $\hat\pi^{B(m)}_i$ for $i\in S_B$ and $\hat pi^B_i$.
\begin{equation}\label{eq:19}
\begin{aligned}
var\left(\hat{\bar y}^{(m)}_{DR}\big|\hat\pi^{B(m)}_i,\hat y^{(m)}_i\right) & =var\left(\frac{1}{\hat N_B}\sum_{i=1}^{n_B}\frac{\left(y_i-\hat y^{(m)}_i\right)}{\hat\pi^B_i}+\frac{1}{\hat N_R}\sum_{i=1}^{n_R}\frac{\hat y^{(m)}_i}{\pi^R_i}\bigg|\hat\pi^{B(m)}_i,\hat y^{(m)}_i\right)\\
&=\frac{1}{\hat N^2_B}\sum_{i=1}^{n_B}\frac{var(y_i)}{\left(\hat\pi^B_i\right)^2}+var\left(\frac{1}{\hat N_R}\sum_{i=1}^{n_R}\frac{\hat y^{(m)}_i}{\pi^R_i}\bigg|\hat y^{(m)}_i\right)
\end{aligned}
\end{equation}
For the first component, which equals $var(y)\sum_{i=1}^{n_B}\left(\hat\pi^B_i\right)^{-2}/\hat N^2_B$, it suffices to estimate the variance of $y$. The second component, however, deals with the variance of a ratio estimator, which requires linearization techniques. Let's define $\hat t_R=\sum_{i=1}^{n_R}\hat y^{(m)}_i/\pi^R_i$, Taylor-series approximation of the variance is given by
\begin{equation}\label{eq:19}
\begin{aligned}
var\left(\frac{1}{\hat N_R}\sum_{i=1}^{n_R}\frac{\hat y^{(m)}_i}{\pi^R_i}\bigg|\hat y^{(m)}_i\right)&=var\left(\frac{\hat t_R}{\hat N_R}\bigg|\hat y^{(m)}_i\right)\\
&\approx\frac{1}{\hat N^2_R}\bigg\{var\left(\hat t_R\big|\hat y^{(m)}_i\right) + \left(\frac{\hat t_R}{\hat N_R}\right)^2var(\hat N_R)-2\left(\frac{\hat t_R}{\hat N_R}\right)cov\left(\hat t_R, \hat N_R\big|\hat y^{(m)}_i\right) \bigg\}
\end{aligned}
\end{equation}
Since $\hat t_R$ depends on $\hat y^{(m)}_i$, we have
\begin{equation}
var\left(\hat t_R\big|\hat y^{(m)}_i\right)=\sum_{i=1}^{n_R}\left(\hat y^{(m)}_i\right)^2var\left(\frac{1}{\pi^R_i}\right)
\end{equation}
\begin{equation}
cov\left(\hat t_R, \hat N_R\big|\hat y^{(m)}_i\right)=\sum_{i=1}^{n_R}\hat y^{(m)}_ivar\left(\frac{1}{\pi^R_i}\right)
\end{equation}
Therefore, the variance of the ratio estimator is approximated by
\begin{equation}\label{eq:19}
var\left(\frac{1}{\hat N_R}\sum_{i=1}^{n_R}\frac{\hat y^{(m)}_i}{\pi^R_i}\bigg|\hat y^{(m)}_i\right)\approx\frac{1}{\hat N^2_R}var\left(\frac{1}{\pi^R_i}\right)\bigg\{\sum_{i=1}^{n_R}\left(\hat y^{(m)}_i\right)^2 + n_R\left(\frac{\hat t_R}{\hat N_R}\right)^2-2\sum_{i=1}^{n_R}\hat y^{(m)}_i \bigg\}
\end{equation}
And the final within-imputation variance can be given by
\begin{equation}
var\left(\hat{\bar y}^{(m)}_{DR}\big|\hat\pi^{B(m)}_i,\hat y^{(m)}_i\right)\approx\frac{1}{\hat N^2_B}\sum_{i=1}^{n_B}\frac{var(y_i)}{\left(\hat\pi^B_i\right)^2}+\frac{1}{\hat N^2_R}var\left(\frac{1}{\pi^R_i}\right)\bigg\{\sum_{i=1}^{n_R}\left(\hat y^{(m)}_i\right)^2 + n_R\left(\frac{\hat t_R}{\hat N_R}\right)^2-2\sum_{i=1}^{n_R}\hat y^{(m)}_i \bigg\}
\end{equation}
Note that in situations where either $S_R$ or $S_B$ is a clustered sample, the derivation of the within-imputation variance would remain the same, but $y_i$, $\pi^R_i$, $\hat\pi^{B(m)}_i$, and $\hat y^{(m)}_i$ will represent the total for cluster $i$, and $n_R$ and $n_B$ are the number of clusters in $S_R$ and $S_B$, respectively.
\pagebreak

\subsection{Bayesian Additive Regression Trees}\label{S:7.2}
\begin{figure}[htp]
\centering\includegraphics[width=0.4\linewidth]{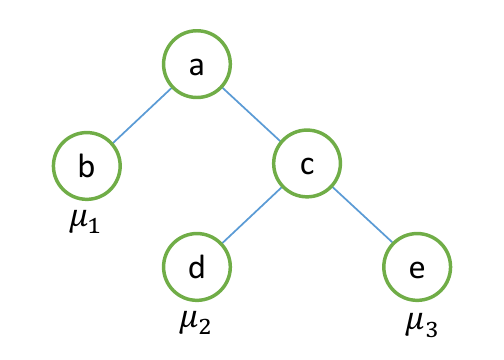}
\caption{Example of a binary-structured trees model}\label{fig:13}
\end{figure}

BART is a flexible ensemble of trees method, which allows handling non-linear relationships as well as multi-way interaction effects. The idea of BART is based on the sum-of-trees, where trees are sequentially modified on the basis of residuals from the other trees. In a tree-based method, the variation in the response variable is explained by hierarchically splitting the sample into more homogeneous subgroups \citep{green2012modeling}. As illustrated in Figure~\ref{fig:13}, a binary-structured tree consists of a root node, a set of interior nodes, a set of terminal nodes associated with parameters and decision rules that links these nodes \citep{abu2008bayesian}.

\subsubsection{BART for continuous outcomes}
Suppose $y=f(x)+\epsilon$ as is the case in every statistical model, where $y\in\mathbb{R}$ is a continuous outcome, $x$ denotes an $n \times p$ matrix of covariates, and $\epsilon \sim N(0, \sigma^2)$ is the error term. BART will then approximate the outcome as below:
\begin{equation}
\label{eq:6.1}
y \approx \sum_{j=1}^m f(x, T_j, M_j)
\end{equation}
where $T_j$ is the $j$-th tree with $b_j$ terminal nodes, and associated $M_j=(\mu_{1j}, \mu_{2j}, … , \mu_{{b_j}j} )^T$ parameters. BART is a Bayesian approach, since it assigns prior distributions to $T$, $M$, and $\sigma$ \citep{chipman2010bart, tan2016predicting}. Assuming an independence structure between trees, we can define the prior as follows:
\begin{equation}\label{eq:6.2}
p[(T_1, M_1), . . . , (T_m, M_m), \sigma^{-2}]=[\prod_{j=1}^m p(T_j, M_j)] p(\sigma^{-2})
\end{equation}
Using the multiplication law of probability, the joint distribution of $p(T_j, M_j)$ can be written as:
\begin{equation}\label{eq:6.3}
\begin{aligned}
p(T_j, M_j) &= p(M_j | T_j) p(T_j)\\ &=\prod_{i=1}^{b_j} p(\mu_{ij} | T_j) p(T_j)
\end{aligned}
\end{equation}
where $i=1, ... , b_j$ denotes the terminal node parameters for tree $j$. Therefore, the joint distribution in~\ref{eq:6.2} can be factored as below:
\begin{equation}\label{eq:6.4}
p\left[(T_1, M_1), . . . , (T_m, M_m), \sigma^{-2}\right]=\left[\prod_{j=1}^m\bigg\{\prod_{i=1}^{b_j}p(\mu_{ij}|T_j)\bigg\}p(T_j)\right] p(\sigma^{-2})
\end{equation}
Suggested by \cite{chipman2007bayesian}, the following distributions can be used for $\mu_{ij} |T_j$ and $\sigma ^{-2}$:
\begin{equation}
\mu_{ij}|T_j \sim N(\mu_{\mu}, \sigma_{\mu}^2)
\end{equation}
\begin{equation}
\sigma^{-2} \sim G(\frac{\nu}{2}, \frac{\nu \lambda}{2})
\end{equation}
The prior for $T_j$ involves three components of the tree structure: length of the tree, decision rules and the choice of covariate at a given node. However, prior specification for $T_j$ depends on several factors, and detailed discussions can be found in \cite{chipman2010bart}. Given the data, these parameters are updated through a combination of ``Bayesian backfitting'' and MCMC Gibbs sampler method. The trained trees are then summed up to approximate the outcome variable. Finally, $m$ is typically assumed to be fixed, but can be assessed by cross-validation. 
\par

\subsubsection{BART for binary outcomes}
For the binary outcome, a $probit$ link function is usually employed in the sense that $y$ is an indicator variable dichotomizing a normally distributed latent continuous outcome like $y^*$ at a real value $c$ so that:
\begin{equation}   
y = 
     \begin{cases}
       \text{1} &\quad\text{y}^\text{*} > \text{c}\\
       \text{0} &\quad\text{y}^{*} \leq \text{c} \\
     \end{cases}, \hspace{5mm} y^* \sim N(0, 1)
\end{equation}
Therefore, the new model will be given by:
\begin{equation}
G(x)=\Phi^{-1}[p(y=1 | x)] = \sum_{j=1}^{m} f(x, T_j, M_j)
\end{equation}
where $\Phi^{-1}[.]$ is the inverse of standard normal CDF. Since we implicitly assumed $\sigma \equiv 1$, the only priors we need to specify are  $p(\mu_{ij} | T_j)$ and $p(T_j)$. 
In order to be able to draw the posterior distribution of $T_j$ and $\mu_{ij}$, we need to generate the latent continuous variable, $y^*$, given $y_k$. \cite{chipman2010bart} recommends a data augmentation method based on the following algorithm:
\begin{equation}
y_k^* = 
\begin{cases}
max(N(G(x_k), 1), 0) & \text{if  } y_k=1\\
max(N(G(x_k), 1), 0) & \text{if  } y_k=0
\end{cases}
\end{equation}
Since the structure of priors is very similar to BART for continuous outcomes \citep{tan2016predicting}, we update the estimates $G(x_k)$ after drawing samples from $T_j$'s and $\mu_{ij}$'s. 
To apply BART, in this research, we utilize the `\textit{BayesTree}' and `\textit{BART}' packages in \textit{R}.\par

\clearpage

\subsection{Further extensions of the simulation study}\label{S:7.3}

\subsubsection{Simulation study I}\label{S:7.3.1}
This subsection provides additional results associated with Simulation I. Table~\ref{tab:110} and Table~\ref{tab:111} summarize the findings of the simulation in~\ref{S:3.1} under the frequentist approach when $n_B=100$, and $n_B=10,000$. We report the corresponding results under the two-step Bayesian approach in Table~\ref{tab:112} and Table~\ref{tab:113}, respectively.

\begin{table}[hbt!]
\caption{Comparing the performance of the bias adjustment methods and associated asymptotic variance estimator under the frequentist approach in the first simulation study for $n_R=100$ and $n_B=100$}\label{tab:110}
\begin{threeparttable}
\setlength{\tabcolsep}{5pt}
\scriptsize{\begin{tabular}{l l l l l l l l l l l l l l l l}
\toprule
 & \multicolumn{4}{c}{\textbf{$\rho=0.3$}}  &  &  \multicolumn{4}{c}{\textbf{$\rho=0.5$}}  &  &  \multicolumn{4}{c}{\textbf{$\rho=0.8$}}\\\cline{2-5}\cline{7-10}\cline{12-15}
\textbf{Method} & rBias & rMSE & crCI   & rSE &  & rBias & rMSE & crCI   & rSE &  & rBias & rMSE & crCI   & rSE \\
\midrule
\multicolumn{10}{l}{\textbf{Probability sample ($S_R$)}}\\
\hline
\hspace{2mm} Unweighted   & 8.528 & 19.248 & 92.6 & 1.009 &  & 8.647 & 11.065 & 77.4 & 1.018 &  & 8.682 & 9.719 & 50.9 & 1.02\\
\hspace{2mm} Fully weighted   & -0.029 & 20.276 & 94.7 & 1.001 &  & 0.006 & 8.035 & 95.1 & 1.010 &  & 0.015 & 5.008 & 94.9 & 1.008\\
\hline
\multicolumn{10}{l}{\textbf{Non-probability sample ($S_B$)}}  &    &    &    &  \\
\hline
\hspace{2mm} Unweighted   & 31.895 & 36.418 & 57.0 & 1.014 &  & 32.213 & 33.2 & 1.740 & 1.008 &  & 32.310 & 32.853 & 0.0 & 0.995\\
\hspace{2mm} Fully weighted   & 0.171 & 21.078 & 94.8 & 0.996 &  & 0.247 & 8.265 & 94.9 & 0.999 &  & 0.268 & 4.994 & 94.2 & 0.995\\
\hline
\multicolumn{10}{l}{\textbf{Non-robust adjustment}}   &    &    &    &  \\
\hline
\multicolumn{10}{l}{Model specification: True}   &    &    &    &  \\
\hline
\hspace{2mm} PAPW   & -1.192 & 23.466 & 95.2 & 1.018 &  & -1.205 & 9.452 & 95.3 & 1.015 &  & -1.211 & 5.982 & 95.8 & 1.007\\
\hspace{2mm} IPSW   & -2.917 & 26.505 & 97.3 & 1.386 &  & -3.036 & 12.700 & 97.0 & 1.355 &  & -3.075 & 9.470 & 97.0 & 1.308\\
\hline
\hspace{2mm} PM   & 0.372 & 20.989 & 94.6 & 0.994 &  & 0.148 & 8.351 & 94.9 & 0.995 &  & 0.077 & 5.160 & 95.0 & 0.992\\
\hline
\multicolumn{10}{l}{Model specification: False}   &    &    &    &  \\
\hline
\hspace{2mm} PAPW   & 27.140 & 33.436 & 75.6 & 1.059 &  & 27.393 & 28.814 & 16.6 & 1.043 &  & 27.470 & 28.276 & 2.5 & 1.025\\
\hspace{2mm} IPSW   & 28.372 & 33.972 & 67.9 & 1.012 &  & 28.711 & 29.951 & 8.3 & 1.002 &  & 28.815 & 29.515 & 0.5 & 0.99\\
\hline
\hspace{2mm} PM   & 28.199 & 33.790 & 68.4 & 1.011 &  & 28.541 & 29.771 & 8.3 & 1.001 &  & 28.645 & 29.337 & 0.3 & 0.988\\
\hline
\multicolumn{10}{l}{\textbf{Doubly robust adjustment}}   &    &    &    &  \\
\hline
\multicolumn{10}{l}{Model specification: QR--True, PM--True}   &    &    &    &  \\
\hline
\hspace{2mm} AIPW--PAPW   & -0.084 & 22.973 & 96.4 & 1.047 &  & -0.014 & 8.996 & 96.2 & 1.038 &  & 0.007 & 5.368 & 95.5 & 1.017\\
\hspace{2mm} AIPW--IPSW   & -0.184 & 22.449 & 96.3 & 1.046 &  & -0.049 & 8.826 & 96.1 & 1.038 &  & -0.009 & 5.314 & 95.9 & 1.016\\
\hline
\multicolumn{10}{l}{Model specification: QR--True, PM--False}   &    &    &    &  \\
\hline
\hspace{2mm} AIPW--PAPW   & -0.436 & 23.709 & 96.4 & 1.038 &  & -0.286 & 9.866 & 96.6 & 1.062 &  & -0.241 & 6.520 & 97.2 & 1.101\\
\hspace{2mm} AIPW--IPSW   & -0.427 & 23.083 & 96.4 & 1.039 &  & -0.227 & 9.570 & 96.6 & 1.070 &  & -0.166 & 6.298 & 97.5 & 1.119\\
\hline
\multicolumn{10}{l}{Model specification: QR--False, PM--True}   &    &    &    &  \\
\hline
\hspace{2mm} AIPW--PAPW   & -0.045 & 29.068 & 97.3 & 1.107 &  & 0.011 & 11.113 & 96.9 & 1.097 &  & 0.026 & 6.073 & 96.2 & 1.068\\
\hspace{2mm} AIPW--IPSW   & -0.194 & 28.208 & 97.5 & 1.104 &  & -0.044 & 10.825 & 97.1 & 1.094 &  & 0.001 & 5.974 & 96.5 & 1.062\\
\hline
\multicolumn{10}{l}{Model specification: QR--False, PM--False}   &    &    &    &  \\
\hline
\hspace{2mm} AIPW--PAPW   & 28.301 & 34.194 & 71.3 & 1.037 &  & 28.570 & 29.868 & 10.9 & 1.028 &  & 28.652 & 29.379 & 0.7 & 1.016\\
\hspace{2mm} AIPW--IPSW   & 28.178 & 33.806 & 70.4 & 1.035 &  & 28.525 & 29.764 & 9.4 & 1.025 &  & 28.631 & 29.326 & 0.5 & 1.013\\
\bottomrule
\end{tabular}}
    \begin{tablenotes}
      \footnotesize
      \item PAPW: propensity adjusted probability weighting; IPSW: Inverse propensity score weighting; QR: quasi-randomization; PM: prediction model; AIPW: augmented inverse propensity weighting. Fully weighted implies the weighted means if the true sampling weights are known.
    \end{tablenotes}
  \end{threeparttable}
\end{table}

\begin{table}[hbt!]
\caption{Comparing the performance of the bias adjustment methods and associated asymptotic variance estimator under the frequentist approach in the first simulation study for $n_R=100$ and $n_B=10,000$}\label{tab:111}
\begin{threeparttable}
\setlength{\tabcolsep}{5pt}
\scriptsize{\begin{tabular}{l l l l l l l l l l l l l l l l}
\toprule
 & \multicolumn{4}{c}{\textbf{$\rho=0.2$}}  &  &  \multicolumn{4}{c}{\textbf{$\rho=0.5$}}  &  &  \multicolumn{4}{c}{\textbf{$\rho=0.8$}}\\\cline{2-5}\cline{7-10}\cline{12-15}
\textbf{Method} & rBias & rMSE & crCI   & rSE &  & rBias & rMSE & crCI   & rSE &  & rBias & rMSE & crCI   & rSE \\
\midrule
\multicolumn{10}{l}{\textbf{Probability sample ($S_R$)}}\\
\hline
\hspace{2mm} Unweighted   & 8.528 & 19.248 & 92.6 & 1.009 &  & 8.647 & 11.065 & 77.4 & 1.018 &  & 8.682 & 9.719 & 50.9 & 1.02\\
\hspace{2mm} Fully weighted   & -0.029 & 20.276 & 94.7 & 1.001 &  & 0.006 & 8.035 & 95.1 & 1.010 &  & 0.015 & 5.008 & 94.9 & 1.008\\
\hline
\multicolumn{10}{l}{\textbf{Non-probability sample ($S_B$)}}  &    &    &    &  \\
\hline
\hspace{2mm} Unweighted   & 30.014 & 30.066 & 0.0 & 1.008 &  & 30.197 & 30.207 & 0.0 & 1.019 &  & 30.252 & 30.257 & 0.0 & 1.033\\
\hspace{2mm} Fully weighted   & 0.032 & 2.083 & 95.3 & 1.005 &  & 0.018 & 0.816 & 95.1 & 1.007 &  & 0.012 & 0.490 & 95.1 & 1.007\\
\hline
\multicolumn{10}{l}{\textbf{Non-robust adjustment}}   &    &    &    &  \\
\hline
\multicolumn{10}{l}{Model specification: True}   &    &    &    &  \\
\hline
\hspace{2mm} PAPW   & -2.067 & 4.582 & 94.9 & 1.108 &  & -2.145 & 4.120 & 92.8 & 1.107 &  & -2.170 & 4.072 & 92.2 & 1.107\\
\hspace{2mm} IPSW   & -2.618 & 7.717 & 94.5 & 0.958 &  & -2.673 & 7.334 & 91.1 & 0.923 &  & -2.692 & 7.308 & 90.6 & 0.979\\
\hline
\hspace{2mm} PM   & 0.296 & 4.515 & 95.2 & 0.994 &  & 0.121 & 4.134 & 94.8 & 0.986 &  & 0.065 & 4.095 & 94.6 & 0.985\\
\hline
\multicolumn{10}{l}{Model specification: False}   &    &    &    &  \\
\hline
\hspace{2mm} PAPW   & 24.493 & 24.616 & 0.0 & 1.126 &  & 24.592 & 24.651 & 0.0 & 1.153 &  & 24.621 & 24.673 & 0.0 & 1.161\\
\hspace{2mm} IPSW   & 26.675 & 26.804 & 0.0 & 0.992 &  & 26.871 & 26.949 & 0.0 & 0.970 &  & 26.930 & 27.002 & 0.0 & 0.964\\
\hline
\hspace{2mm} PM   & 26.509 & 26.645 & 0.0 & 1.003 &  & 26.717 & 26.800 & 0.0 & 0.989 &  & 26.779 & 26.856 & 0.0 & 0.986\\
\hline
\multicolumn{10}{l}{\textbf{Doubly robust adjustment}}   &    &    &    &  \\
\hline
\multicolumn{10}{l}{Model specification: QR--True, PM--True}   &    &    &    &  \\
\hline
\hspace{2mm} AIPW--PAPW   & 0.180 & 4.633 & 95.1 & 0.994 &  & 0.080 & 4.162 & 94.8 & 0.986 &  & 0.047 & 4.104 & 94.7 & 0.985\\
\hspace{2mm} AIPW--IPSW   & 0.052 & 4.582 & 95.2 & 0.995 &  & 0.035 & 4.152 & 94.6 & 0.987 &  & 0.028 & 4.101 & 94.5 & 0.985\\
\hline
\multicolumn{10}{l}{Model specification: QR--True, PM--False}   &    &    &    &  \\
\hline
\hspace{2mm} AIPW--PAPW   & 0.262 & 4.719 & 95.1 & 1.000 &  & 0.163 & 4.250 & 94.9 & 0.997 &  & 0.130 & 4.191 & 94.7 & 0.996\\
\hspace{2mm} AIPW--IPSW   & 0.188 & 4.652 & 95.4 & 1.002 &  & 0.171 & 4.225 & 95.0 & 0.998 &  & 0.164 & 4.174 & 94.8 & 0.998\\
\hline
\multicolumn{10}{l}{Model specification: QR--False, PM--True}   &    &    &    &  \\
\hline
\hspace{2mm} AIPW--PAPW   & 1.376 & 8.569 & 94.5 & 0.953 &  & 0.503 & 4.829 & 95.1 & 0.995 &  & 0.231 & 4.215 & 95.2 & 0.992\\
\hspace{2mm} AIPW--IPSW   & 0.864 & 7.648 & 94.7 & 0.948 &  & 0.322 & 4.643 & 95.3 & 0.990 &  & 0.152 & 4.182 & 95.0 & 0.989\\
\hline
\multicolumn{10}{l}{Model specification: QR--False, PM--False}   &    &    &    &  \\
\hline
\hspace{2mm} AIPW--PAPW   & 26.696 & 26.835 & 0.0 & 0.998 &  & 26.779 & 26.862 & 0.0 & 0.987 &  & 26.803 & 26.880 & 0.0 & 0.985\\
\hspace{2mm} AIPW--IPSW   & 26.520 & 26.655 & 0.0 & 1.001 &  & 26.718 & 26.801 & 0.0 & 0.989 &  & 26.777 & 26.854 & 0.0 & 0.986\\
\bottomrule
\end{tabular}}
    \begin{tablenotes}
      \footnotesize
      \item PAPW: propensity adjusted probability weighting; IPSW: Inverse propensity score weighting; QR: quasi-randomization; PM: prediction model; AIPW: augmented inverse propensity weighting. Fully weighted implies the weighted means if the true sampling weights are known.
    \end{tablenotes}
  \end{threeparttable}
\end{table}

\begin{table}[hbt!]
\caption{Comparing the performance of the bias adjustment methods and associated variance estimator under the two-step Bayesian approach in the first simulation study for $n_R=100$ and $n_B=100$}\label{tab:112}
\begin{threeparttable}
\setlength{\tabcolsep}{3pt}
\scriptsize{\begin{tabular}{l l l l l l l l l l l l l l l l}
\toprule
 & \multicolumn{4}{c}{\textbf{$\rho=0.3$}}  &  &  \multicolumn{4}{c}{\textbf{$\rho=0.5$}}  &  &  \multicolumn{4}{c}{\textbf{$\rho=0.8$}}\\\cline{2-5}\cline{7-10}\cline{12-15}
\textbf{Method} & rBias & rMSE & crCI   & rSE &  & rBias & rMSE & crCI   & rSE &  & rBias & rMSE & crCI   & rSE \\
\midrule
\multicolumn{10}{l}{\textbf{Probability sample ($S_R$)}}\\
\hline
\hspace{2mm} Unweighted   & 8.528 & 19.248 & 92.6 & 1.009 &  & 8.647 & 11.065 & 77.4 & 1.018 &  & 8.682 & 9.719 & 50.9 & 1.020\\
\hspace{2mm} Fully weighted   & -0.029 & 20.276 & 94.7 & 1.001 &  & 0.006 & 8.035 & 95.1 & 1.010 &  & 0.015 & 5.008 & 95.0 & 1.008\\
\hline
\multicolumn{10}{l}{\textbf{Non-probability sample ($S_B$)}}   &    &    &    &  \\
\hline
\hspace{2mm} Unweighted   & 32.238 & 36.815 & 56.3 & 1.003 &  & 32.303 & 33.3 & 1.620 & 1.003 &  & 32.322 & 32.865 & 0.0 & 0.996\\
\hspace{2mm} Fully weighted   & 0.494 & 21.398 & 94.3 & 0.981 &  & 0.329 & 8.400 & 94.0 & 0.981 &  & 0.276 & 5.057 & 93.6 & 0.979\\
\hline
\multicolumn{10}{l}{\textbf{Non-robust adjustment}}   &    &    &    &  \\
\hline
\multicolumn{10}{l}{Model specification: True}   &    &    &    &  \\
\hline
\hspace{2mm} PAPW   & -0.589 & 24.195 & 97.4 & 1.117 &  & -0.755 & 9.795 & 99.0 & 1.326 &  & -0.801 & 6.178 & 99.8 & 1.653\\
\hspace{2mm} IPSW   & 1.169 & 22.844 & 97.2 & 1.118 &  & 1.016 & 9.163 & 98.6 & 1.345 &  & 0.976 & 5.719 & 99.8 & 1.701\\
\hline
\hspace{2mm} PM   & 0.709 & 21.489 & 95.280 & 1.029 &  & 0.272 & 8.545 & 95.580 & 1.020 &  & 0.140 & 5.245 & 94.640 & 1.000\\
\hline
\multicolumn{10}{l}{Model specification: False}   &    &    &    &  \\
\hline
\hspace{2mm} PAPW   & 28.008 & 34.396 & 76.3 & 1.091 &  & 28.027 & 29.477 & 19.6 & 1.116 &  & 28.022 & 28.840 & 3.4 & 1.141\\
\hspace{2mm} IPSW   & 29.763 & 35.215 & 70.2 & 1.083 &  & 29.827 & 31.032 & 10.0 & 1.106 &  & 29.841 & 30.519 & 0.8 & 1.125\\
\hline
\hspace{2mm} PM   & 28.588 & 34.226 & 70.9 & 1.055 &  & 28.658 & 29.895 & 10.6 & 1.050 &  & 28.691 & 29.380 & 0.7 & 1.042\\
\hline
\multicolumn{10}{l}{\textbf{Doubly robust adjustment}}   &    &    &    &  \\
\hline
\multicolumn{10}{l}{Model specification: QR--True, PM--True}   &    &    &    &  \\
\hline
\hspace{2mm} AIPW--PAPW   & 0.320 & 23.802 & 97.8 & 1.154 &  & 0.125 & 9.306 & 99.1 & 1.357 &  & 0.067 & 5.493 & 99.9 & 1.731\\
\hspace{2mm} AIPW--IPSW   & 0.249 & 22.778 & 97.4 & 1.142 &  & 0.099 & 8.976 & 99.1 & 1.339 &  & 0.056 & 5.387 & 99.9 & 1.688\\
\hline
\multicolumn{10}{l}{Model specification: QR--True, PM--False}   &    &    &    &  \\
\hline
\hspace{2mm} AIPW--PAPW   & 0.304 & 23.858 & 97.7 & 1.156 &  & 0.126 & 9.386 & 99.2 & 1.389 &  & 0.065 & 5.661 & 99.9 & 1.781\\
\hspace{2mm} AIPW--IPSW   & 0.226 & 22.814 & 97.5 & 1.146 &  & 0.096 & 9.041 & 99.1 & 1.376 &  & 0.052 & 5.543 & 99.8 & 1.747\\
\hline
\multicolumn{10}{l}{Model specification: QR--False, PM--True}   &    &    &    &  \\
\hline
\hspace{2mm} AIPW--PAPW   & 0.881 & 22.077 & 96.8 & 1.126 &  & 0.333 & 8.742 & 98.6 & 1.281 &  & 0.153 & 5.303 & 99.8 & 1.558\\
\hspace{2mm} AIPW--IPSW   & 0.762 & 21.483 & 96.6 & 1.103 &  & 0.290 & 8.554 & 98.4 & 1.251 &  & 0.135 & 5.246 & 99.7 & 1.509\\
\hline
\multicolumn{10}{l}{Model specification: QR--False, PM--False}   &    &    &    &  \\
\hline
\hspace{2mm} AIPW--PAPW   & 28.659 & 34.756 & 77.6 & 1.135 &  & 28.660 & 30.013 & 17.4 & 1.142 &  & 28.649 & 29.399 & 2.1 & 1.151\\
\hspace{2mm} AIPW--IPSW   & 28.575 & 34.237 & 74.7 & 1.115 &  & 28.656 & 29.903 & 13.7 & 1.124 &  & 28.674 & 29.368 & 1.1 & 1.132\\
\bottomrule
\end{tabular}}
    \begin{tablenotes}
      \footnotesize
      \item PAPW: propensity adjusted probability weighting; IPSW: Inverse propensity score weighting; QR: quasi-randomization; PM: prediction model; AIPW: augmented inverse propensity weighting. Fully weighted implies the weighted means if the true sampling weights are known.
    \end{tablenotes}
  \end{threeparttable}
\end{table}

\begin{table}[hbt!]
\caption{Comparing the performance of the bias adjustment methods and associated variance estimator under the two-step Bayesian approach in the first simulation study for $n_R=100$ and $n_B=10,000$}\label{tab:113}
\begin{threeparttable}
\setlength{\tabcolsep}{5pt}
\scriptsize{\begin{tabular}{l l l l l l l l l l l l l l l l}
\toprule
 & \multicolumn{4}{c}{\textbf{$\rho=0.2$}}  &  &  \multicolumn{4}{c}{\textbf{$\rho=0.5$}}  &  &  \multicolumn{4}{c}{\textbf{$\rho=0.8$}}\\\cline{2-5}\cline{7-10}\cline{12-15}
\textbf{Method} & rBias & rMSE & crCI   & rSE &  & rBias & rMSE & crCI   & rSE &  & rBias & rMSE & crCI   & rSE \\
\midrule
\multicolumn{10}{l}{\textbf{Probability sample ($S_R$)}}\\
\hline
\hspace{2mm} Unweighted   & 8.528 & 19.248 & 92.6 & 1.009 &  & 8.647 & 11.065 & 77.4 & 1.018 &  & 8.682 & 9.719 & 50.9 & 1.020\\
\hspace{2mm} Fully weighted   & -0.029 & 20.276 & 94.7 & 1.001 &  & 0.006 & 8.035 & 95.1 & 1.010 &  & 0.015 & 5.008 & 94.9 & 1.008\\
\hline
\multicolumn{10}{l}{\textbf{Non-probability sample ($S_B$)}}   &    &    &    &  \\
\hline
\hspace{2mm} Unweighted   & 30.014 & 30.066 & 0.0 & 1.008 &  & 30.197 & 30.207 & 0.0 & 1.019 &  & 30.252 & 30.257 & 0.0 & 1.033\\
\hspace{2mm} Fully weighted   & 0.032 & 2.083 & 95.3 & 1.005 &  & 0.018 & 0.816 & 95.1 & 1.007 &  & 0.012 & 0.490 & 95.1 & 1.007\\
\hline
\multicolumn{10}{l}{\textbf{Non-robust adjustment}}   &    &    &    &  \\
\hline
\multicolumn{10}{l}{Model specification: True}   &    &    &    &  \\
\hline
\hspace{2mm} PAPW   & -2.032 & 4.578 & 93.0 & 1.031 &  & -2.106 & 4.111 & 90.9 & 1.032 &  & -2.138 & 4.062 & 90.2 & 1.035\\
\hspace{2mm} IPSW   & -0.015 & 4.094 & 95.2 & 1.011 &  & -0.036 & 3.605 & 95.1 & 1.004 &  & -0.042 & 3.547 & 95.2 & 1.002\\
\hline
\hspace{2mm} PM   & 0.297 & 4.517 & 81.6 & 0.679 &  & 0.120 & 4.136 & 75.3 & 0.579 &  & 0.065 & 4.094 & 73.1 & 0.563\\
\hline
\multicolumn{10}{l}{Model specification: False}   &    &    &    &  \\
\hline
\hspace{2mm} PAPW   & 24.524 & 24.647 & 0.0 & 1.042 &  & 24.618 & 24.678 & 0.0 & 1.062 &  & 24.650 & 24.702 & 0.0 & 1.069\\
\hspace{2mm} IPSW   & 26.406 & 26.518 & 0.0 & 0.982 &  & 26.602 & 26.662 & 0.0 & 0.940 &  & 26.663 & 26.717 & 0.0 & 0.931\\
\hline
\hspace{2mm} PM   & 26.512 & 26.648 & 0.0 & 0.851 &  & 26.715 & 26.798 & 0.0 & 0.728 &  & 26.779 & 26.856 & 0.0 & 0.700\\
\hline
\multicolumn{10}{l}{\textbf{Doubly robust adjustment}}   &    &    &    &  \\
\hline
\multicolumn{10}{l}{Model specification: QR--True, PM--True}   &    &    &    &  \\
\hline
\hspace{2mm} AIPW--PAPW   & 0.178 & 4.635 & 84.7 & 0.721 &  & 0.079 & 4.160 & 77.3 & 0.607 &  & 0.047 & 4.103 & 75.7 & 0.588\\
\hspace{2mm} AIPW--IPSW   & 0.058 & 4.574 & 83.6 & 0.705 &  & 0.036 & 4.149 & 77.0 & 0.601 &  & 0.028 & 4.100 & 75.5 & 0.585\\
\hline
\multicolumn{10}{l}{Model specification: QR--True, PM--False}   &    &    &    &  \\
\hline
\hspace{2mm} AIPW--PAPW   & 0.151 & 4.273 & 94.5 & 0.971 &  & 0.050 & 3.734 & 93.7 & 0.943 &  & 0.025 & 3.660 & 93.9 & 0.941\\
\hspace{2mm} AIPW--IPSW   & 0.106 & 4.245 & 94.4 & 0.966 &  & 0.083 & 3.767 & 93.7 & 0.945 &  & 0.075 & 3.712 & 93.7 & 0.941\\
\hline
\multicolumn{10}{l}{Model specification: QR--False, PM--True}   &    &    &    &  \\
\hline
\hspace{2mm} AIPW--PAPW   & 0.496 & 4.566 & 83.7 & 0.709 &  & 0.193 & 4.142 & 76.8 & 0.599 &  & 0.096 & 4.096 & 75.2 & 0.581\\
\hspace{2mm} AIPW--IPSW   & 0.312 & 4.514 & 82.7 & 0.695 &  & 0.127 & 4.133 & 76.7 & 0.595 &  & 0.068 & 4.094 & 74.9 & 0.579\\
\hline
\multicolumn{10}{l}{Model specification: QR--False, PM--False}   &    &    &    &  \\
\hline
\hspace{2mm} AIPW--PAPW   & 26.709 & 26.849 & 0.0 & 0.893 &  & 26.786 & 26.869 & 0.0 & 0.751 &  & 26.808 & 26.885 & 0.0 & 0.717\\
\hspace{2mm} AIPW--IPSW   & 26.521 & 26.656 & 0.0 & 0.870 &  & 26.718 & 26.800 & 0.0 & 0.740 &  & 26.777 & 26.854 & 0.0 & 0.709\\
\bottomrule
\end{tabular}}
    \begin{tablenotes}
      \footnotesize
      \item PAPW: propensity adjusted probability weighting; IPSW: Inverse propensity score weighting; QR: quasi-randomization; PM: prediction model; AIPW: augmented inverse propensity weighting. Fully weighted implies the weighted means if the true sampling weights are known.
    \end{tablenotes}
  \end{threeparttable}
\end{table}

\newpage
\clearpage
\subsubsection{Simulation study III}\label{S:7.3.2}

Table~\ref{tab:120} and Table~\ref{tab:121} exhibits the numerical results associated with the plots of Simulation III.

\begin{table}[hbt!]
\caption{Comparing the performance of the bias adjustment methods in the third simulation study for $\rho=0.8$}\label{tab:120}
\begin{threeparttable}
\begin{tabular}{l l l l l l l l l l l}
\toprule
 & \multicolumn{4}{c}{\textbf{Continuous outcome ($Y_c$)}}  &  &  \multicolumn{4}{c}{\textbf{Binary outcome ($Y_b$)}}\\\cline{2-5}\cline{7-10}
\textbf{Model-method} & rBias & rMSE & crCI   & rSE &  & rBias & rMSE & crCI   & rSE \\
\midrule
\multicolumn{10}{l}{\textbf{Probability sample ($S_R$)}}\\
\hline
\hspace{2mm} Unweighted  & 48.705 & 52.900 & 30.7 & 1.015 &  & 11.304 & 16.881 & 88.2 & 1.022\\
\hspace{2mm} Fully weighted  & 0.080 & 15.400 & 96.2 & 1.025 &  & 0.131 & 13.858 & 95.3 & 1.026\\
\hline
\multicolumn{10}{l}{\textbf{Non-probability sample ($S_B$)}}\\
\hline
\hspace{2mm} Unweighted  & 68.309 & 70.415 & 0.0 & 0.156 &  & 21.763 & 22.794 & 0.5 & 0.181\\
\hspace{2mm} Fully weighted  & 0.137 & 7.581 & 95.7 & 1.023 &  & 0.074 & 6.512 & 94.7 & 0.99\\
\hline
\multicolumn{10}{l}{\textbf{Non-robust adjustment}}\\
\hline
\multicolumn{10}{l}{Model specification: True}\\
\hline
\hspace{2mm} GLM--PAPW  & 0.448 & 10.994 & 94.7 & 1.036 &  & 0.072 & 7.266 & 96.2 & 1.034\\
\hspace{2mm} GLM--PAPP  & 0.204 & 11.192 & 93.9 & 1.037 &  & 0.080 & 7.188 & 96.2 & 1.031\\
\hspace{2mm} GLM--IPSW  & 0.839 & 18.138 & 96.0 & 1.275 &  & -0.838 & 9.458 & 97.3 & 1.116\\
\hline
\hspace{2mm} GLM--PM  & 0.110 & 11.157 & 94.2 & 1.015 &  & 0.055 & 7.401 & 94.4 & 0.995\\
\hline
\multicolumn{10}{l}{Model specification: False}\\
\hline
\hspace{2mm} GLM--PAPW  & 7.337 & 13.187 & 94.2 & 1.033 &  & 5.115 & 8.502 & 90.4 & 1.02\\
\hspace{2mm} GLM--PAPP  & 6.762 & 13.546 & 94.2 & 1.032 &  & 5.046 & 8.471 & 88.5 & 1.035\\
\hspace{2mm} GLM--IPSW  & 22.513 & 35.600 & 99.5 & 1.155 &  & 9.390 & 13.098 & 89.5 & 1.099\\
\hspace{2mm}BART--PAPW  & 2.272 & 10.468 & 100.0 & 2.487 &  & 1.633 & 7.391 & 99.5 & 1.436\\
\hspace{2mm}BART--PAPP  & 3.990 & 11.469 & 100.0 & 2.299 &  & 0.313 & 7.243 & 99.3 & 1.342\\
\hline
\hspace{2mm} GLM--PM  & 37.071 & 42.523 & 53.0 & 1.006 &  & 12.600 & 14.932 & 63.6 & 1.003\\
\hspace{2mm} BART--PM  & 0.286 & 11.581 & 92.7 & 0.996 &  & 0.594 & 9.102 & 81.2 & 0.688\\
\hline
\multicolumn{10}{l}{\textbf{Doubly robust adjustment}}\\
\hline
\multicolumn{10}{l}{Model specification: QR--True, PM--True}\\
\hline
\hspace{2mm} GLM--AIPW--PAPW  & 0.307 & 11.186 & 95.0 & 1.019 &  & 0.083 & 7.459 & 94.2 & 1.001\\
\hspace{2mm} GLM--AIPW--PAPP  & 0.295 & 11.187 & 94.5 & 1.019 &  & 0.089 & 7.439 & 94.0 & 0.998\\
\hspace{2mm} GLM--AIPW--IPSW  & 0.372 & 11.193 & 95.8 & 1.037 &  & 0.120 & 7.478 & 94.4 & 1.003\\
\hline
\multicolumn{10}{l}{Model specification: QR--True, PM--False}\\
\hline
\hspace{2mm} GLM--AIPW--PAPW  & 0.381 & 12.774 & 95.5 & 1.035 &  & 0.047 & 7.487 & 96.2 & 1.04\\
\hspace{2mm} GLM--AIPW--PAPP  & 0.424 & 11.934 & 94.7 & 1.041 &  & 0.155 & 7.275 & 96.0 & 1.032\\
\hspace{2mm} GLM--AIPW--IPSW  & -8.223 & 17.625 & 92.3 & 1.181 &  & -2.842 & 9.086 & 95.2 & 1.047\\
\hline
\multicolumn{10}{l}{Model specification: QR--False, PM--True}\\
\hline
\hspace{2mm} GLM--AIPW--PAPW  & 0.127 & 11.177 & 94.7 & 1.020 &  & 0.067 & 7.451 & 94.0 & 0.997\\
\hspace{2mm} GLM--AIPW--PAPP  & 0.122 & 11.172 & 94.7 & 1.019 &  & 0.054 & 7.438 & 94.2 & 0.997\\
\hspace{2mm} GLM--AIPW--IPSW  & 0.117 & 11.167 & 94.8 & 1.020 &  & 0.055 & 7.433 & 94.0 & 0.998\\
\hline
\multicolumn{10}{l}{Model specification: QR--False, PM--False}\\
\hline
\hspace{2mm}GLM--AIPW--PAPW  & 50.327 & 53.922 & 21.9 & 1.002 &  & 15.651 & 17.552 & 50.3 & 1.007\\
\hspace{2mm}GLM--AIPW--PAPP  & 50.793 & 54.215 & 20.9 & 1.002 &  & 15.834 & 17.605 & 47.8 & 1.003\\
\hspace{2mm}GLM--AIPW--IPSW  & 47.867 & 51.106 & 27.9 & 1.163 &  & 15.112 & 16.884 & 53.8 & 1.051\\
\hspace{2mm}BART--AIPW--PAPW  & 0.276 & 11.593 & 94.4 & 1.035 &  & 0.701 & 9.186 & 81.9 & 0.698\\
\hspace{2mm}BART--AIPW--PAPP  & 0.261 & 11.591 & 94.2 & 1.031 &  & 0.682 & 9.155 & 81.7 & 0.697\\
\bottomrule
\end{tabular}
    \begin{tablenotes}
      \footnotesize
      \item PAPW: propensity adjusted probability weighting; PAPP: propensity adjusted probability Prediction; IPSW: Inverse propensity score weighting; QR: quasi-randomization; PM: prediction model; AIPW: augmented inverse propensity weighting.
    \end{tablenotes}
  \end{threeparttable}
\end{table}

\begin{table}[hbt!]
\caption{Comparing the values of rBias and rMSE for different methods across different values of $\rho$.}\label{tab:121}
\begin{threeparttable}
\setlength{\tabcolsep}{3pt}
\scriptsize{\begin{tabular}{l l l l l l l l l l l l l l l l l}
\toprule
 & \multicolumn{7}{c}{\textbf{Continuous outcome ($Y_c$)}}  &  &  \multicolumn{7}{c}{\textbf{Binary outcome ($Y_b$)}}\\\cline{2-8}\cline{10-16}
 & \multicolumn{4}{c}{Non-robust} & \multicolumn{3}{c}{Doubly robust}  &  &  \multicolumn{4}{c}{Non-robust} & \multicolumn{3}{c}{Doubly robust}\\\cline{2-8}\cline{10-16}
\textbf{$\rho$} & PAPW & PAPP & IPSW & PM & PAPW & PAPP & IPSW &  & PAPW & PAPP & IPSW & PM & PAPW & PAPP & IPSW\\
\midrule
 & \multicolumn{15}{c}{\textbf{rBias}}\\
\hline
\textbf{0.0}  & 0.545 & 0.870 & -6.791 & 0.259 & 0.447 & 0.443 & 0.511 &    & -0.186 & 0.128 & -3.248 & -0.016 & -0.006 & -0.005 & 0.004\\
\textbf{0.1}  & -0.179 & 0.022 & -4.772 & -0.224 & -0.215 & -0.218 & -0.235 &    & -0.537 & -0.220 & -2.345 & -0.399 & -0.464 & -0.475 & -0.489\\
\textbf{0.2}  & -0.195 & 0.493 & -4.406 & 0.048 & -0.161 & -0.160 & -0.250 &    & -0.329 & 0.095 & -2.071 & -0.144 & -0.159 & -0.149 & -0.172\\
\textbf{0.3}  & 0.288 & 0.668 & -3.832 & 0.420 & 0.459 & 0.449 & 0.435 &    & -0.244 & 0.069 & -1.980 & 0.160 & 0.108 & 0.114 & 0.111\\
\textbf{0.4}  & 0.212 & 0.361 & -2.332 & 0.425 & 0.237 & 0.254 & 0.233 &    & -0.097 & 0.150 & -1.372 & -0.031 & 0.031 & 0.048 & 0.085\\
\textbf{0.5}  & 0.248 & 0.173 & -2.257 & 0.175 & 0.227 & 0.216 & 0.239 &    & -0.169 & -0.067 & -1.817 & 0.117 & 0.068 & 0.065 & -0.010\\
\textbf{0.6}  & 0.286 & 0.516 & -1.010 & 0.411 & 0.404 & 0.404 & 0.420 &    & 0.128 & 0.231 & -1.146 & 0.133 & 0.162 & 0.156 & 0.198\\
\textbf{0.7}  & 0.072 & -0.052 & -0.217 & -0.027 & 0.084 & 0.084 & 0.100 &    & -0.019 & 0.062 & -1.029 & 0.021 & -0.001 & 0.009 & -0.001\\
\textbf{0.8}  & 0.538 & 0.527 & 0.652 & 0.623 & 0.509 & 0.517 & 0.498 &    & 0.012 & 0.175 & -1.017 & 0.015 & 0.053 & 0.048 & 0.078\\
\textbf{0.9}  & 0.343 & 0.424 & 2.090 & 0.469 & 0.496 & 0.495 & 0.466 &    & 0.079 & 0.122 & -0.932 & 0.155 & 0.158 & 0.144 & 0.164\\
\hline
  &  \multicolumn{15}{c}{\textbf{rMSE}}\\
\hline
\textbf{0.0}  & 13.916 & 14.724 & 20.949 & 14.934 & 14.964 & 14.934 & 14.994 &  & 8.702 & 8.702 & 11.773 & 9.214 & 9.214 & 9.214 & 9.214\\
\textbf{0.1}  & 12.979 & 13.640 & 19.378 & 13.760 & 13.790 & 13.790 & 13.850 &  & 8.443 & 8.188 & 10.746 & 8.699 & 8.699 & 8.699 & 8.699\\
\textbf{0.2}  & 12.297 & 13.220 & 18.877 & 13.161 & 13.220 & 13.220 & 13.250 &  & 8.237 & 8.237 & 10.811 & 8.494 & 8.494 & 8.494 & 8.494\\
\textbf{0.3}  & 12.187 & 13.049 & 18.132 & 13.019 & 13.049 & 13.019 & 13.049 &  & 7.859 & 7.859 & 9.887 & 8.113 & 8.113 & 8.113 & 8.366\\
\textbf{0.4}  & 11.823 & 12.392 & 17.330 & 12.452 & 12.511 & 12.511 & 12.511 &  & 7.910 & 7.654 & 9.951 & 8.165 & 8.165 & 8.165 & 8.165\\
\textbf{0.5}  & 11.745 & 12.101 & 17.647 & 12.190 & 12.190 & 12.190 & 12.190 &  & 7.576 & 7.576 & 9.849 & 7.829 & 7.829 & 7.829 & 7.829\\
\textbf{0.6}  & 11.691 & 12.145 & 18.748 & 12.085 & 12.115 & 12.115 & 12.115 &  & 7.673 & 7.673 & 9.975 & 7.929 & 7.929 & 7.929 & 7.929\\
\textbf{0.7}  & 10.927 & 11.166 & 15.567 & 11.196 & 11.226 & 11.226 & 11.226 &  & 7.332 & 7.080 & 8.849 & 7.332 & 7.585 & 7.332 & 7.585\\
\textbf{0.8}  & 10.769 & 10.918 & 17.888 & 10.918 & 10.918 & 10.918 & 10.948 &  & 7.359 & 7.105 & 9.389 & 7.359 & 7.612 & 7.359 & 7.612\\
\textbf{0.9}  & 10.951 & 11.042 & 22.084 & 10.981 & 11.012 & 11.012 & 11.012 &  & 6.935 & 6.935 & 8.990 & 7.192 & 7.192 & 7.192 & 7.192\\
\bottomrule
\end{tabular}}
    \begin{tablenotes}
      \small
      \item NOTE: GLM has been used for prediction, and the underlying models in each method have been correctly specified.
    \end{tablenotes}
  \end{threeparttable}
\end{table}

\newpage
\clearpage
\subsection{Supplemental results of the application on SHRP2 data}\label{S:7.4}

\begin{table}[hbt!]
\caption{Mean daily trip duration (min) and associated 95\% CIs by different covariates across DR adjustment methods}\label{tab:5}
\begin{threeparttable}
\setlength{\tabcolsep}{3pt}
\scriptsize{\begin{tabular}{l l l l l l}
\toprule
\textbf{ } & \textbf{ } & \textbf{Unweighted} & \textbf{GLM-AIPW-PAPP} & \textbf{GLM-AIPW-PMLE} & \textbf{BART--AIPW-PAPP}\\
\textbf{Covariate} & \textbf{n} & \textbf{(95\%CI)} & \textbf{(95\%CI)} & \textbf{(95\%CI)} & \textbf{(95\%CI)}\\
\midrule
\textbf{Total}  & 837,061 & 68.94 (67.955,69.925) & 71.603 (66.565,76.641) & 70.058 (67.902,72.214) & 69.582 (66.117,73.047)\\
\hline
\textbf{Gender}  &    &    &    &    &  \\
Male  & 407,312 & 70.289 (68.809,71.77) & 72.411 (63.583,81.238) & 70.97 (67.971,73.97) & 70.61 (66.131,75.088)\\
Female  & 429,749 & 67.662 (66.355,68.968) & 70.79 (67.353,74.226) & 69.107 (66.683,71.531) & 68.522 (64.432,72.611)\\
\hline
\textbf{Age group}  &    &    &    &    &  \\
16-24  & 311,106 & 70 (68.514,71.485) & 72.889 (69.435,76.342) & 72.318 (69.636,74.999) & 71.937 (66.79,77.085)\\
25-34  & 117,758 & 73.889 (71.099,76.679) & 72.669 (67.713,77.625) & 71.562 (67.688,75.435) & 72.511 (66.132,78.889)\\
35-44  & 61,908 & 75.4 (71.304,79.496) & 71.215 (64.668,77.762) & 75.72 (69.882,81.559) & 71.919 (63.874,79.964)\\
45-54  & 77,903 & 74.666 (71.734,77.599) & 71.803 (61.432,82.175) & 70.437 (66.525,74.349) & 73.237 (67.727,78.747)\\
55-64  & 63,891 & 70.823 (67.027,74.62) & 66.99 (60.85,73.13) & 67.054 (62.252,71.855) & 67.518 (60.885,74.152)\\
65-74  & 88,762 & 67.122 (64.13,70.113) & 84.262 (52.155,116.369) & 64.475 (59.374,69.576) & 64.286 (59.779,68.794)\\
75+  & 115,733 & 54.103 (51.965,56.241) & 49.358 (46.14,52.576) & 51.359 (47.896,54.822) & 51.442 (46.894,55.99)\\
\hline
\textbf{Race}  &    &    &    &    &  \\
White  & 745,596 & 67.845 (66.833,68.858) & 71.687 (65.246,78.128) & 68.183 (65.836,70.529) & 67.861 (64.386,71.336)\\
Black  & 43,109 & 86.294 (80.759,91.83) & 74.42 (66.374,82.466) & 81.587 (75.046,88.127) & 79.728 (68.019,91.437)\\
Asian  & 26,265 & 68.723 (63.684,73.761) & 66.792 (58.089,75.495) & 66.777 (60.785,72.769) & 65.958 (53.748,78.169)\\
Other  & 22,091 & 72.284 (66.895,77.674) & 79.723 (69.505,89.942) & 75.924 (69.729,82.118) & 75.314 (63.089,87.539)\\
\hline
\textbf{Ethnicity}  &    &    &    &    &  \\
Non-Hisp  & 808,098 & 68.699 (67.697,69.701) & 71.999 (66.066,77.933) & 69.337 (67.166,71.507) & 68.555 (64.866,72.244)\\
Hispanic  & 28,963 & 75.681 (70.488,80.873) & 72.068 (63.599,80.536) & 74.545 (69.145,79.944) & 75.449 (66.582,84.316)\\
\hline
\textbf{Education}  &    &    &    &    &  \\
<High school  & 50,943 & 61.108 (58.134,64.083) & 67.647 (58.129,77.165) & 67.32 (61.588,73.051) & 68.246 (56.385,80.108)\\
HS completed  & 78,045 & 69.025 (65.979,72.071) & 86.848 (58.569,115.128) & 69.752 (64.868,74.637) & 70.399 (61.472,79.325)\\
College  & 237,206 & 68.997 (67.153,70.841) & 70.312 (64.184,76.44) & 70.712 (66.638,74.785) & 70.896 (65.722,76.069)\\
Graduate  & 326,860 & 70.859 (69.188,72.529) & 71.314 (68.333,74.296) & 71.313 (69.073,73.554) & 69.984 (65.783,74.186)\\
Post-grad  & 144,007 & 67.218 (64.984,69.451) & 64.26 (60.143,68.377) & 68.713 (64.864,72.562) & 66.496 (62.395,70.597)\\
\hline
\textbf{HH income}  &    &    &    &    &  \\
0-49  & 332,586 & 68.105 (66.553,69.658) & 75.441 (62.136,88.745) & 69.441 (65.872,73.009) & 69.049 (65.13,72.968)\\
50-99  & 309,387 & 69.755 (68.089,71.421) & 70.608 (63.639,77.578) & 70.359 (67.276,73.442) & 69.836 (66.552,73.12)\\
100-149  & 132,757 & 69.487 (66.999,71.975) & 68.685 (63.743,73.626) & 70.276 (66.911,73.642) & 69.55 (60.835,78.265)\\
150+  & 62,331 & 68.187 (65.109,71.264) & 69.772 (66.389,73.154) & 69.9 (66.158,73.643) & 70.352 (64.31,76.394)\\
\hline
\textbf{HH size}  &    &    &    &    &  \\
1 & 177,140 & 66.779 (64.452,69.106) & 80.258 (54.973,105.544) & 66.501 (62.817,70.186) & 67.607 (63.28,71.934)\\
2 & 286,106 & 67.608 (65.994,69.223) & 65.532 (61.489,69.574) & 66.781 (63.894,69.667) & 67.282 (63.371,71.193)\\
3 & 152,684 & 71.233 (68.836,73.631) & 72.398 (66.412,78.384) & 74.177 (69.507,78.848) & 71.127 (67.04,75.214)\\
4 & 143,442 & 70.161 (67.969,72.352) & 69.794 (65.273,74.315) & 69.944 (66.494,73.395) & 70.839 (65.417,76.261)\\
5+  & 77,689 & 72.012 (68.913,75.11) & 74.664 (64.68,84.648) & 76.567 (71.368,81.765) & 73.321 (68.479,78.163)\\
\hline
\textbf{Urban size}  &    &    &    &    &  \\
<50k  & 34,987 & 67.602 (62.771,72.432) & 79.22 (59.18,99.26) & 65.75 (59.749,71.751) & 66.109 (57.069,75.149)\\
50-200k  & 119,970 & 62.608 (60.337,64.879) & 65.759 (61.25,70.268) & 65.151 (62.164,68.138) & 67.211 (61.409,73.014)\\
200-500k  & 44,578 & 68.576 (63.52,73.632) & 87.248 (73.018,101.477) & 68.884 (63.664,74.104) & 69.636 (61.746,77.526)\\
500-1000k  & 276,629 & 68.017 (66.289,69.745) & 66.524 (61.364,71.685) & 68.123 (65.323,70.923) & 70.338 (64.971,75.704)\\
1000k+  & 360,897 & 71.928 (70.451,73.404) & 70.91 (67.926,73.894) & 73.567 (71.441,75.693) & 72.962 (68.493,77.43)\\
\hline
\textbf{Vehicle make}  &    &    &    &    &  \\
American  & 290,228 & 66.507 (64.905,68.108) & 71.826 (59.917,83.734) & 68.256 (65.302,71.21) & 69.04 (63.968,74.113)\\
Asian  & 528,810 & 70.265 (69,71.53) & 72.7 (69.653,75.747) & 71.602 (69.436,73.768) & 70.211 (66.415,74.007)\\
European  & 18,023 & 69.261 (63.898,74.624) & 66.191 (59.703,72.679) & 71.403 (65.95,76.855) & 69.836 (60.506,79.166)\\
\hline
\textbf{Vehicle type}  &    &    &    &    &  \\
Car  & 610,245 & 68.686 (67.539,69.834) & 73.853 (65.931,81.776) & 69.706 (67.4,72.012) & 70.236 (66.799,73.673)\\
Van  & 27,866 & 69.2 (64.432,73.968) & 68.389 (61.064,75.714) & 73.096 (66.388,79.804) & 64.905 (54.298,75.512)\\
SUV  & 158,202 & 68.993 (66.851,71.134) & 68.424 (62.145,74.704) & 69.291 (66.318,72.263) & 69.469 (64.351,74.587)\\
Pickup  & 40,748 & 72.361 (66.713,78.008) & 69.934 (59.062,80.805) & 74.495 (64.949,84.04) & 70.256 (58.87,81.643)\\
\hline
\textbf{Fuel type}  &    &    &    &    &  \\
Gas/D  & 761,292 & 68.637 (67.61,69.664) & 71.334 (66.221,76.446) & 69.895 (67.66,72.131) & 69.443 (65.954,72.931)\\
Other  & 75,769 & 71.986 (68.598,75.373) & 82.674 (72.987,92.361) & 77.039 (72.37,81.708) & 75.696 (67.822,83.571)\\
\hline
\textbf{Weekend}  &    &    &    &    &  \\
Weekday  & 712,411 & 67.671 (66.701,68.64) & 70.362 (65.734,74.991) & 68.72 (66.616,70.824) & 68.348 (64.806,71.89)\\
Weekend  & 124,650 & 76.196 (75.001,77.392) & 78.646 (71.128,86.164) & 77.649 (75.099,80.199) & 76.577 (73.08,80.074)\\
\bottomrule
\end{tabular}}
    \begin{tablenotes}
      \small
      \item
    \end{tablenotes}
  \end{threeparttable}
\end{table}

\begin{table}[hbt!]
\caption{Mean daily trip distance (mile) and associated 95\% CIs by different covariates across DR adjustment methods}\label{tab:6}
\begin{threeparttable}
\setlength{\tabcolsep}{3pt}
\scriptsize{\begin{tabular}{l l l l l l}
\toprule
\textbf{ } & \textbf{ } & \textbf{Unweighted} & \textbf{GLM-AIPW-PAPP} & \textbf{GLM-AIPW-PMLE} & \textbf{BART--AIPW-PAPP}\\
\textbf{Covariate} & \textbf{n} & \textbf{(95\%CI)} & \textbf{(95\%CI)} & \textbf{(95\%CI)} & \textbf{(95\%CI)}\\
\midrule
\textbf{Total}  & 837,061 & 32.418 (31.823,33.013) & 33.76 (31.806,35.715) & 33.39 (32.22,34.56) & 32.926 (31.185,34.667)\\
\hline
\textbf{Gender}  &    &    &    &    &  \\
Male  & 407,312 & 33.852 (32.963,34.741) & 35.51 (32.247,38.773) & 34.782 (33.146,36.418) & 34.358 (32.254,36.461)\\
Female  & 429,749 & 31.06 (30.27,31.849) & 31.901 (29.932,33.871) & 31.947 (30.601,33.293) & 31.428 (28.965,33.89)\\
\hline
\textbf{Age group}  &    &    &    &    &  \\
16-24  & 311,106 & 32.828 (32,33.657) & 34.904 (33.085,36.723) & 34.864 (33.358,36.369) & 34.491 (32.804,36.178)\\
25-34  & 117,758 & 36.246 (34.603,37.888) & 36.546 (33.364,39.728) & 34.841 (32.742,36.94) & 35.324 (32.837,37.81)\\
35-44  & 61,908 & 35.958 (33.318,38.597) & 31.774 (28.585,34.962) & 35.067 (32.321,37.813) & 33.9 (30.173,37.627)\\
45-54  & 77,903 & 36.103 (34.231,37.976) & 35.721 (31.46,39.981) & 34.301 (31.843,36.759) & 34.578 (30.108,39.049)\\
55-64  & 63,891 & 35.037 (32.735,37.34) & 33.159 (29.352,36.966) & 33.138 (30.097,36.178) & 32.135 (29.896,34.375)\\
65-74  & 88,762 & 31.548 (29.552,33.544) & 33.875 (24.893,42.856) & 29.948 (26.934,32.962) & 29.028 (26.593,31.464)\\
75+  & 115,733 & 22.269 (21.044,23.493) & 20.184 (18.108,22.259) & 21.037 (19.304,22.77) & 21.421 (18.979,23.863)\\
\hline
\textbf{Race}  &    &    &    &    &  \\
White  & 745,596 & 32.189 (31.554,32.824) & 33.173 (30.862,35.484) & 33.426 (32.11,34.743) & 32.85 (31.118,34.582)\\
Black  & 43,109 & 37.275 (34.577,39.973) & 35.696 (31.364,40.029) & 34.187 (31.146,37.228) & 34.131 (28.834,39.427)\\
Asian  & 26,265 & 30.638 (28.095,33.181) & 29.601 (25.527,33.675) & 28.311 (25.238,31.383) & 28.289 (22.62,33.958)\\
Other  & 22,091 & 32.789 (29.699,35.879) & 37.919 (30.975,44.862) & 35.518 (30.553,40.484) & 35.058 (27.876,42.239)\\
\hline
\textbf{Ethnicity}  &    &    &    &    &  \\
Non-Hisp  & 808,098 & 32.328 (31.723,32.933) & 32.852 (30.823,34.881) & 33.217 (32.085,34.349) & 32.362 (30.845,33.879)\\
Hispanic  & 28,963 & 34.935 (31.713,38.158) & 36.882 (32.766,40.997) & 35.126 (31.226,39.027) & 36.344 (29.859,42.828)\\
\hline
\textbf{Education}  &    &    &    &    &  \\
<High school  & 50,943 & 25.659 (23.905,27.412) & 28.248 (24.014,32.482) & 28.977 (25.986,31.967) & 30.351 (22.902,37.8)\\
HS completed  & 78,045 & 32.04 (30.14,33.939) & 36.853 (29.19,44.515) & 33.596 (30.989,36.203) & 33.497 (30.336,36.658)\\
College  & 237,206 & 31.848 (30.812,32.885) & 33.666 (30.509,36.824) & 33.038 (31.177,34.899) & 32.956 (30.622,35.29)\\
Graduate  & 326,860 & 33.879 (32.85,34.908) & 35.56 (32.73,38.389) & 35.093 (33.48,36.706) & 34.091 (31.938,36.244)\\
Post-grad  & 144,007 & 32.637 (31.235,34.039) & 29.935 (27.6,32.269) & 32.801 (30.877,34.725) & 31.414 (29.555,33.273)\\
\hline
\textbf{HH income}  &    &    &    &    &  \\
0-49  & 332,586 & 31.185 (30.273,32.097) & 32.845 (28.788,36.901) & 31.979 (30.333,33.626) & 31.538 (28.932,34.144)\\
50-99  & 309,387 & 33.024 (32.004,34.043) & 35.811 (32.755,38.866) & 33.907 (32.201,35.613) & 33.744 (32.011,35.476)\\
100-149  & 132,757 & 33.765 (32.235,35.295) & 33.354 (30.228,36.479) & 34.433 (32.472,36.394) & 33.532 (30.736,36.329)\\
150+  & 62,331 & 33.124 (31.231,35.017) & 31.693 (29.605,33.781) & 33.795 (31.147,36.442) & 33.428 (29.886,36.97)\\
\hline
\textbf{HH size}  &    &    &    &    &  \\
1 & 177,140 & 30.588 (29.231,31.945) & 34.322 (27.864,40.779) & 31.133 (28.899,33.366) & 30.768 (28.385,33.152)\\
2 & 286,106 & 32.415 (31.372,33.458) & 31.701 (29.362,34.039) & 32.742 (30.989,34.494) & 32.301 (30.95,33.651)\\
3 & 152,684 & 33.786 (32.452,35.12) & 34.54 (30.838,38.242) & 34.806 (32.647,36.966) & 34.421 (31.549,37.293)\\
4 & 143,442 & 32.95 (31.524,34.376) & 32.048 (29.257,34.84) & 33.04 (31.09,34.99) & 32.898 (30.012,35.784)\\
5+  & 77,689 & 32.934 (31.314,34.554) & 36.522 (32.397,40.647) & 36.383 (33.774,38.993) & 34.731 (32.103,37.359)\\
\hline
\textbf{Urban size}  &    &    &    &    &  \\
<50k  & 34,987 & 36.147 (32.885,39.408) & 34.93 (28.343,41.518) & 34.077 (30.506,37.648) & 32.945 (30.263,35.628)\\
50-200k  & 119,970 & 31.028 (29.388,32.668) & 32.379 (29.47,35.288) & 32.032 (29.692,34.372) & 32.636 (30.058,35.215)\\
200-500k  & 44,578 & 36.416 (33.616,39.216) & 44.143 (36.054,52.231) & 35.585 (33.228,37.942) & 36.461 (32.151,40.771)\\
500-1000k  & 276,629 & 31.973 (30.952,32.994) & 32.453 (27.672,37.234) & 31.005 (29.331,32.679) & 32.781 (28.04,37.522)\\
1000k+  & 360,897 & 32.366 (31.497,33.236) & 32.475 (30.577,34.373) & 32.371 (31.117,33.626) & 32.302 (30.134,34.471)\\
\hline
\textbf{Vehicle make}  &    &    &    &    &  \\
American  & 290,228 & 30.948 (29.995,31.9) & 35.285 (31.317,39.254) & 32.784 (31.234,34.335) & 32.956 (30.909,35.004)\\
Asian  & 528,810 & 33.249 (32.476,34.022) & 33.339 (31.554,35.124) & 34.022 (32.623,35.42) & 33.124 (31.156,35.092)\\
European  & 18,023 & 31.719 (28.896,34.542) & 29.905 (26.439,33.37) & 32.979 (30.006,35.951) & 32.05 (27.154,36.946)\\
\hline
\textbf{Vehicle type}  &    &    &    &    &  \\
Car  & 610,245 & 32.126 (31.428,32.823) & 33.619 (31.253,35.985) & 32.916 (31.691,34.141) & 32.518 (30.426,34.611)\\
Van  & 27,866 & 31.212 (28.225,34.199) & 29.109 (24.54,33.679) & 32.682 (28.052,37.312) & 31.109 (24.519,37.699)\\
SUV  & 158,202 & 32.848 (31.558,34.137) & 33.857 (30.466,37.249) & 33.15 (31.374,34.926) & 32.559 (29.986,35.133)\\
Pickup  & 40,748 & 35.958 (32.813,39.103) & 35.086 (30.375,39.798) & 36.557 (32.917,40.197) & 36.352 (31.036,41.668)\\
\hline
\textbf{Fuel type}  &    &    &    &    &  \\
Gas/D  & 761,292 & 32.121 (31.502,32.739) & 33.524 (31.522,35.526) & 33.18 (32.006,34.354) & 32.813 (31.021,34.605)\\
Other  & 75,769 & 35.409 (33.302,37.515) & 43.864 (37.513,50.214) & 39.259 (35.942,42.576) & 37.388 (34.271,40.505)\\
\hline
\textbf{Weekend}  &    &    &    &    &  \\
Weekday  & 712,411 & 31.895 (31.307,32.482) & 33.181 (31.327,35.036) & 32.817 (31.666,33.968) & 32.41 (30.68,34.14)\\
Weekend  & 124,650 & 35.41 (34.689,36.132) & 37.037 (34.351,39.724) & 36.64 (35.09,38.19) & 35.853 (33.806,37.899)\\
\bottomrule
\end{tabular}}
    \begin{tablenotes}
      \small
      \item
    \end{tablenotes}
  \end{threeparttable}
\end{table}

\begin{table}[hbt!]
\caption{Mean daily average speed (MPH) of trips and associated 95\% CIs by different covariates across DR adjustment methods}\label{tab:7}
\begin{threeparttable}
\setlength{\tabcolsep}{3pt}
\scriptsize{\begin{tabular}{l l l l l l}
\toprule
\textbf{ } & \textbf{ } & \textbf{Unweighted} & \textbf{GLM-AIPW-PAPP} & \textbf{GLM-AIPW-PMLE} & \textbf{BART--AIPW-PAPP}\\
\textbf{Covariate} & \textbf{n} & \textbf{(95\%CI)} & \textbf{(95\%CI)} & \textbf{(95\%CI)} & \textbf{(95\%CI)}\\
\midrule
\textbf{Total}  & 837,061 & 25.03 (24.8,25.261) & 25.775 (24.277,27.274) & 25.562 (25.063,26.06) & 25.39 (24.309,26.471)\\
\hline
\textbf{Gender}  &    &    &    &    &  \\
Male  & 407,312 & 25.474 (25.137,25.811) & 26.964 (24.667,29.261) & 26.199 (25.578,26.82) & 25.999 (24.877,27.12)\\
Female  & 429,749 & 24.61 (24.297,24.923) & 24.463 (23.239,25.688) & 24.906 (24.32,25.492) & 24.76 (23.59,25.93)\\
\hline
\textbf{Age group}  &    &    &    &    &  \\
16-24  & 311,106 & 25.239 (24.893,25.586) & 25.876 (24.878,26.873) & 25.921 (25.287,26.555) & 25.831 (24.724,26.938)\\
25-34  & 117,758 & 26.951 (26.318,27.583) & 27.242 (26.062,28.422) & 26.571 (25.723,27.419) & 27.07 (26.055,28.085)\\
35-44  & 61,908 & 26.065 (25.183,26.947) & 25.045 (23.196,26.894) & 25.696 (24.675,26.716) & 25.516 (23.327,27.705)\\
45-54  & 77,903 & 26.527 (25.727,27.328) & 27.731 (22.197,33.265) & 26.412 (25.419,27.405) & 25.665 (23.242,28.088)\\
55-64  & 63,891 & 26.22 (25.471,26.969) & 26.075 (24.48,27.67) & 26.275 (25.152,27.398) & 25.525 (24.078,26.971)\\
65-74  & 88,762 & 23.956 (23.339,24.572) & 22.601 (20.487,24.716) & 23.618 (22.683,24.554) & 23.216 (22.2,24.232)\\
75+  & 115,733 & 21.12 (20.559,21.681) & 20.545 (19.251,21.838) & 21.317 (20.539,22.094) & 20.728 (19.29,22.167)\\
\hline
\textbf{Race}  &    &    &    &    &  \\
White  & 745,596 & 25.109 (24.863,25.354) & 25.225 (24.255,26.196) & 26.086 (25.565,26.608) & 25.656 (24.95,26.363)\\
Black  & 43,109 & 24.227 (23.188,25.265) & 27.223 (22.295,32.151) & 23.198 (22.202,24.194) & 23.433 (20.47,26.397)\\
Asian  & 26,265 & 24.35 (23.278,25.423) & 25.203 (23.46,26.947) & 23.038 (21.674,24.402) & 24.508 (21.082,27.934)\\
Other  & 22,091 & 24.76 (23.473,26.046) & 25.049 (23.008,27.091) & 24.68 (23.128,26.232) & 25.956 (22.168,29.744)\\
\hline
\textbf{Ethnicity}  &    &    &    &    &  \\
Non-Hisp  & 808,098 & 25.039 (24.805,25.274) & 25.023 (24.156,25.89) & 25.674 (25.197,26.152) & 25.246 (24.437,26.055)\\
Hispanic  & 28,963 & 24.777 (23.526,26.028) & 27.808 (24.042,31.574) & 25.233 (23.873,26.593) & 26.284 (22.881,29.688)\\
\hline
\textbf{Education}  &    &    &    &    &  \\
<High school  & 50,943 & 23.16 (22.31,24.01) & 23.142 (21.364,24.919) & 23.825 (22.322,25.328) & 24.791 (21.638,27.943)\\
HS completed  & 78,045 & 25.192 (24.438,25.945) & 24.851 (22.707,26.995) & 26.025 (25.019,27.031) & 25.165 (22.538,27.793)\\
College  & 237,206 & 24.506 (24.09,24.921) & 25.888 (22.586,29.189) & 24.988 (24.184,25.793) & 24.7 (23.717,25.683)\\
Graduate  & 326,860 & 25.426 (25.043,25.81) & 26.769 (25.764,27.775) & 26.363 (25.795,26.93) & 26.368 (25.077,27.659)\\
Post-grad  & 144,007 & 25.569 (25.027,26.11) & 25.031 (23.955,26.107) & 25.446 (24.775,26.117) & 25.555 (24.399,26.711)\\
\hline
\textbf{HH income}  &    &    &    &    &  \\
0-49  & 332,586 & 24.333 (23.956,24.709) & 23.975 (22.766,25.183) & 24.659 (23.851,25.467) & 24.401 (22.918,25.884)\\
50-99  & 309,387 & 25.25 (24.878,25.623) & 27.547 (24.479,30.615) & 25.971 (25.316,26.627) & 25.744 (24.828,26.66)\\
100-149  & 132,757 & 25.963 (25.411,26.515) & 25.892 (24.611,27.174) & 26.281 (25.569,26.994) & 25.981 (24.275,27.687)\\
150+  & 62,331 & 22.937 (22.564,23.31) & 25.096 (21.747,28.444) & 23.419 (22.632,24.206) & 23.288 (22.044,24.533)\\
\hline
\textbf{HH size}  &    &    &    &    &  \\
1 & 177,140 & 23.837 (23.337,24.337) & 23.986 (22.176,25.797) & 24.355 (23.538,25.173) & 24.024 (22.597,25.452)\\
2 & 286,106 & 25.155 (24.746,25.563) & 25.606 (24.679,26.532) & 25.778 (25.128,26.428) & 25.55 (24.735,26.365)\\
3 & 152,684 & 25.77 (25.223,26.316) & 26.042 (24.785,27.3) & 25.645 (24.886,26.403) & 26.035 (24.807,27.262)\\
4 & 143,442 & 25.423 (24.895,25.952) & 25.025 (23.669,26.381) & 25.766 (25.015,26.517) & 25.622 (24.254,26.991)\\
5+  & 77,689 & 25.112 (24.48,25.745) & 27.472 (22.365,32.58) & 26.14 (24.981,27.3) & 25.155 (23.23,27.08)\\
\hline
\textbf{Urban size}  &    &    &    &    &  \\
<50k  & 34,987 & 28.437 (27.061,29.813) & 25.595 (22.943,28.247) & 27.951 (26.354,29.548) & 27.097 (25.536,28.659)\\
50-200k  & 119,970 & 24.455 (23.814,25.096) & 25.081 (23.851,26.31) & 24.784 (23.965,25.603) & 25.031 (24.155,25.907)\\
200-500k  & 44,578 & 27.64 (26.634,28.645) & 27.073 (23.546,30.601) & 27.024 (26.049,27.999) & 26.931 (24.582,29.279)\\
500-1000k  & 276,629 & 25.758 (25.355,26.162) & 26.189 (23.701,28.678) & 25.05 (24.289,25.812) & 25.513 (24.121,26.904)\\
1000k+  & 360,897 & 20.451 (18.941,21.961) & 23.557 (21.359,25.755) & 21.9 (20.41,23.389) & 23.488 (20.639,26.336)\\
\hline
\textbf{Vehicle make}  &    &    &    &    &  \\
American  & 290,228 & 24.799 (24.402,25.195) & 27.212 (24.331,30.094) & 25.766 (25.047,26.485) & 25.353 (24.079,26.627)\\
Asian  & 528,810 & 25.174 (24.884,25.464) & 24.771 (23.69,25.853) & 25.509 (25.004,26.015) & 25.464 (24.358,26.569)\\
European  & 18,023 & 24.534 (23.553,25.514) & 24.974 (23.083,26.866) & 24.291 (22.942,25.64) & 25.307 (23.285,27.329)\\
\hline
\textbf{Vehicle type}  &    &    &    &    &  \\
Car  & 610,245 & 24.893 (24.622,25.164) & 25.115 (24.327,25.904) & 25.313 (24.794,25.832) & 25.357 (24.467,26.247)\\
Van  & 27,866 & 23.562 (22.539,24.586) & 23.064 (21.378,24.75) & 23.484 (22.376,24.591) & 23.527 (20.832,26.223)\\
SUV  & 158,202 & 25.398 (24.87,25.925) & 26.495 (22.622,30.369) & 25.635 (24.887,26.384) & 25.008 (23.511,26.505)\\
Pickup  & 40,748 & 26.43 (25.484,27.375) & 26.245 (23.43,29.059) & 26.628 (25.453,27.804) & 25.788 (23.842,27.733)\\
\hline
\textbf{Fuel type}  &    &    &    &    &  \\
Gas/D  & 761,292 & 24.955 (24.711,25.199) & 25.727 (24.205,27.249) & 25.507 (25.005,26.01) & 25.361 (24.277,26.446)\\
Other  & 75,769 & 25.784 (25.091,26.476) & 27.804 (26.114,29.493) & 27.052 (25.991,28.113) & 26.676 (24.691,28.66)\\
\hline
\textbf{Weekend}  &    &    &    &    &  \\
Weekday  & 712,411 & 25.077 (24.847,25.308) & 25.744 (24.351,27.138) & 25.598 (25.1,26.096) & 25.425 (24.36,26.49)\\
Weekend  & 124,650 & 24.76 (24.518,25.003) & 25.939 (23.843,28.034) & 25.356 (24.811,25.901) & 25.194 (23.987,26.401)\\
\bottomrule
\end{tabular}}
    \begin{tablenotes}
      \small
      \item
    \end{tablenotes}
  \end{threeparttable}
\end{table}

\begin{table}[hbt!]
\caption{Mean start time of the first daytrips and associated 95\% CIs by different covariates across DR adjustment methods}\label{tab:8}
\begin{threeparttable}
\setlength{\tabcolsep}{3pt}
\scriptsize{\begin{tabular}{l l l l l l}
\toprule
\textbf{ } & \textbf{ } & \textbf{Unweighted} & \textbf{GLM-AIPW-PAPP} & \textbf{GLM-AIPW-PMLE} & \textbf{BART--AIPW-PAPP}\\
\textbf{Covariate} & \textbf{n} & \textbf{(95\%CI)} & \textbf{(95\%CI)} & \textbf{(95\%CI)} & \textbf{(95\%CI)}\\
\midrule
\textbf{Total}  & 837,061 & 13.811 (13.763,13.859) & 13.564 (13.391,13.737) & 13.553 (13.427,13.68) & 13.5 (13.364,13.636)\\
\hline
\textbf{Gender}  &    &    &    &    &  \\
Male  & 407,312 & 13.824 (13.751,13.898) & 13.556 (13.304,13.807) & 13.572 (13.418,13.725) & 13.486 (13.304,13.667)\\
Female  & 429,749 & 13.799 (13.736,13.861) & 13.578 (13.362,13.793) & 13.533 (13.386,13.681) & 13.515 (13.389,13.64)\\
\hline
\textbf{Age group}  &    &    &    &    &  \\
16-24  & 311,106 & 14.411 (14.354,14.468) & 14.396 (14.254,14.537) & 14.351 (14.218,14.485) & 14.266 (14.13,14.402)\\
25-34  & 117,758 & 13.999 (13.891,14.106) & 13.864 (13.562,14.165) & 13.923 (13.734,14.112) & 13.843 (13.65,14.037)\\
35-44  & 61,908 & 13.57 (13.399,13.741) & 13.694 (13.178,14.211) & 13.467 (13.164,13.77) & 13.448 (13.117,13.78)\\
45-54  & 77,903 & 13.489 (13.368,13.61) & 13.414 (13.028,13.8) & 13.389 (13.187,13.592) & 13.335 (13.043,13.626)\\
55-64  & 63,891 & 13.344 (13.185,13.503) & 13.59 (13.248,13.933) & 13.279 (13.004,13.555) & 13.27 (12.902,13.637)\\
65-74  & 88,762 & 13.244 (13.091,13.397) & 12.529 (12.082,12.975) & 13.131 (12.874,13.388) & 13.026 (12.663,13.388)\\
75+  & 115,733 & 13.047 (12.9,13.193) & 13.412 (13.089,13.736) & 13.051 (12.815,13.287) & 13.216 (12.904,13.528)\\
\hline
\textbf{Race}  &    &    &    &    &  \\
White  & 745,596 & 13.77 (13.719,13.821) & 13.522 (13.331,13.712) & 13.506 (13.372,13.64) & 13.46 (13.318,13.602)\\
Black  & 43,109 & 14.065 (13.887,14.242) & 13.632 (13.387,13.876) & 13.695 (13.49,13.901) & 13.662 (13.21,14.114)\\
Asian  & 26,265 & 14.351 (14.135,14.567) & 14.281 (13.405,15.157) & 14.051 (13.522,14.58) & 13.841 (13.51,14.172)\\
Other  & 22,091 & 14.055 (13.786,14.324) & 13.349 (12.959,13.739) & 13.585 (13.254,13.917) & 13.492 (12.695,14.289)\\
\hline
\textbf{Ethnicity}  &    &    &    &    &  \\
Non-Hisp  & 808,098 & 13.803 (13.754,13.851) & 13.563 (13.359,13.767) & 13.547 (13.419,13.675) & 13.49 (13.362,13.617)\\
Hispanic  & 28,963 & 14.049 (13.81,14.288) & 13.602 (13.303,13.901) & 13.544 (13.278,13.81) & 13.558 (13.084,14.031)\\
\hline
\textbf{Education}  &    &    &    &    &  \\
<High school  & 50,943 & 14.3 (14.177,14.424) & 13.601 (13.416,13.786) & 13.604 (13.394,13.814) & 13.513 (13.106,13.921)\\
HS completed  & 78,045 & 13.895 (13.73,14.06) & 13.425 (12.957,13.893) & 13.509 (13.236,13.781) & 13.478 (13.072,13.884)\\
College  & 237,206 & 14.003 (13.913,14.092) & 13.641 (13.36,13.922) & 13.611 (13.433,13.788) & 13.532 (13.339,13.725)\\
Graduate  & 326,860 & 13.695 (13.617,13.774) & 13.68 (13.378,13.982) & 13.65 (13.5,13.799) & 13.558 (13.42,13.696)\\
Post-grad  & 144,007 & 13.539 (13.431,13.648) & 13.307 (12.878,13.737) & 13.369 (13.197,13.542) & 13.399 (13.122,13.677)\\
\hline
\textbf{HH income}  &    &    &    &    &  \\
0-49  & 332,586 & 13.891 (13.809,13.973) & 13.62 (13.357,13.882) & 13.641 (13.465,13.817) & 13.612 (13.339,13.886)\\
50-99  & 309,387 & 13.745 (13.669,13.822) & 13.573 (13.341,13.805) & 13.55 (13.395,13.705) & 13.469 (13.323,13.615)\\
100-149  & 132,757 & 13.777 (13.671,13.882) & 13.383 (13.062,13.704) & 13.424 (13.243,13.605) & 13.415 (13.247,13.584)\\
150+  & 62,331 & 13.531 (13.437,13.625) & 13.201 (12.949,13.454) & 13.457 (13.277,13.636) & 13.342 (13.14,13.544)\\
\hline
\textbf{HH size}  &    &    &    &    &  \\
1 & 177,140 & 13.649 (13.533,13.765) & 13.337 (12.98,13.694) & 13.518 (13.349,13.688) & 13.489 (13.276,13.703)\\
2 & 286,106 & 13.6 (13.513,13.687) & 13.462 (13.164,13.761) & 13.469 (13.275,13.663) & 13.383 (13.215,13.551)\\
3 & 152,684 & 14.02 (13.918,14.122) & 13.718 (13.351,14.085) & 13.58 (13.395,13.765) & 13.5 (13.336,13.664)\\
4 & 143,442 & 14.033 (13.941,14.125) & 13.514 (13.189,13.838) & 13.64 (13.491,13.788) & 13.542 (13.362,13.723)\\
5+  & 77,689 & 14.138 (14.017,14.259) & 13.819 (13.433,14.206) & 13.581 (13.321,13.841) & 13.73 (13.474,13.985)\\
\hline
\textbf{Urban size}  &    &    &    &    &  \\
<50k  & 34,987 & 13.52 (13.266,13.773) & 13.383 (12.795,13.972) & 13.337 (12.845,13.829) & 13.328 (13.031,13.625)\\
50-200k  & 119,970 & 13.928 (13.794,14.062) & 13.747 (13.489,14.005) & 13.842 (13.672,14.011) & 13.698 (13.522,13.873)\\
200-500k  & 44,578 & 13.918 (13.705,14.13) & 13.518 (12.933,14.103) & 13.817 (13.571,14.064) & 13.66 (13.276,14.045)\\
500-1000k  & 276,629 & 13.759 (13.678,13.84) & 13.395 (13.213,13.576) & 13.564 (13.45,13.679) & 13.503 (13.305,13.7)\\
1000k+  & 360,897 & 14.286 (13.859,14.713) & 13.451 (12.893,14.01) & 13.654 (13.317,13.992) & 13.385 (12.683,14.087)\\
\hline
\textbf{Vehicle make}  &    &    &    &    &  \\
American  & 290,228 & 13.8 (13.714,13.886) & 13.339 (13.067,13.61) & 13.455 (13.258,13.651) & 13.426 (13.221,13.631)\\
Asian  & 528,810 & 13.799 (13.741,13.858) & 13.646 (13.485,13.808) & 13.627 (13.517,13.736) & 13.561 (13.43,13.692)\\
European  & 18,023 & 14.337 (14.094,14.58) & 14.19 (13.492,14.888) & 13.684 (13.334,14.034) & 13.552 (13.171,13.933)\\
\hline
\textbf{Vehicle type}  &    &    &    &    &  \\
Car  & 610,245 & 13.849 (13.792,13.906) & 13.649 (13.445,13.854) & 13.658 (13.53,13.787) & 13.611 (13.476,13.747)\\
Van  & 27,866 & 13.588 (13.336,13.84) & 13.414 (13.064,13.764) & 13.472 (13.172,13.772) & 13.514 (13.168,13.861)\\
SUV  & 158,202 & 13.773 (13.675,13.871) & 13.623 (13.279,13.966) & 13.497 (13.336,13.657) & 13.528 (13.264,13.793)\\
Pickup  & 40,748 & 13.714 (13.536,13.893) & 13.725 (13.41,14.04) & 13.544 (13.305,13.783) & 13.502 (13.079,13.925)\\
\hline
\textbf{Fuel type}  &    &    &    &    &  \\
Gas/D  & 761,292 & 13.841 (13.791,13.891) & 13.565 (13.389,13.741) & 13.556 (13.428,13.685) & 13.498 (13.36,13.636)\\
Other  & 75,769 & 13.51 (13.338,13.683) & 13.525 (13.217,13.833) & 13.463 (13.254,13.672) & 13.56 (13.264,13.856)\\
\hline
\textbf{Weekend}  &    &    &    &    &  \\
Weekday  & 712,411 & 13.824 (13.775,13.872) & 13.576 (13.408,13.744) & 13.558 (13.431,13.684) & 13.502 (13.364,13.64)\\
Weekend  & 124,650 & 13.74 (13.685,13.794) & 13.496 (13.22,13.773) & 13.531 (13.397,13.664) & 13.486 (13.334,13.637)\\
\bottomrule
\end{tabular}}
    \begin{tablenotes}
      \small
      \item
    \end{tablenotes}
  \end{threeparttable}
\end{table}

\begin{table}[hbt!]
\caption{Mean daily maximum speed (MPH) and associated 95\% CIs by different covariates across DR adjustment methods}\label{tab:9}
\begin{threeparttable}
\setlength{\tabcolsep}{3pt}
\scriptsize{\begin{tabular}{l l l l l l}
\toprule
\textbf{ } & \textbf{ } & \textbf{Unweighted} & \textbf{GLM-AIPW-PAPP} & \textbf{GLM-AIPW-PMLE} & \textbf{BART--AIPW-PAPP}\\
\textbf{Covariate} & \textbf{n} & \textbf{(95\%CI)} & \textbf{(95\%CI)} & \textbf{(95\%CI)} & \textbf{(95\%CI)}\\
\midrule
\textbf{Total}  & 837,061 & 59.808 (59.467,60.149) & 61.547 (59.717,63.377) & 60.447 (59.833,61.062) & 59.947 (58.623,61.27)\\
\hline
\textbf{Gender}  &    &    &    &    &  \\
Male  & 407,312 & 60.187 (59.706,60.669) & 62.687 (59.483,65.89) & 60.847 (60.045,61.649) & 60.677 (58.953,62.402)\\
Female  & 429,749 & 59.448 (58.969,59.928) & 60.28 (59.195,61.366) & 60.023 (59.205,60.84) & 59.193 (58.034,60.353)\\
\hline
\textbf{Age group}  &    &    &    &    &  \\
16-24  & 311,106 & 61.475 (60.97,61.98) & 62.484 (61.235,63.733) & 62.212 (61.442,62.981) & 62.078 (60.788,63.368)\\
25-34  & 117,758 & 63.41 (62.567,64.253) & 62.907 (61.082,64.733) & 62.359 (61.171,63.546) & 62.373 (60.598,64.148)\\
35-44  & 61,908 & 62.617 (61.358,63.878) & 62.761 (60.791,64.731) & 63.986 (62.235,65.737) & 62.039 (59.911,64.166)\\
45-54  & 77,903 & 60.872 (59.853,61.89) & 64.943 (58.295,71.591) & 60.117 (58.688,61.545) & 59.738 (57.435,62.041)\\
55-64  & 63,891 & 59.478 (58.406,60.55) & 59.666 (57.225,62.107) & 59.611 (57.872,61.35) & 58.797 (56.332,61.262)\\
65-74  & 88,762 & 55.91 (55.068,56.753) & 56.915 (55.026,58.805) & 55.693 (54.645,56.742) & 55.5 (53.256,57.744)\\
75+  & 115,733 & 52.613 (51.8,53.426) & 52.602 (50.493,54.71) & 52.88 (51.871,53.889) & 52.262 (50.92,53.604)\\
\hline
\textbf{Race}  &    &    &    &    &  \\
White  & 745,596 & 59.449 (59.092,59.806) & 60.312 (59.478,61.145) & 60.254 (59.581,60.929) & 59.586 (58.217,60.954)\\
Black  & 43,109 & 64.628 (62.991,66.264) & 68.229 (60.656,75.802) & 62.22 (60.3,64.14) & 62.152 (58.85,65.453)\\
Asian  & 26,265 & 61.08 (59.323,62.836) & 60.573 (58.414,62.733) & 59.081 (56.873,61.288) & 59.464 (55.818,63.11)\\
Other  & 22,091 & 61.008 (59.099,62.917) & 60.986 (58.585,63.387) & 59.968 (58.45,61.485) & 60.988 (55.744,66.231)\\
\hline
\textbf{Ethnicity}  &    &    &    &    &  \\
Non-Hisp  & 808,098 & 59.718 (59.37,60.066) & 60.221 (59.395,61.048) & 60.232 (59.595,60.869) & 59.528 (58.485,60.571)\\
Hispanic  & 28,963 & 62.31 (60.707,63.914) & 66.308 (60.971,71.645) & 61.976 (60.418,63.533) & 62.437 (58.69,66.184)\\
\hline
\textbf{Education}  &    &    &    &    &  \\
<High school  & 50,943 & 58.103 (56.954,59.251) & 59.199 (57.013,61.386) & 58.949 (57.325,60.572) & 60.162 (57.18,63.145)\\
HS completed  & 78,045 & 59.865 (58.83,60.901) & 61.116 (59.657,62.576) & 60.812 (59.457,62.166) & 60.382 (57.673,63.092)\\
College  & 237,206 & 59.874 (59.25,60.497) & 62.623 (58.482,66.764) & 59.982 (58.9,61.065) & 59.491 (57.762,61.219)\\
Graduate  & 326,860 & 60.185 (59.62,60.751) & 61.474 (60.049,62.898) & 61.405 (60.379,62.43) & 60.718 (59.63,61.807)\\
Post-grad  & 144,007 & 59.414 (58.566,60.262) & 59.809 (58.019,61.598) & 60.113 (58.801,61.426) & 59.221 (57.561,60.881)\\
\hline
\textbf{HH income}  &    &    &    &    &  \\
0-49  & 332,586 & 59.127 (58.575,59.68) & 59.271 (57.864,60.679) & 59.263 (58.317,60.208) & 58.757 (57.339,60.175)\\
50-99  & 309,387 & 60.031 (59.461,60.6) & 64.22 (60.289,68.151) & 60.94 (60.026,61.853) & 60.508 (58.903,62.113)\\
100-149  & 132,757 & 60.663 (59.901,61.425) & 60.409 (58.784,62.035) & 61.507 (60.192,62.822) & 60.611 (59.169,62.052)\\
150+  & 62,331 & 60.513 (59.305,61.721) & 61.386 (59.529,63.244) & 60.484 (58.966,62.004) & 60.549 (58.951,62.147)\\
\hline
\textbf{HH size}  &    &    &    &    &  \\
1 & 177,140 & 57.902 (57.123,58.682) & 58.973 (57.29,60.655) & 58.243 (57.246,59.24) & 57.958 (56.619,59.298)\\
2 & 286,106 & 59.02 (58.421,59.619) & 60.033 (58.585,61.48) & 59.371 (58.389,60.353) & 59.371 (57.722,61.019)\\
3 & 152,684 & 61.35 (60.569,62.132) & 60.399 (58.548,62.25) & 61.373 (59.998,62.748) & 60.541 (58.797,62.285)\\
4 & 143,442 & 61.214 (60.476,61.951) & 61.592 (60.031,63.152) & 61.488 (60.377,62.599) & 61.068 (59.616,62.52)\\
5+  & 77,689 & 61.428 (60.57,62.286) & 66.669 (60.605,72.733) & 62.759 (60.967,64.551) & 60.989 (58.922,63.057)\\
\hline
\textbf{Urban size}  &    &    &    &    &  \\
<50k  & 34,987 & 60.422 (58.928,61.917) & 61.118 (58.986,63.251) & 59.964 (58.006,61.921) & 59.665 (57.238,62.093)\\
50-200k  & 119,970 & 56.12 (55.22,57.021) & 57.621 (55.544,59.698) & 57.162 (56.071,58.254) & 57.313 (55.844,58.782)\\
200-500k  & 44,578 & 62.847 (61.27,64.423) & 64.07 (60.75,67.39) & 62.507 (60.978,64.037) & 62.711 (60.338,65.086)\\
500-1000k  & 276,629 & 60.193 (59.62,60.766) & 61.073 (57.067,65.078) & 59.886 (58.928,60.846) & 60.296 (58.82,61.773)\\
1000k+  & 360,897 & 60.303 (59.8,60.807) & 62.075 (59.415,64.736) & 60.647 (59.977,61.317) & 60.287 (59.099,61.475)\\
\hline
\textbf{Vehicle make}  &    &    &    &    &  \\
American  & 290,228 & 59.36 (58.773,59.946) & 63.24 (59.63,66.85) & 60.218 (59.339,61.096) & 59.877 (58.058,61.695)\\
Asian  & 528,810 & 60.013 (59.588,60.438) & 60.796 (59.891,61.701) & 60.751 (60.095,61.407) & 60.203 (59.036,61.371)\\
European  & 18,023 & 61.016 (59.074,62.958) & 58.842 (56.171,61.514) & 58.984 (56.944,61.025) & 59.049 (55.175,62.922)\\
\hline
\textbf{Vehicle type}  &    &    &    &    &  \\
Car  & 610,245 & 59.744 (59.338,60.149) & 60.92 (60.083,61.757) & 60.44 (59.765,61.115) & 60.119 (58.854,61.383)\\
Van  & 27,866 & 57.722 (56.154,59.289) & 58.36 (55.812,60.907) & 58.674 (56.088,61.26) & 58.263 (55.586,60.94)\\
SUV  & 158,202 & 60.093 (59.36,60.825) & 62.613 (57.431,67.795) & 60.444 (59.297,61.59) & 59.511 (57.678,61.345)\\
Pickup  & 40,748 & 61.092 (59.557,62.627) & 62.97 (61.201,64.739) & 61.359 (59.346,63.371) & 60.707 (57.51,63.905)\\
\hline
\textbf{Fuel type}  &    &    &    &    &  \\
Gas/D  & 761,292 & 59.878 (59.516,60.239) & 61.537 (59.67,63.404) & 60.473 (59.842,61.105) & 59.937 (58.565,61.309)\\
Other  & 75,769 & 59.105 (58.131,60.079) & 61.685 (58.505,64.865) & 61.082 (59.975,62.189) & 60.645 (58.056,63.234)\\
\hline
\textbf{Weekend}  &    &    &    &    &  \\
Weekday  & 712,411 & 59.684 (59.344,60.023) & 61.322 (59.663,62.982) & 60.295 (59.687,60.902) & 59.801 (58.483,61.119)\\
Weekend  & 124,650 & 60.517 (60.151,60.883) & 62.809 (60.044,65.575) & 61.312 (60.601,62.023) & 60.768 (59.333,62.204)\\
\bottomrule
\end{tabular}}
    \begin{tablenotes}
      \small
      \item
    \end{tablenotes}
  \end{threeparttable}
\end{table}

\begin{table}[hbt!]
\caption{Mean daily frequency of brakes per driven mile and associated 95\% CIs by different covariates across DR adjustment methods}\label{tab:10}
\begin{threeparttable}
\setlength{\tabcolsep}{3pt}
\scriptsize{\begin{tabular}{l l l l l l}
\toprule
\textbf{ } & \textbf{ } & \textbf{Unweighted} & \textbf{GLM-AIPW-PAPP} & \textbf{GLM-AIPW-PMLE} & \textbf{BART--AIPW-PAPP}\\
\textbf{Covariate} & \textbf{n} & \textbf{(95\%CI)} & \textbf{(95\%CI)} & \textbf{(95\%CI)} & \textbf{(95\%CI)}\\
\midrule
\textbf{Total}  & 837,061 & 4.499 (4.387,4.611) & 4.356 (3.887,4.825) & 4.644 (4.345,4.942) & 4.426 (3.984,4.867)\\
\hline
\textbf{Gender}  &    &    &    &    &  \\
Male  & 407,312 & 4.415 (4.247,4.583) & 3.835 (3.139,4.531) & 4.456 (4.129,4.784) & 4.345 (3.789,4.902)\\
Female  & 429,749 & 4.579 (4.43,4.728) & 4.957 (4.518,5.396) & 4.825 (4.471,5.179) & 4.508 (4.04,4.977)\\
\hline
\textbf{Age group}  &    &    &    &    &  \\
16-24  & 311,106 & 4.283 (4.114,4.451) & 4.368 (3.735,5.001) & 4.417 (4.068,4.766) & 4.173 (3.777,4.57)\\
25-34  & 117,758 & 4.085 (3.819,4.351) & 4.201 (3.59,4.812) & 4.609 (4.084,5.133) & 3.984 (3.288,4.681)\\
35-44  & 61,908 & 4.422 (4.052,4.792) & 4.575 (3.779,5.371) & 4.643 (4.05,5.236) & 4.574 (3.905,5.243)\\
45-54  & 77,903 & 4.14 (3.846,4.435) & 3.764 (2.22,5.308) & 4.174 (3.719,4.628) & 3.997 (3.359,4.636)\\
55-64  & 63,891 & 4.565 (4.136,4.995) & 5.003 (4.102,5.904) & 4.589 (4.042,5.136) & 4.62 (3.365,5.875)\\
65-74  & 88,762 & 4.801 (4.41,5.193) & 3.522 (1.749,5.296) & 4.799 (4.195,5.402) & 4.596 (3.372,5.821)\\
75+  & 115,733 & 5.518 (5.147,5.888) & 6.902 (5.723,8.081) & 5.857 (5.25,6.464) & 6.44 (5.548,7.332)\\
\hline
\textbf{Race}  &    &    &    &    &  \\
White  & 745,596 & 4.521 (4.401,4.641) & 4.518 (4.018,5.018) & 4.583 (4.272,4.895) & 4.402 (4,4.805)\\
Black  & 43,109 & 4.366 (3.885,4.848) & 3.807 (2.117,5.496) & 5.074 (4.442,5.706) & 5.053 (4.117,5.988)\\
Asian  & 26,265 & 4.256 (3.675,4.837) & 4.349 (3.393,5.304) & 5.222 (4.061,6.383) & 4.483 (3.405,5.561)\\
Other  & 22,091 & 4.319 (3.732,4.907) & 4.197 (3.011,5.382) & 4.438 (3.574,5.302) & 3.705 (2.167,5.243)\\
\hline
\textbf{Ethnicity}  &    &    &    &    &  \\
Non-Hisp  & 808,098 & 4.488 (4.375,4.601) & 4.56 (4.117,5.003) & 4.597 (4.323,4.871) & 4.442 (4.051,4.832)\\
Hispanic  & 28,963 & 4.813 (4.037,5.588) & 3.842 (2.609,5.075) & 4.84 (3.782,5.898) & 4.3 (2.899,5.701)\\
\hline
\textbf{Education}  &    &    &    &    &  \\
<High school  & 50,943 & 4.942 (4.522,5.362) & 4.779 (3.58,5.979) & 5.209 (4.526,5.893) & 5.204 (3.91,6.498)\\
HS completed  & 78,045 & 4.163 (3.8,4.526) & 3.495 (1.656,5.335) & 4.241 (3.662,4.819) & 4.179 (3.275,5.084)\\
College  & 237,206 & 4.561 (4.347,4.775) & 4.466 (3.609,5.323) & 4.713 (4.269,5.158) & 4.604 (4.062,5.145)\\
Graduate  & 326,860 & 4.347 (4.165,4.528) & 4.174 (3.765,4.583) & 4.564 (4.201,4.927) & 4.17 (3.59,4.749)\\
Post-grad  & 144,007 & 4.769 (4.509,5.029) & 5.031 (4.514,5.548) & 4.788 (4.387,5.19) & 4.541 (3.857,5.224)\\
\hline
\textbf{HH income}  &    &    &    &    &  \\
0-49  & 332,586 & 4.542 (4.353,4.731) & 4.639 (3.669,5.608) & 4.728 (4.325,5.131) & 4.752 (4.157,5.347)\\
50-99  & 309,387 & 4.386 (4.207,4.566) & 3.75 (2.847,4.653) & 4.482 (4.098,4.866) & 4.196 (3.682,4.71)\\
100-149  & 132,757 & 4.579 (4.32,4.838) & 4.8 (4.301,5.3) & 4.743 (4.228,5.258) & 4.564 (3.933,5.194)\\
150+  & 62,331 & 4.66 (4.265,5.056) & 4.662 (3.942,5.383) & 4.721 (4.213,5.229) & 4 (3.02,4.98)\\
\hline
\textbf{HH size}  &    &    &    &    &  \\
1 & 177,140 & 4.644 (4.38,4.908) & 4.13 (2.528,5.733) & 4.852 (4.438,5.265) & 4.658 (4.058,5.258)\\
2 & 286,106 & 4.674 (4.46,4.888) & 4.855 (4.259,5.451) & 4.782 (4.302,5.262) & 4.515 (3.92,5.111)\\
3 & 152,684 & 4.328 (4.097,4.558) & 4.425 (3.937,4.913) & 4.593 (4.123,5.064) & 4.321 (3.785,4.857)\\
4 & 143,442 & 4.294 (4.064,4.524) & 4.356 (3.762,4.95) & 4.475 (4.051,4.9) & 4.201 (3.71,4.692)\\
5+  & 77,689 & 4.241 (3.949,4.533) & 3.851 (2.313,5.388) & 4.396 (3.893,4.899) & 4.459 (4.017,4.901)\\
\hline
\textbf{Urban size}  &    &    &    &    &  \\
<50k  & 34,987 & 4.051 (3.567,4.535) & 4.313 (2.858,5.768) & 4.293 (3.57,5.015) & 4.24 (3.746,4.734)\\
50-200k  & 119,970 & 4.789 (4.49,5.089) & 4.921 (4.454,5.389) & 4.761 (4.265,5.257) & 4.696 (4.02,5.372)\\
200-500k  & 44,578 & 4.241 (3.825,4.657) & 4.177 (3.147,5.206) & 4.567 (4.075,5.059) & 4.231 (3.049,5.413)\\
500-1000k  & 276,629 & 3.969 (3.793,4.145) & 3.977 (3.342,4.612) & 4.21 (3.867,4.552) & 4.171 (3.549,4.793)\\
1000k+  & 360,897 & 4.884 (4.703,5.066) & 4.539 (3.985,5.093) & 4.948 (4.599,5.297) & 4.626 (4.025,5.227)\\
\hline
\textbf{Vehicle make}  &    &    &    &    &  \\
American  & 290,228 & 4.762 (4.548,4.975) & 4.239 (3.381,5.097) & 5.007 (4.549,5.466) & 4.914 (4.347,5.482)\\
Asian  & 528,810 & 4.392 (4.261,4.524) & 4.569 (4.244,4.894) & 4.451 (4.181,4.721) & 4.206 (3.78,4.632)\\
European  & 18,023 & 3.401 (2.946,3.856) & 3.161 (2.359,3.963) & 3.664 (3.006,4.322) & 2.898 (0.958,4.837)\\
\hline
\textbf{Vehicle type}  &    &    &    &    &  \\
Car  & 610,245 & 4.504 (4.368,4.641) & 4.281 (3.65,4.913) & 4.564 (4.225,4.903) & 4.248 (3.785,4.711)\\
Van  & 27,866 & 4.435 (4.064,4.806) & 5.298 (4.287,6.31) & 4.855 (4.41,5.3) & 4.569 (3.345,5.792)\\
SUV  & 158,202 & 4.351 (4.148,4.555) & 4.222 (3.078,5.365) & 4.56 (4.215,4.904) & 4.381 (3.73,5.032)\\
Pickup  & 40,748 & 5.043 (4.391,5.696) & 4.611 (3.881,5.34) & 5.022 (4.255,5.789) & 5.226 (4.061,6.391)\\
\hline
\textbf{Fuel type}  &    &    &    &    &  \\
Gas/D  & 761,292 & 4.435 (4.319,4.551) & 4.345 (3.865,4.825) & 4.649 (4.34,4.959) & 4.426 (3.992,4.859)\\
Other  & 75,769 & 5.145 (4.737,5.553) & 4.718 (3.688,5.747) & 4.934 (4.308,5.561) & 4.388 (3.347,5.429)\\
\hline
\textbf{Weekend}  &    &    &    &    &  \\
Weekday  & 712,411 & 4.492 (4.379,4.605) & 4.365 (3.91,4.82) & 4.639 (4.339,4.938) & 4.413 (3.963,4.864)\\
Weekend  & 124,650 & 4.54 (4.427,4.654) & 4.305 (3.738,4.871) & 4.675 (4.374,4.975) & 4.497 (4.073,4.921)\\
\bottomrule
\end{tabular}}
    \begin{tablenotes}
      \small
      \item
    \end{tablenotes}
  \end{threeparttable}
\end{table}

\begin{table}[hbt!]
\caption{Mean daily percentage of stop time and associated 95\% CIs by different covariates across DR adjustment methods}\label{tab:11}
\begin{threeparttable}
\setlength{\tabcolsep}{3pt}
\scriptsize{\begin{tabular}{l l l l l l}
\toprule
\textbf{ } & \textbf{ } & \textbf{Unweighted} & \textbf{GLM-AIPW-PAPP} & \textbf{GLM-AIPW-PMLE} & \textbf{BART--AIPW-PAPP}\\
\textbf{Covariate} & \textbf{n} & \textbf{(95\%CI)} & \textbf{(95\%CI)} & \textbf{(95\%CI)} & \textbf{(95\%CI)}\\
\midrule
\textbf{Total}  & 837,061 & 25.518 (25.202,25.834) & 25.515 (24.043,26.987) & 24.949 (24.217,25.681) & 0.251 (0.242,0.26)\\
\hline
\textbf{Gender}  &    &    &    &    &  \\
Male  & 407,312 & 24.618 (24.157,25.079) & 24.048 (21.863,26.234) & 24.06 (23.158,24.961) & 0.242 (0.231,0.252)\\
Female  & 429,749 & 26.371 (25.945,26.797) & 27.107 (25.226,28.988) & 25.873 (24.968,26.779) & 0.261 (0.25,0.271)\\
\hline
\textbf{Age group}  &    &    &    &    &  \\
16-24  & 311,106 & 26.713 (26.221,27.204) & 26.551 (25.177,27.925) & 25.913 (25.109,26.716) & 0.258 (0.245,0.271)\\
25-34  & 117,758 & 25.199 (24.385,26.014) & 24.178 (22.653,25.704) & 25.013 (23.946,26.08) & 0.247 (0.232,0.262)\\
35-44  & 61,908 & 25.575 (24.528,26.621) & 27.6 (23.466,31.735) & 26.828 (25.017,28.639) & 0.265 (0.247,0.284)\\
45-54  & 77,903 & 23.406 (22.257,24.555) & 24.908 (20.144,29.672) & 22.926 (21.57,24.281) & 0.239 (0.22,0.258)\\
55-64  & 63,891 & 22.879 (21.906,23.852) & 22.949 (21.035,24.862) & 23.408 (21.72,25.095) & 0.235 (0.211,0.259)\\
65-74  & 88,762 & 24.425 (23.448,25.402) & 27.099 (23.644,30.554) & 24.739 (23.395,26.084) & 0.262 (0.246,0.279)\\
75+  & 115,733 & 26.315 (25.367,27.264) & 27.682 (25.755,29.609) & 26.185 (25.039,27.332) & 0.28 (0.264,0.296)\\
\hline
\textbf{Race}  &    &    &    &    &  \\
White  & 745,596 & 25.216 (24.882,25.55) & 25.693 (24.172,27.213) & 24.071 (23.292,24.849) & 0.245 (0.237,0.253)\\
Black  & 43,109 & 29.711 (28.421,31.001) & 27.666 (25.29,30.042) & 29.697 (28.071,31.324) & 0.294 (0.267,0.32)\\
Asian  & 26,265 & 25.989 (24.582,27.396) & 23.631 (20.325,26.937) & 26.098 (24.013,28.184) & 0.252 (0.216,0.289)\\
Other  & 22,091 & 26.955 (24.952,28.958) & 26.442 (23.616,29.269) & 26.484 (23.923,29.045) & 0.252 (0.21,0.294)\\
\hline
\textbf{Ethnicity}  &    &    &    &    &  \\
Non-Hisp  & 808,098 & 25.438 (25.118,25.758) & 25.622 (24.061,27.183) & 24.532 (23.79,25.274) & 0.251 (0.242,0.26)\\
Hispanic  & 28,963 & 27.746 (25.946,29.546) & 26.345 (24.033,28.657) & 27.102 (25.369,28.836) & 0.251 (0.224,0.277)\\
\hline
\textbf{Education}  &    &    &    &    &  \\
<High school  & 50,943 & 27.86 (26.587,29.134) & 28.96 (26.302,31.618) & 27.223 (25.326,29.119) & 0.262 (0.233,0.292)\\
HS completed  & 78,045 & 26.136 (25.022,27.249) & 28.155 (25.07,31.241) & 25.534 (24.179,26.889) & 0.263 (0.24,0.287)\\
College  & 237,206 & 26.881 (26.288,27.474) & 27.043 (24.085,30.001) & 26.128 (24.88,27.377) & 0.264 (0.253,0.276)\\
Graduate  & 326,860 & 24.96 (24.472,25.448) & 22.543 (21.102,23.983) & 23.625 (22.803,24.447) & 0.236 (0.218,0.254)\\
Post-grad  & 144,007 & 23.375 (22.656,24.094) & 24.105 (22.558,25.651) & 23.845 (22.697,24.992) & 0.237 (0.219,0.254)\\
\hline
\textbf{HH income}  &    &    &    &    &  \\
0-49  & 332,586 & 26.578 (26.059,27.098) & 27.376 (25.379,29.374) & 26.485 (25.414,27.557) & 0.265 (0.254,0.276)\\
50-99  & 309,387 & 25.205 (24.717,25.694) & 24.47 (21.599,27.341) & 24.537 (23.514,25.559) & 0.246 (0.232,0.26)\\
100-149  & 132,757 & 24.218 (23.441,24.996) & 24.214 (22.662,25.766) & 23.707 (22.615,24.799) & 0.241 (0.226,0.256)\\
150+  & 62,331 & 24.176 (22.962,25.39) & 25.297 (20.664,29.931) & 23.96 (22.384,25.536) & 0.243 (0.223,0.264)\\
\hline
\textbf{HH size}  &    &    &    &    &  \\
1 & 177,140 & 26.133 (25.442,26.823) & 26.166 (22.88,29.452) & 25.409 (24.247,26.571) & 0.257 (0.247,0.266)\\
2 & 286,106 & 24.412 (23.871,24.953) & 24.289 (23.042,25.537) & 23.693 (22.808,24.578) & 0.244 (0.234,0.254)\\
3 & 152,684 & 25.507 (24.748,26.267) & 24.228 (21.958,26.498) & 25.177 (23.648,26.705) & 0.243 (0.231,0.256)\\
4 & 143,442 & 26.236 (25.522,26.951) & 26.843 (23.645,30.042) & 25.755 (24.608,26.901) & 0.259 (0.244,0.273)\\
5+  & 77,689 & 26.882 (25.884,27.88) & 27.356 (22.56,32.152) & 25.719 (24.458,26.98) & 0.263 (0.244,0.282)\\
\hline
\textbf{Urban size}  &    &    &    &    &  \\
<50k  & 34,987 & 20.874 (19.097,22.651) & 26.022 (21.85,30.194) & 21.679 (19.496,23.862) & 0.228 (0.21,0.247)\\
50-200k  & 119,970 & 23.798 (22.902,24.694) & 23.87 (22.364,25.376) & 24.287 (22.881,25.692) & 0.239 (0.222,0.257)\\
200-500k  & 44,578 & 22.435 (21.355,23.515) & 24.487 (18.513,30.461) & 23.436 (22.035,24.837) & 0.236 (0.209,0.263)\\
500-1000k  & 276,629 & 25.334 (24.785,25.882) & 25.695 (24.531,26.86) & 26.128 (25.326,26.93) & 0.263 (0.249,0.278)\\
1000k+  & 360,897 & 27.062 (26.615,27.508) & 26.883 (25.813,27.953) & 27.43 (26.73,28.131) & 0.271 (0.26,0.283)\\
\hline
\textbf{Vehicle make}  &    &    &    &    &  \\
American  & 290,228 & 26.28 (25.728,26.831) & 24.962 (22.25,27.673) & 24.957 (23.906,26.007) & 0.255 (0.244,0.265)\\
Asian  & 528,810 & 25.075 (24.682,25.468) & 26.283 (24.65,27.917) & 24.94 (24.125,25.755) & 0.249 (0.239,0.258)\\
European  & 18,023 & 26.233 (24.82,27.646) & 23.294 (19.732,26.857) & 25.298 (22.99,27.607) & 0.243 (0.21,0.276)\\
\hline
\textbf{Vehicle type}  &    &    &    &    &  \\
Car  & 610,245 & 25.632 (25.267,25.997) & 25.738 (24.14,27.335) & 25.054 (24.321,25.788) & 0.25 (0.24,0.261)\\
Van  & 27,866 & 27.205 (25.523,28.887) & 28.585 (24.943,32.227) & 28.5 (25.898,31.103) & 0.277 (0.237,0.316)\\
SUV  & 158,202 & 25.274 (24.518,26.031) & 25.441 (22.085,28.796) & 25.053 (23.917,26.189) & 0.254 (0.241,0.267)\\
Pickup  & 40,748 & 23.596 (22.247,24.945) & 23.906 (20.704,27.107) & 23.839 (21.462,26.217) & 0.239 (0.211,0.267)\\
\hline
\textbf{Fuel type}  &    &    &    &    &  \\
Gas/D  & 761,292 & 25.777 (25.444,26.11) & 25.638 (24.128,27.147) & 25.059 (24.318,25.801) & 0.252 (0.243,0.261)\\
Other  & 75,769 & 22.913 (21.982,23.844) & 20.192 (18.427,21.956) & 22.057 (20.348,23.765) & 0.216 (0.195,0.237)\\
\hline
\textbf{Weekend}  &    &    &    &    &  \\
Weekday  & 712,411 & 25.395 (25.079,25.712) & 25.465 (24.053,26.876) & 24.83 (24.105,25.554) & 0.25 (0.241,0.259)\\
Weekend  & 124,650 & 26.218 (25.892,26.545) & 25.812 (23.822,27.803) & 25.623 (24.805,26.442) & 0.258 (0.248,0.268)\\
\bottomrule
\end{tabular}}
    \begin{tablenotes}
      \small
      \item
    \end{tablenotes}
  \end{threeparttable}
\end{table}

\clearpage

\begin{figure}[ht]
  \begin{adjustbox}{addcode={\begin{minipage}{\width}}{\caption{%
      Comparing the distribution of common auxiliary variables in SHRP2 with weighted NHTS
      }\label{fig:10}\end{minipage}},rotate=90,center}\vspace{-5mm}
      \includegraphics[width=1.3\linewidth, height=15cm]{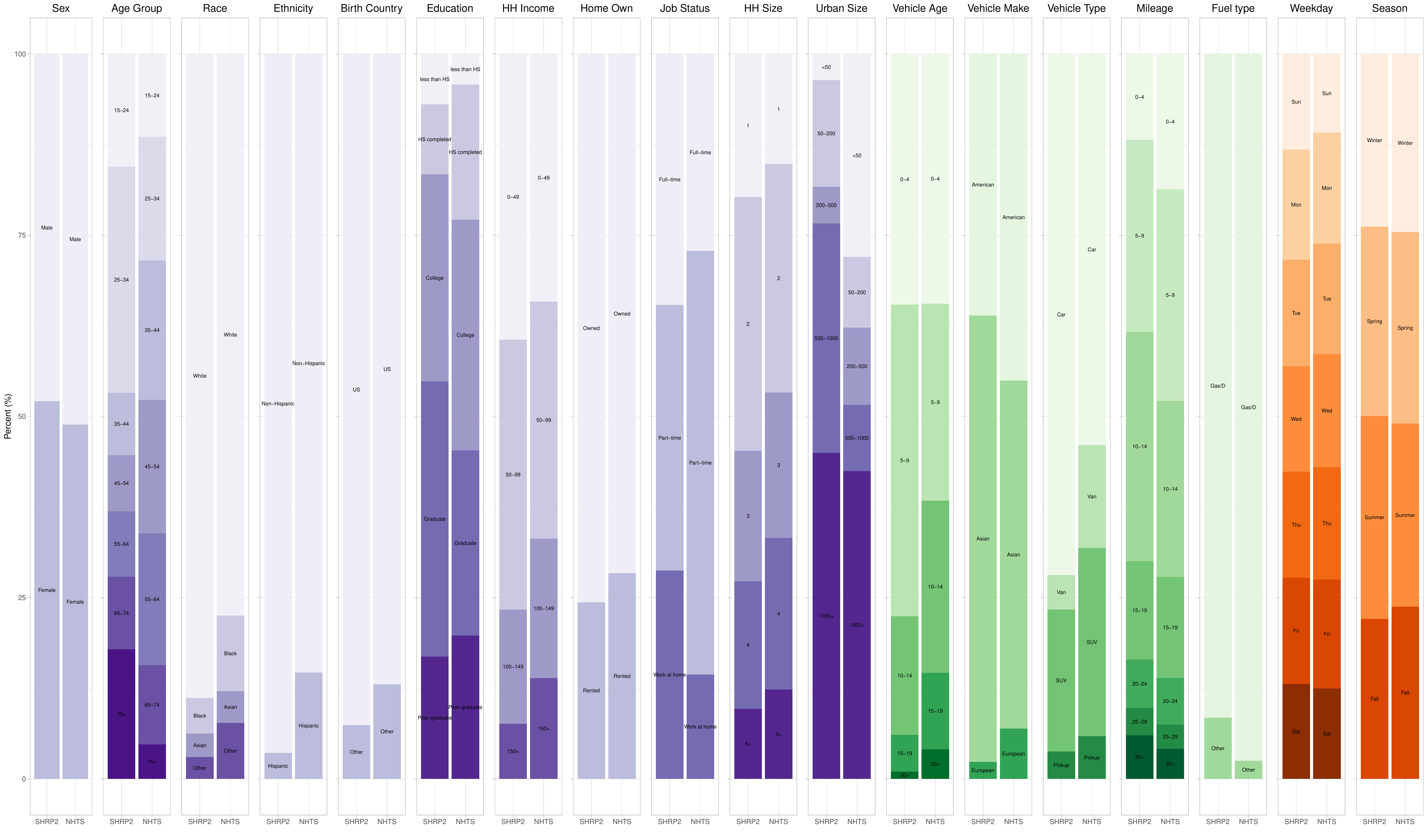}%
  \end{adjustbox}
  \vspace{-25mm}
\end{figure}

\begin{figure}[htp]
\centering\includegraphics[scale=0.61]{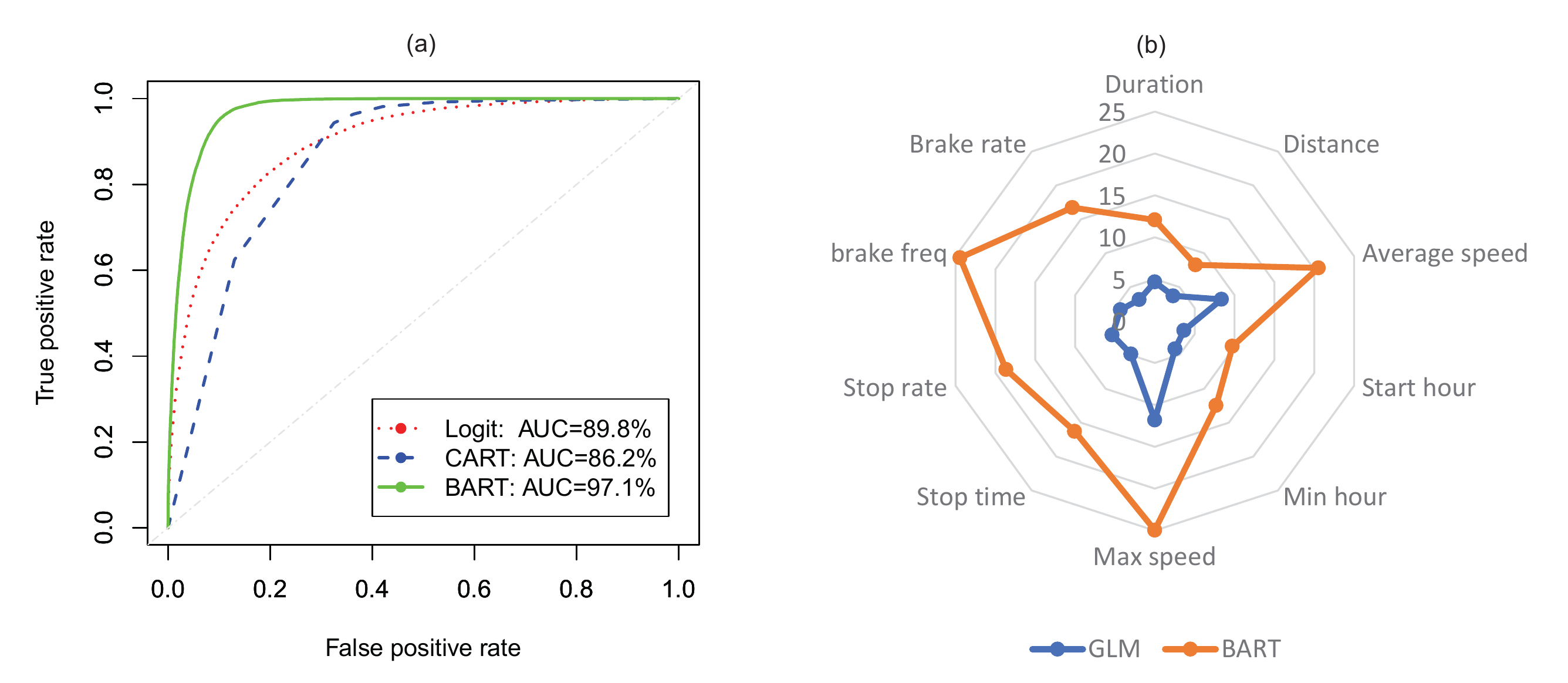}
\caption{Comparing the performance of BART vs GLM in both estimating propensity scores and predicting some trip-related outcomes. The radar plot on the right side displays the values of (pseudo-)$R^2$ between BART and GLM. AUC: area under curve; CART: classification and regression trees}\label{fig:6}
\end{figure}

\begin{figure}[ht]
  \begin{adjustbox}{addcode={\begin{minipage}{\width}}{\caption{%
      Comparing the distribution of common auxiliary variables in pseudo-weighted SHRP2 (PAPP--BART) with weighted NHTS
      }\label{fig:11}\end{minipage}},rotate=90,center}\vspace{-5mm}
      \includegraphics[width=1.3\linewidth, height=15cm]{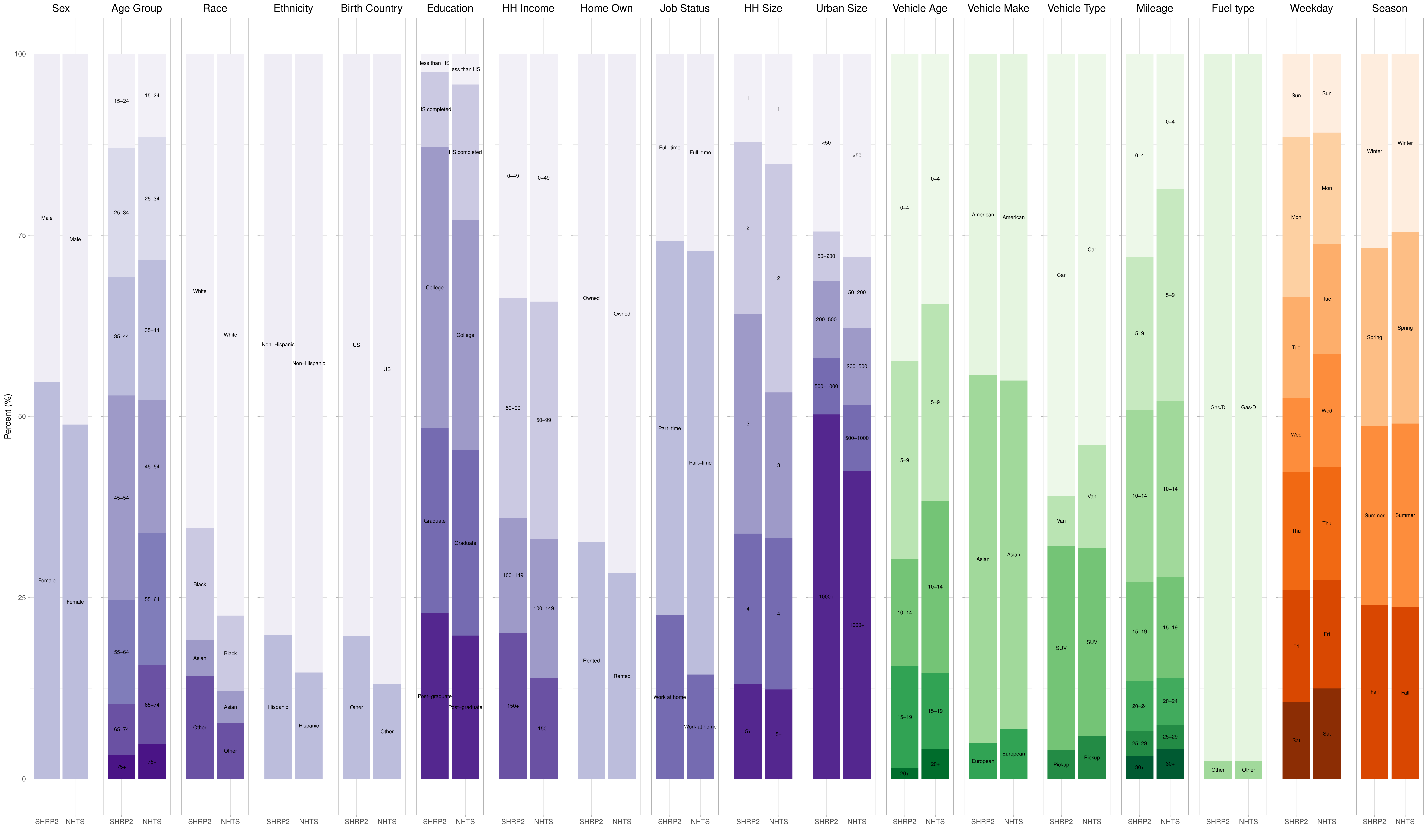}%
  \end{adjustbox}
  \vspace{-10mm}
\end{figure}

\end{document}